\documentclass[journal]{IEEEtran}
\usepackage[pdftex]{graphicx}
\usepackage{todonotes}
\usepackage[utf8]{inputenc}
\usepackage{changepage}
\usepackage{cite}
\usepackage{amsmath}
\usepackage{url}
\usepackage[acronym]{glossaries}
\usepackage{glossary-mcols}
\usepackage{glossary-tree}
\usepackage{longtable}
\usepackage{supertabular}
\usepackage{tikz}

\usepackage{array}
\newcolumntype{P}[1]{>{\centering\arraybackslash}p{#1}}
\def\checkmark{\tikz\fill[scale=0.4](0,.35) -- (.25,0) -- (1,.7) -- (.25,.15) -- cycle;} 

\makeglossaries

\newglossarystyle{clong}{%
     {\begin{itemize}}%
     {\end{itemize}}%
  \renewcommand*{\glsgroupheading}[1]{}%
}

\newacronym{mmimo}{mMIMO}{Massive Multiple-Input Multiple-Output}
\newacronym{mimo}{MIMO}{Multiple-Input Multiple-Output}
\newacronym{ue}{UE}{User Equipment}
\newacronym{mmwave}{mmWave}{Millimeter Wave}
\newacronym{haps}{HAPS}{High Altitude Platform Station}
\newacronym{uav}{UAV}{Unmanned Aerial Vehicle}
\newacronym{embb}{eMBB}{Enhanced Mobile Broadband}
\newacronym{mmtc}{mMTC}{Massive Machine Type Communication}
\newacronym{urllc}{URLLC}{Ultra Reliable Low Latency Communication}
\newacronym{iab}{IAB}{Integrated Acces- Backhaul}
\newacronym{gnb}{gNB}{Next Generation Node-B}
\newacronym{enb}{eNB}{Evolved Node-B}
\newacronym{lte}{LTE}{Long Term Evolution}
\newacronym{5g}{5G}{Fifth Generation}
\newacronym{6g}{6G}{Sixth Generation}
\newacronym{3g}{3G}{Third Generation}
\newacronym{2g}{2G}{Second Generation}
\newacronym{3gpp}{3GPP}{Third Generation Partnership Project}
\newacronym{sbs}{SBS}{Small Base Station}
\newacronym{bs}{BS}{Base Station}
\newacronym{mbs}{MBS}{Macro Base Station}
\newacronym{gprs}{GPRS}{General Packet Radio Service}
\newacronym{gsm}{GSM}{Global System for Mobile Communications}
\newacronym{tdm}{TDM}{Time Division Multiplexing}
\newacronym{bts}{BTS}{Base Transceiver Station}
\newacronym{bsc}{BSC}{Base Station Controller}
\newacronym{msc}{MSC}{Mobile Switching Center}
\newacronym{umts}{UMTS}{Universal Mobile Telecommunications System}
\newacronym{dsl}{DSL}{Digital Subscriber Line}
\newacronym{atm}{ATM}{Asynchronous Transfer Mode}
\newacronym{pon}{PON}{Passive Optical Networks}
\newacronym{rrm}{RRM}{Radio Resource Management}
\newacronym{hetnet}{HetNet}{Heterogeneous Network}
\newacronym{nr}{NR}{New Radio}
\newacronym{los}{LoS}{Line of Sight}
\newacronym{nlos}{NLoS}{Non-Line of Sight}
\newacronym{fso}{FSO}{Free Space Optics}
\newacronym{ptp}{PtP}{Point-to-Point}
\newacronym{ptmp}{PtMP}{Point-to-Multipoint}
\newacronym{csi}{CSI}{Channel State Information}
\newacronym{ofdm}{OFDM}{Orthogonal Frequency Division Multiplexing}
\newacronym{sinr}{SINR}{Signal-to-Interference+Noise Ratio}
\newacronym{ibfd}{IBFD}{In-band Full-Duplex}
\newacronym{obfd}{OBFD}{Out-of-band Full-Duplex}
\newacronym{ra}{RA}{Resource Allocation}
\newacronym{ua}{UA}{User Association}
\newacronym{mido}{MIDO}{Multiple-Input Distributed-Output}
\newacronym{fdd}{FDD}{Frequency Division Duplexing}
\newacronym{tdd}{TDD}{Time Division Duplexing}
\newacronym{kpi}{KPI}{Key Performance Indicator}
\newacronym{oab}{OAB}{Orthogonal Access \& Backhaul}
\newacronym{tdma}{TDMA}{Time Division Multiple Access}
\newacronym{vsat}{VSAT}{Very Small Aperture Terminal}
\newacronym{qos}{QoS}{Quality of Service}
\newacronym{minlp}{MINLP}{Mixed-Integer Nonlinear Programming}
\newacronym{ber}{BER}{Bit-Error Rate}
\newacronym{mec}{MEC}{Mobile Edge Computing}
\newacronym{capex}{CapEx}{Capital Expenditure}
\newacronym{opex}{OpEx}{Operating Expenses}
\newacronym{leo}{LEO}{Low Earth Orbit}
\newacronym{meo}{MEO}{Medium Earth Orbit}
\newacronym{geo}{GEO}{Geostationary Orbit}
\newacronym{sagin}{SAGIN}{Space-Air-Ground Integrated Network}
\newacronym{sdn}{SDN}{Software Defined Network}
\newacronym{nfv}{NFV}{Network Function Virtualization}
\newacronym{cp}{CP}{Content Provider}
\newacronym{roi}{ROI}{Return of Investments}
\newacronym{gsma}{GSMA}{GSM Association}
\newacronym{vnf}{VNF}{Virtual Network Function}
\newacronym{mno}{MNO}{Mobile Network Operator}
\newacronym{cu}{CU}{Centralized Unit}
\newacronym{du}{DU}{Distributed Unit}
\newacronym{ap}{AP}{Access Point}
\newacronym{irs}{IRS}{Intelligent Reflecting Surfaces}
\newacronym{hcran}{H-CRAN}{Heterogeneous Cloud Radio Access Network}
\newacronym{d2d}{D2D}{Device to Device}
\newacronym{comp}{CoMP}{Coordinated Multi-Point}
\newacronym{noma}{NOMA}{Non-Orthogonal Multiple Access}
\newacronym{awgn}{AWGN}{Additive White Gaussian Noise}
\newacronym{qam}{QAM}{Quadrature Amplitude Modulation}

\UseRawInputEncoding

\begin{document}

\title{Wireless Backhaul in 5G and Beyond:\\Issues, Challenges and Opportunities}

\author{Berke~Tezergil,
        Ertan~Onur,~\IEEEmembership{Member,~IEEE}
\thanks{B. Tezergil and E. Onur are with the Department
of Computer Engineering, Middle East Technical University, Turkey,  e-mail: \{berke.tezergil,eronur\}@metu.edu.tr.}
\thanks{A preprint of this paper can be accessed at {\tt arXiv:{2103.08234} [cs.NI]}}
}

\maketitle

\begin{abstract}
With the introduction of new technologies such as \acrfull{uav}, \acrfull{haps}, \acrfull{mmwave} frequencies, \acrfull{mmimo}, and beamforming, wireless backhaul is expected to be an integral part of the 5G networks. While this concept is nothing new, it was shortcoming in terms of performance compared to the fiber backhauling. However, with these new technologies, fiber is no longer the foremost technology for backhauling. With the projected densification of networks, wireless backhaul has become mandatory to use. There are still challenges to be tackled if wireless backhaul is to be used efficiently. Resource allocation, deployment, scheduling, power management and energy efficiency are some of these problems. Wireless backhaul also acts as an enabler for new technologies and improves some of the existing ones significantly. To name a few, rural connectivity, satellite communication, and mobile edge computing are some concepts for which wireless backhauling acts as an enabler. Small cell usage with wireless backhaul presents different security challenges. Governing bodies of cellular networks have standardization efforts going on especially for the \acrfull{iab} concept, and this is briefly mentioned. Finally, wireless backhaul is also projected to be an important part of the beyond 5G networks, and newly developed concepts such as cell-free networking, ultra-massive MIMO, and extremely dense network show this trend as well. In this survey, we present the aforementioned issues, challenges, opportunities, and applications of wireless backhaul in 5G, while briefly mentioning concepts related to wireless backhaul beyond 5G alongside with security and standardization issues.
\end{abstract}


\setlength\glsdescwidth{3in}
\printglossary[type=\acronymtype,style=clong,nopostdot,nonumberlist, nogroupskip]


\section{Introduction}
\IEEEPARstart{F}{ifth} generation (5G) networks offer 1000x more capacity and 10-100x data rates compared to the \acrfull{lte}-Advanced \cite{gupta2015survey}. This will be possible with the introduction of \acrshort{mmwave} frequencies, massive MIMO systems, directional beams and ultra-dense network structures, to name a few. With the introduction of these new technologies and concepts, wireless backhauling has become a prominent figure in future networks enabling densification with high performance than previous approaches. 

We define wireless backhauling as the use of wireless connection for core network connectivity for which fiber cables are traditionally used. Effectively, for MBS and SBS's, employing wireless backhaul means wireless connectivity in both downlink and uplink connections. Wireless backhauling is not a new concept; it was used in pre-5G networks by microwave connections or relays. In \acrshort{lte}-Advanced, self-backhauling concept was introduced enabling the backhauling of traffic using the \acrshort{lte}-Advanced radio resources.

While the concepts are defined in the previous generations, relays and self-backhauled cells did not see widespread use. This is mostly because its characteristic challenges (inter-cell interference, scheduling) were not fully solved. Furthermore, since radio link is a very scarce resource for networks using sub-6 GHz frequencies, most network operators did not want to divide their already strained spectral resources. Another point is that the projected densification did not take place until now, which allowed the traditional network structure with MBS's and User Equipments' \acrshort{ue} to persist. Also, sub-6 GHz frequencies did not meet the performance of fiber cables in terms of data transfer, which was a limiting factor. 

However, with the introduction of \acrshort{mmwave} frequencies, \acrshort{mmimo} and beamforming, the performance requirements ceased to be a limiting factor. Moreover, \acrshort{mmwave} being a new frequency band eased the frequency resources in disposal, faciliating wireless backhauling. Combined with the simultaneous connection targets of 1 million per square kilometer for \acrshort{mmtc} use-case \cite{3gpp.22.261}, network densification became a necessity and a new network architecture was required. At this point, small cells came into play. Dense deployments of small cells, whose transmission power are lower than that of the macro cells \cite{nie2016forward}, can tackle the problem of high connection volume. It is almost impossible to deploy a dense network with fiber connectivity, which necessitates the use of wireless backhauling, and hence its significance.

Even though beamforming with mMIMO solves the inter-cell interference problem, there are still significant problems that do not allow the maximum gains from wireless backhaul use. Resource allocation and user association is a problem in wireless backhauling dramatically affecting the overall performance. With the high number of SBS's, deployment is an important problem which allows massive gains from the beginning if solved correctly. Scheduling of access and backhaul traffic is still a persisting problem, and with the dense networks, its solution is even more critical to the network performance. With the green initiatives gaining momentum, reducing power consumption is not only important for the climate, but it also increases the overall performance of the network. Finally, wireless backhauling usage enables coverage improvements if the network is configured for that purpose. These challenges sometimes have conflicting repercussions, making research in this area laborious but rewarding. 

Finally, if wireless backhaul usage can be widely realized, there are various use-cases that can significantly gain from it. The usage of \acrshort{uav}/HAPSs and satellite backhauling are directly dependent on wireless backhauling performance. If wireless backhauling can enable these, then the solution to rural connectivity is within reach. 


\subsection{Scope}

We present a comprehensive overview of pre-5G wireless backhauling developments. Then, we introduce the concepts related to the wireless backhauling developments in 5G. Later, we present the challenges related to wireless backhaul in 5G. We also mention the opportunities made possible by wireless backhauling, and some applications of it. We briefly mention the security aspects of wireless backhauling and standardization efforts on it. Finally, since the research about this topic moves on to beyond-5G, we present the works on beyond-5G and \acrshort{6g} networks that may be linked with wireless backhauling.

\subsection{Motivation}

We have chosen the topic of wireless backhaul since we believe that it will be an integral part of future networks. With the introduction of mobile/nomadic base stations, \acrshort{uav}, and \acrshort{haps}, wireless backhaul use will be mandatory due to aforementioned platforms' nature. Moreover, the \acrshort{iab} concept is envisioned as a new breakthrough that will change how future networks will operate. 

Because of the importance of this topic, there are many works that touch on this issue from one perspective or another. However, there are no works we know of that comparatively gives all the available perspectives of this topic. Furthermore, the surveys in this area are mostly outdated. We wanted to present a comprehensive work on this concept covering all the angles, be it the enabled technologies, future use and current status. 

\subsection{Difference from Other Surveys and Our Contributions}

There are some previous survey papers on similar topics. For backhauling, in \cite{jaber20165g}, a discussion is made about the wireless backhaul in 5G from different perspectives, such as the evolution of the cellular network prior to this idea, the state-of-the-art that enables this concept, challenges that can be solved and available configurations that can enable the use of wireless backhaul. In \cite{siddique2015wireless}, the existing backhaul solutions in terms of different frequencies are analyzed thoroughly, along with their benefits, limitations and applications. Moreover, small cells and their backhauling is also considered from different perspectives such as deployment, flexibility, interference management, delay management and signaling overhead. An argument about the usage of half-duplexing and full-duplexing is also made. In \cite{bojic2013advanced}, Bojic et al. analyze the wireless backhaul of small cells. Wireless and optical mobile backhaul technologies are comparatively analyzed. Wang, Hossain and Bhargava have explored the small cell backhauling in 5G in \cite{wang2015backhauling}. General aspects of the backhauling are explored and some existing approaches and technologies are briefly analyzed in this work. In \cite{ahamed20185g}, Ahamed and Faruque provide a general image of 5G backhaul, with strengths and weaknesses of every alternative such as wired (e.g. fiber) and wireless (e.g. \acrshort{mmwave} or sub-6 GHz) backhaul. Polese et al. explain in detail the integrated access and backhaul concept, its internals such as architecture, network procedures, and scheduling, and through simulations, evaluated the end to end performance of an \acrshort{iab} \acrshort{mmwave} network with different scenarios \cite{polese2020integrated}. Kamel, Hamouda and Youssef present the concept of ultra-dense networks \cite{kamel2016ultra}. Kamel et al. discuss the modeling and performance of UDNs alongside the enabling technologies. 

For moving networks, UAVs, and satellites, Jaffry et al. explore the moving networks concept thoroughly in their work \cite{jaffry2020comprehensive}. This work is also up-to-date and covers the concepts related to mobile 5G systems. However, the scope of this work is limited only to land-based moving networks based on public transport vehicles. They analyze how these moving networks will improve the performance metrics and give information about the use-cases and applications of moving networks. Kurt et al. focus on the high altitude platform stations and highlight HAPS' potential on many use-cases \cite{kurt2021vision}. The authors mention how reconfigurable smart surfaces can be used alogside HAPS' and present a detailed overview of radio resource management, handover management, machine learning for HAPS', and latest developments in energy and payload systems.  Dao et al. present an overview of aerial radio access networks (ARAN) in \cite{dao2021survey}. Dao et al. describe the system model of ARANs and analyze its transmission propagation, energy consumption, latency, and mobility. Kodheli et al. present the main innovation drivers, applications and challenges of satellite communications \cite{kodheli2020satellite}. Liu et al. review the space-air-ground integrated network \cite{liu2018space}. Liu et al. discuss the network design, resource allocation, performance analysis, and technology challenges of SAGINs.

Our survey briefly introduces the backhaul concept and wireless backhauling. Then, we detail the concept that forms the basis for wireless backhauling research and highlight the crucial topics in this research area. Later, we introduce the areas that wireless backhaul can serve as an enabler. We briefly mention the security and standardization issues about wireless backhauling and finish with beyond 5G vision of the wireless backhauling. To our knowledge, there exist no surveys that are this comprehensive and touch on multiple issues as we do. The surveys mentioned above are detailed in one or more parts that we cover, but they are not as comprehensive as ours are in terms of the aforementioned perspectives. 

\subsection{Outline of the Survey}

The rest of the paper is organized as follows: In Section II, we define the backhaul concept and give brief information on its evolution in cellular networks, in Section III, we highlight the importance of wireless backhaul in 5G as an enabler to various functions, in Section IV we give the current state of wireless backhaul and the concepts that are prevalent in its research, in Section V we introduce the main part of the survey, in Sections VI-XI we present the challenges of wireless backhauling, in Section XII we show the applications for which the wireless backhaul concept serves as an enabler, in Section XIII we give brief information about the standardization efforts done on the wireless backhaul, in Section VIII we analyze the security aspects of the wireless backhaul, in Section XIV we briefly mention the evolution of wireless backhaul beyond \acrshort{5g}, and conclude the paper in Section XV.

\section{The Backhaul Concept and Its Evolution}

\subsection{An Overview of pre-5G Cellular Networks}

In the first 2G cellular networks, the mobile devices connected to the base transceiver stations (\acrshort{bts}) wirelessly. These BTS's were then connected to the base station controllers (\acrshort{bsc}), which had control plane functionality for the air interface, and a connection to the core network via mobile switching centers (\acrshort{msc}). All of this connection except the one between the mobile devices and BTS's were wired E-1 based or IP-based connections. Initially, GSM networks had the task of carrying voice but with the introduction of \acrshort{gprs}, \acrshort{gsm} networks also started carrying data. The voice used circuit-switching, which meant that required resources for the call were allocated on the call setup and then released after the call. On the contrary, data was transported on the Internet using packet switching, meaning that 2G networks had to implement and use both stacks on their network elements.

In 3G, base stations of \acrshort{umts} (which are called Node-B's) were connected with high-speed \acrshort{dsl} connections, microwave radio, fiber and microwave Ethernet lines. Moreover, the transition was to the IP-based protocols instead of legacy SS7. Like its predecessor, 3G networks also carried voice and data using the same infrastructure. This duality imposed more and more challenges in network design and operation. For example, any network element had to implement one stack for voice traffic with its own protocols and another one for data traffic. Moreover, scheduling and handling two different stacks added complexity to network management and operation. Because of these reasons, especially during 3G, the transition from circuit-switching to packet-switching became very common due to the increasing effect of data services over the whole network. With a packet-switched network, all services can be implemented on a unified core, resulting in better resource utilization and improved cost efficiency \cite{chia2009next}. 

Despite the difference in voice and data service, networks usually used the same backhaul lines to the controllers and to the core network. The ATM-based transports provided constant throughput, which was suitable for the voice traffic but not for data traffic since it had a bursty nature. Ethernet was a technology that has proven itself time and again for the data communications. In spite of this reliability, Ethernet could not be used directly in the telecommunication networks due to some inherent differences, such as the lack of synchronization signals, even though it delivered higher capacity at a lower cost per bit \cite{briggs2010carrier}. Finally, with the introduction of Carrier Ethernet, these shortcomings were solved and operators were able to unify the voice and data communication on the same network without any additional supporting protocols \cite{chia2009next}.

With the introduction of LTE, the network adopted an all-IP approach, which finally unified the voice and data duality. This unification is realized by transitioning the voice telephony to the IP side. Network design was also greatly simplified with this unification, as the network now implemented and operated only one stack. Perhaps the biggest advantage of IP architecture was that all interfaces became IP compliant, and physical infrastructure became transparent and interchangeable. These developments made it possible to change the backhaul medium quite easily and opened new possibilities to implement wireless backhauling. To give an example, using the same interface for all connections effectively meant that theoretically, a BS could connect to another BS since both used the same protocols and procedures. In the next subsection, we will detail the wireless backhaul usage for the mentioned networks.

\begin{figure*}[!t]
\includegraphics[width=7.8in]{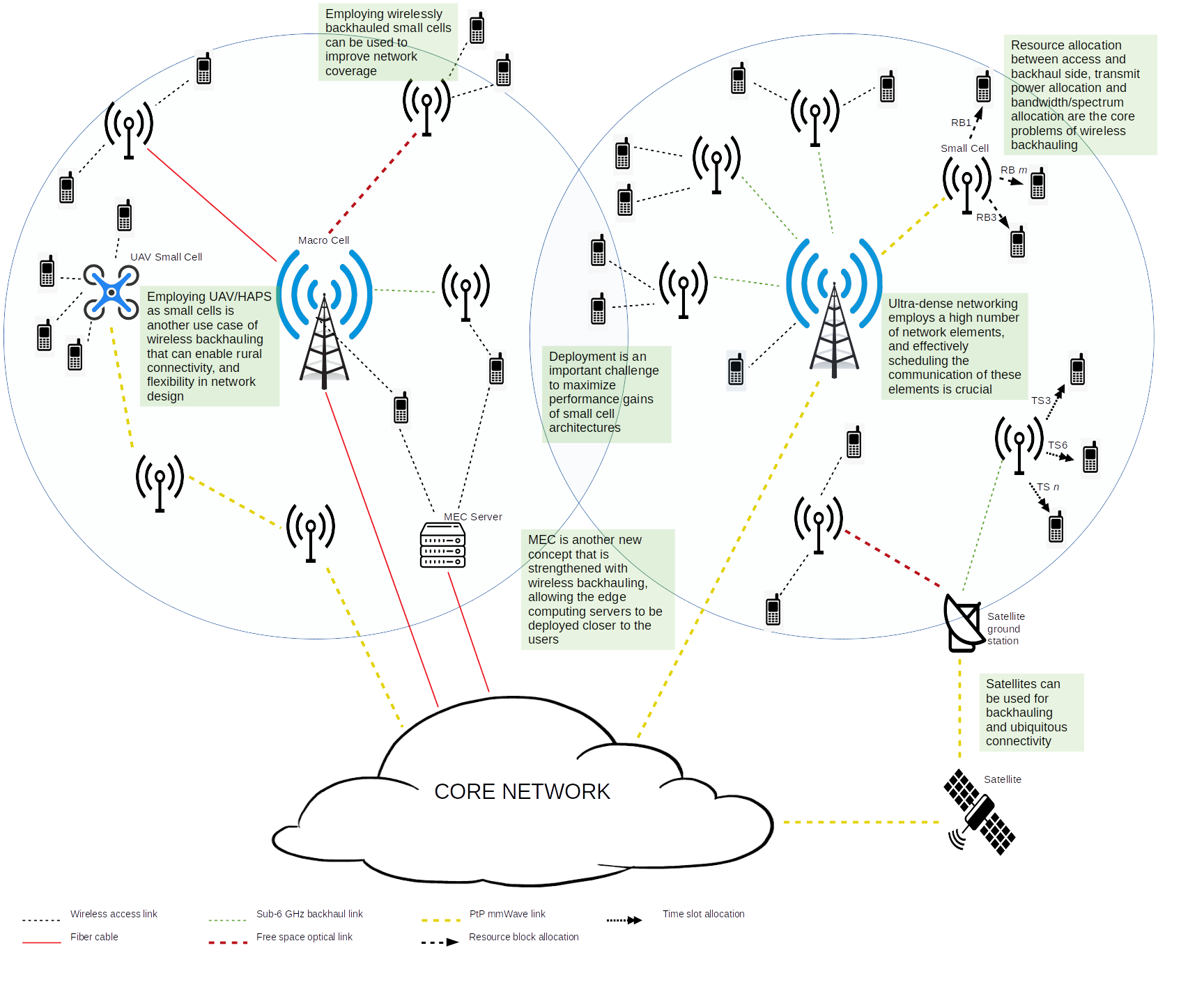}
\caption{A 5G cellular network employing different technologies for backhauling.}
\label{bh-conc:architecture}
\end{figure*}

\subsection{Wireless Backhaul in pre-5G Networks}

For GSM/GPRS networks, only wireless backhaul solution was microwave radio at this time. Although backhauling using satellite was possible, it was too expensive and cumbersome to be of wide use. At that time, passive optical networks (\acrshort{pon}) and \acrshort{mmwave} radio were seen as the future backhaul technologies enabling higher datarates with better cost-efficiency \cite{chia2009next}. However, while the \acrshort{mmwave} technology has the potential to support Gbps-level speeds, it was not fit for cellular use due to its power consumption and hardware characteristics up until 5G era \cite{pi2011introduction}. 

For \acrshort{lte} and \acrshort{lte}-Advanced networks, relays were introduced as a new network element to improve network coverage and range by employing wireless backhauling. 3GPP TR 36.806 \cite{3gpp.36.806} detailed the relay architectures. There were two types of relays: Type 1 relays had cells and identities of their own and had the same radio resource management (\acrshort{rrm}) mechanisms as the eNBs, Type 2 relays did not have identities and some parts of the RRM was controlled by their donor \acrshort{enb} \cite{3gpp.36.814}. Type 2 relays supported L1 (decode-and-forward) and L2 relaying, which did not require the implementation of the whole RRM stack and consequently made these type relays simpler. Type 1 relays also supported L3 relaying and even self-backhauling, which we will mention later.

Both Type 1 and Type 2 relays were dependent on donor eNBs for their operations, and while Type 1 was capable of independent operations with a unique identity, it was still classified as a relay, not as an eNB with wireless backhaul capabilities. This shortcoming also shows the expectation of the network maintainers from relays: they were seen as a way to extend network coverage. This is more evident in the later version of the relays in Release 11 where their operation scenarios were defined as coverage extension and indoor coverage \cite{3gpp.36.826}. 

While the proposed concepts did have some good points and were a potential game-changer for contemporary networks, \acrshort{lte} relays did not see widespread use. There are two major reasons for this: the first reason is that in pre-5G networks, frequency was a very scarce resource. \acrshort{lte} networks were stuck to sub-6 GHz frequencies for access, and this weakened the \acrshort{lte} relay concept significantly simply because using the same frequencies for backhaul meant taking away from the access side which was already strained. The second reason is that the networks did not meet the densification projections that were made for \acrshort{lte} networks. In a sense, the networks did not get dense enough to justify the deployment of wireless relays and allocation of spectrum from access to backhaul side.

To make eNB's capable of wireless backhauling, self-backhauling concept was introduced in \acrshort{lte}-Advanced. This new concept gave eNB's full relay capability: in essence, self-backhauling enabled eNB's to use the same radio resources for both access and backhaul usage. Self-backhauling improved spectrum efficiency through resource reuse, and cost efficiency through hardware and management tools reuse. However, the self-backhaul brought about a few challenges of its own, such as access-backhaul interference, and scheduling between access and backhaul \cite{mourad2016self}. Gamboa and Demirkol \cite{gamboa2018softwarized} have experimented using the \acrshort{lte} self-backhauling solution and have shown improvements in both the coverage and downlink bitrate of the network, despite self-interference and in-band communication reducing the effectiveness of the overall setup. 

The challenges introduced with self-backhauling and relays will be handled with the new technology introduced in 5G. Bhushan et al. highlight network densification as the key mechanism for future wireless evolutions \cite{bhushan2014network}. This densification is twofold: spatial densification and spectral aggregation. The spatial densification is achieved by increasing the number of antennas per node as well as increasing density of base stations deployed in a given area. The spectral aggregation is achieved when larger amounts of bandwidths are used. However, these concepts are beneficial only if the backhaul can support the denser network. The details of these developments and challenges will be given in the next section.

\section{Importance of Wireless Backhaul in 5G}

In the previous section, we mentioned some concepts introduced for wireless backhauling, namely relays and self-backhauling. In this section, we will first mention the network densification concept that goes hand in hand with wireless backhauling. We will mention why the densification is required and the challenges related with realizing it. 

With wireless backhauling, a resource allocation problem is also introduced since access and backhaul effectively uses the same wireless resources. For this problem, fixed access \& backhaul, and integrated access \& backhaul concepts are introduced and we will mention them later in this section. 

\subsection{Network Densification}

The three major use-cases of the 5G, namely \acrfull{embb}, \acrfull{urllc} and \acrshort{mmtc}, has shaped the 5G network architecture. These three use-cases necessitate the development of different technologies and concepts to meet the service requirements, sometimes with conflicting targets. The network architecture has become increasingly denser over the years, from 4-5 \acrshort{bs}/km$^2$ in 3G to an anticipated 40-50 BS/km$^2$ in 5G \cite{ge20165g}. This densification cannot be achieved with the existing macrocell architecture, and small cell architecture is introduced to realize the densification. 

There are some initial challenges related to the densification. First challenge is the cost of deployment for SBSs. Fiber backhauling solutions may be effective in terms of the connection quality it provides, but not only it is costly, but installing fiber is also a tedious process which takes time and effort. Moreover, there may be some infrastructural challenges for cabling certain areas, such as roads or densely populated urban centers. Costs can be significantly reduced by employing wireless backhauling on small cells as they do not require any extra infrastructure. Furthermore, since wireless access and backhaul can use the same medium and protocols, there is further cost reduction through hardware reuse.

Second challenge related to the densification is the performance. Densification can only be realized with wireless backhauling, but the performance of wireless backhauling with sub-6 GHz frequencies is nowhere near fiber. In wireless backhaul, a new type of interference is introduced into the system: namely the access-backhaul interference, which reduces the performance of both links. In releases 10 and 11, enhanced inter-cell interference coordination (eICIC) and further enhanced inter-cell interference coordination (feICIC) were introduced to reduce interference in order to extend the coverage range in heterogeneous networks (HetNets) \cite{nec67r1, docomo2010performance, motorola79r1}. Traditional fiber backhaul solutions do not suffer from any interference problems and perform adequately well in any situation. 

Even if the interference problem is solved, the spectrum used in \acrshort{lte} is not in a capacity to satisfy the requirements of 5G networks. With the introduction of \acrshort{mmwave} spectrum in Release 15, the wireless backhaul finally has the potential to reach the performance of its fiber counterpart. With the beam-based air communication in \acrshort{mmwave} spectra, the access-backhaul interference no longer becomes the limiting problem, as they are noise-limited unlike the interference-limited \acrshort{lte} networks \cite{singh2015tractable}.

\subsection{Access \& Backhaul Resource Management}

With wired backhauling, BSs manage the air interface for effective user access. The resources are allocated in such a way to maximize the users' throughput. However, if a BS employs wireless backhauling, then the air interface management has to balance users trying to access to the network and the BS trying to send the backhaul traffic to network core. In other words, available resources have to be allocated between the access and backhaul sides. 

Initially, two kinds of resource management solutions existed: namely fixed access \& backhaul and \acrshort{iab}. In fixed access\&backhaul, the resources that will be used for access and backhaul are separately fixed beforehand \cite{islam2017integrated}. Although the solution is viable in the wireless backhaul sense, it leaves a considerable room for improvement in some specific cases. For example, especially during rush hours, some SBSs will receive considerably more connections than others, which will strain the access resources that these SBSs have. In this case, if would be much more beneficial for the SBSs to channel the unused backhaul resources to the access side as to increase the service quality. On the other hand, when SBSs stay idle due to the number of connected UEs (or lack thereof), the unused access resources can actually be used in the backhauling of other SBSs which experience heavier loads.

\acrshort{iab} is proposed to replace fixed access \& backhaul, for it allows a flexible split of access and backhaul resources to increase the overall efficiency of the network \cite{polese2018distributed}. Moreover, another advantage of the \acrshort{iab} is that both access side and backhaul side will use exactly the same hardware and technology for their purposes, cutting off hardware costs significantly, while eliminating fiber cabling costs altogether due to employment of wireless backhauling. Specifically, \acrshort{iab} has also other use cases other than supply wireless backhaul, such as bridging the indoor-outdoor coverage, enhancing the system capacity and extending the coverage \cite{ghosh20195g}. 

Due to such advantages, \acrshort{iab} is seen as an enabler to the deployments in FR2 spectrum (24-52 GHz). Aside from being an enabler, \acrshort{iab} also adds reliability and adaptability to the system in cases of unit failure (since an alternative path can be found on the fly), transmission multiplexing capabilities since it is designed to operate at much higher frequencies than \acrshort{lte} (400 MHz per component carrier as a result of this), and multihop transmissions to extend the coverage of \acrshort{gnb}'s \cite{kim2019new}.

3GPP has also prioritized the \acrshort{iab} concept as the enabler for ultra-dense networks, and there is a study item prepared for this topic. This will be mentioned later in Section \ref{standardization-sect}.

Even though the introduction of new technologies have made wireless backhauling or \acrshort{iab} more efficient, there are new problems introduced alongside these new improvements, as well as existing problems that are not fully solved. These must be handled accordingly if network operators are to use \acrshort{iab} in their networks extensively. These concepts and challenges will be mentioned later in Sections \ref{resalloc}-\ref{wbsecurity}.

\section{Current State of Technologies Related to Wireless Backhaul in 5G}

In this section, we will analyze some aspects of the wireless backhaul technologies that are currently in use. We will explore the frequencies, topologies, and frequency use schemes. Moreover, we will also detail some of the widely used concepts that can be considered an enabler to effective wireless backhaul, such as \acrshort{mimo}.

We have chosen to elaborate on these topics before delving into the literature since these topics represent general characterstics of wireless backhaul. These concepts model the characteristics and design considerations to realize wireless backhauling; whereas the literature focuses on the actual problems such as resource allocation or user association. We find it necessary to get through these aspects to understand the research on wireless backhauling.

We will mention the frequencies used widely, from sub-6 GHz to sub-THz frequencies. We will introduce the topologies that wireless backhaul networks widely use, and the effects of different topologies on wireless backhauling. We will mention MIMO, how it works, and why it is important for wireless backhauling. Finally, we will introduce the half-duplex and full-duplex communication modes.

\subsection{Frequencies}

In Release 15 New Radio (\acrshort{nr}), the air interface is expected to perform much better than \acrshort{lte}-Advanced even with the same frequencies. NR supports maximum mobility of 500 km/h (from 350 km/h) while having ten times better minimum latency. Moreover, peak data rates without carrier aggregation for \acrshort{lte}-Advanced are 600 MB/s for downlink and 300 MB/s for uplink, whereas for NR these values are 4.9 GB/s and 2.4 GB/s, respectively. These values for NR are for the FR1 frequencies, which lie between 450 MHz and 7.125 GHz. Suppose we consider the FR2, whose frequencies lie between 24.25 and 52.6 GHz and which allows four times better maximum bandwidth \& minimum subcarrier spacing and two times better maximum subcarrier spacing. In that case, the aforementioned figures go to 10.7 GB/s and 4 GB/s, respectively \cite{kim2019new}. These figures alone show the effect of higher frequencies on the performance. 

FR2 is the first frequency band above the traditional sub-6 GHz band. Above that, there are also various other alternatives such as V-band (57-64 and 64-71 GHz), E-band (71-76 and 81-86 GHz), and D-band (110-170 GHz). From a theoretical point of view, these bands carry even better value than the FR2 due do their higher frequencies, but their use is not without challenges. For example, V-band suffers from severe oxygen absorption, E-band is affected by heavy rainfall, and D-band has relatively high atmospheric attenuation \cite{mesodiakaki2016energy}.

\begin{figure*}[!t]
\centering
\includegraphics[width=4.7in]{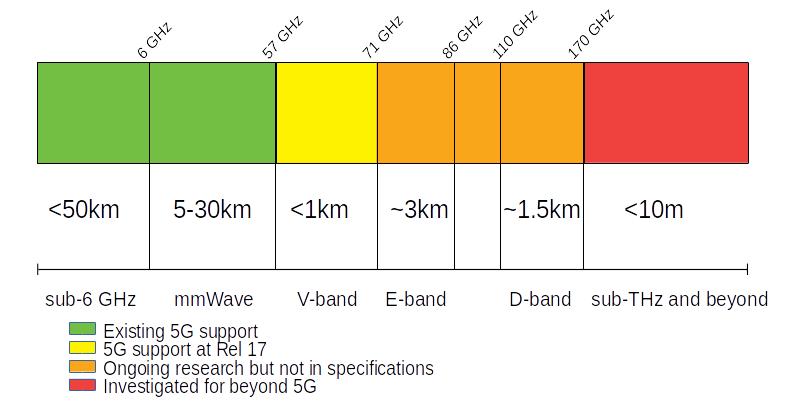}
\caption{5G spectrum from research and standardization viewpoints can be seen in the figure alongside the characteristics of different bands.}
\label{freq:spectr-5g}
\end{figure*}

It needs to be highlighted that this desire to migrate to higher frequencies came out of necessity as the lower frequencies are highly congested and cannot meet the demands of 5G networks due to their characteristics. At microwave frequencies, higher capacities came thanks to complex radio frequency techniques such as multipath propagation, channel aggregation, and cross-channel polarization. In contrast, higher frequencies deliver high throughput due to their generous spectrum \cite{bojic2013advanced}. Fig. \ref{freq:spectr-5g} highlights the spectra that are considered for 5G and beyond. An overall overview of the frequencies mentioned here can be found in Table \ref{curstate:freq-summary}.

Before further elaboration of different frequencies, we have to clarify the licensing approaches for frequency usage as this directly changes their appeal to network operators. There are three approaches for licensing. The first approach is fully licensing a spectrum. This is the most common licensing scheme, especially for sub-6 GHz frequencies. In this usage, the network operator buys a share of spectrum for exclusive use. The exclusive use allows the operator to operate interference-free.

The second licensing scheme is unlicensed use or general authorization. The advantages of this scheme are no payment for frequency use and extra spectrum availability. However, since there is no exclusive use, interference can hamper the performance \cite{gupta2015survey}. Furthermore, potentially this can be an unsolvable problem since with this kind of licensing, none of the users can claim rightful use. WiFi bands of 2.4 and 5GHz are also examples of usage adhering to this scheme. These frequencies are not used for wireless backhaul since these are not considered stable enough to be used for backhauling. Moreover, these bands are not predictable in the sense of characteristics such as jitter and latency can have large variations. 

For 60 GHz frequencies, coordination of transmit beams and channel access are centralized methods that reduce the effects of problems arising in unlicensed spectrum. These solutions can make these frequencies suitable for wireless backhauling. However, these are not adequate for multi-radio access technologies and distributed approaches are required. This unlicensed frequency range is expected to be standardized in NR Release 17 and beyond \cite{lagen2019new}, which can pave the way for wireless backhaul usage as well.

The last licensing scheme is light licensing. This is a new approach developed with the introduction of \acrshort{mmwave} frequencies. While there are no solid definitions for this licensing scheme, this one provides more flexibility and less constraint for the frequency usage, while providing some semblance of protection. Light licensing is anticipated to be used for frequency bands where the interference risk is low \cite{mucalo2012introduction}. \acrshort{mmwave} is such a frequency band due to its propagation characteristics and use of highly directive wave forms in communication.

Light licensing is also implemented differently in various regions. For example, in the United States, an automated interference analysis and verification of compliance to specific rules is made. On the contrary, in the UK, the licensee is responsible for interference analysis, reducing the regulator's role to a minimum \cite{mucalo2012introduction}. As can be seen, light licensing is understood and implemented differently in different countries.

\subsubsection{Microwave Frequencies (Sub-40 GHz)}

\acrshort{lte} systems heavily used the sub-6 GHz and higher microwave frequencies for their backhaul purposes. With developments in processing and signaling technologies, these frequencies managed to keep up with the data demand of pre-5G networks. Sub-6 GHz frequencies can meet throughput demands of up to 500-750 Mbps, while higher microwave frequencies (7-40 GHz) can exceed 1 Gbps. The main difference is that sub-6 GHz frequencies can operate in non-line of sight (\acrshort{nlos}) mode while higher frequencies require line of sight (\acrshort{los}) for seamless operation. The reason for this different operation is the directivity of higher frequencies; highly directive beams are used for communication, which is interrupted when any obstacle gets in the beam's way

Initial mMIMO and ultra-dense developments used 26-28 GHz frequencies as an enabler to get higher rates. For example, \cite{coldrey2014wireless} used 28 GHz frequencies for small cell deployments in different configurations. Similarly, \cite{taori2015point} used the same frequencies for point-to-point in-band backhauling. Moreover, this frequency range is in wide use for backhauling. Throughout the world, different bands in this frequency range are in use for wireless backhaul \cite{gsm2018mobile}. As given above, this frequency band is the first one that can be exploited to meet the metrics during the initial deployment of 5G networks. 

$K_a$-(26.5-40 GHz in IEEE definition) and $K_u$-bands (12-18 GHz) also lie in this range. These two bands are widely used in satellite communications. $K_a$-band is used for the satellite-terrestrial backhaul communication, and numerous literature about the topic can be found \cite{di2019ultradense, artiga2017spectrum, hu2019joint, yaacoub2020key}. Similarly, $K_u$-band is used for the forward links (space to earth) in satellite communications \cite{vazquez2018spectrum}.

\subsubsection{V-band (57-64 and 64-71 GHz)}

V-band consists of two sections of the spectrum whose use is not linked to any licensing. Numerous countries such as United States, Canada, and Australia have allocated a band between 57 and 64 GHz, making this band a common one throughout the world. These frequencies are set aside for unlicensed outdoor usages, which is not expected to suffer from interference as the connections are LoS and point-to-point. Moreover, both V-band and E-band are inherently resistant to outside attacks since the signal's interception will entirely block the communication, which will almost surely be notified by the network operators. This frequency range is well suited to high-capacity, short hops, making them an ideal candidate for wireless small cell backhaul. V-band potentially offers throughputs of up to 10 Gbps with a range of up to 1 km \cite{gsm2018mobile}. 

The first shortcoming of V-band is that it suffers from severe oxygen or dry air absorption, limiting its range considerably. Furthermore, V-band communication relies on LoS since the beams cannot penetrate foliage or reflect from other surfaces \cite{mesodiakaki2016energy}. Finally, since V-band is unlicensed, the vendor interest is not as high as other bands \cite{gsm2018mobile}. 

\subsubsection{E-band (71-76 and 81-86 GHz)}

E-band consists of two spectra that are lightly licensed. Contrary to the V-band, E-band is very lightly affected by the atmospheric attenuation \cite{mesodiakaki2016energy}. This band is also resistant to interference like other \acrshort{mmwave} frequencies as the communication relies on highly directive narrow beams. Spatial frequency reuse is also made possible due to this directivity property. Thanks to wider channel sizes in these frequencies, a throughput of 25+ Gbps can be reached, making this band suitable for 5G operations. Its range is also around 3 km, which is higher than that of the V-band \cite{gsm2018mobile}. 

E-band can be used effectively in different scenarios with the help of different hardware. In suburban scenarios, E-band can supply up to 10 Gbps bandwidths up to 5 miles \cite{nokia20197080ghz}. Whereas in urban scenarios, 10-20 Gbps bandwidths can be achieved up to 2 miles. This flexibility enables true 5G densification by addressing different use cases.

While not as severe, E-band suffers from heavy or excessive rainfall, limiting its range, albeit not as oxygen absorption does to the V-band. Like V-band, E-band also requires LoS communication as the beams cannot penetrate obstacles or reflect from them.

\subsubsection{W-band (92-110 GHz)}

W-band is quite similar to the D-band in terms of its characteristics. It houses multiple bands that span contiguous regions. This results in an advantage enabling high bandwidth usage. For example, two separate subbands in this band can supply 5 GHz and 7.5 GHz, respectively \cite{nokia2017wanddband}. Moreover, W-band performs slightly better than the D-band in terms of attenuation and range. However, D-band is considered a better alternative since it houses even wider subbands, enabling better performance.

\subsubsection{D-band (110-170 GHz)}

D-band is a lightly licensed frequency band consisting of multiple smaller bands. Since D-band frequencies are higher than that of V-band and E-band, a higher level of throughput is easier to achieve. This band can also potentially reach a throughput of 100 Gbit/s with certain configurations \cite{frecassetti2019d}, meaning that this band is the ideal candidate for high throughput requirements. This performance can serve not only small cells, but also make high-throughput backhauling of macro cells possible. Moreover, D-band allows contiguous frequency deployments of up to 12.5 GHz, allowing higher channel sizes along with carrier aggregation \cite{nokia2017wanddband}. 

The D-band disadvantages are as follows: D-band suffers from relatively higher atmospheric attenuation compared to the E-band . Also, appropriate beamforming usage is a must for these frequencies as the omnidirectional free space path loss is higher compared to the lower frequency bands due to Friis' law. Finally, this frequency band is relatively young compared to the lower frequency bands, and as such, vendors are not ready to use this band to its full potential as they do with lower frequency bands \cite{mesodiakaki2016energy}.

\subsubsection{Sub-THz frequencies and beyond}

Beyond D-band, some research uses free space optics (FSO) technology to replace fiber in some areas or use FSO in tandem with RF systems. FSO uses unlicensed frequencies beyond 300 GHz. It has very high bandwidth and supports full-duplex operation, making FSO an effective medium for wireless backhauling. However, FSO requires LoS for operation, and is affected by weather phenomena such as fog and dust storms. Because of such characteristics, researchers have focused on developing hybrid systems with FSO and RF, employing the best of both worlds i.e. high-capacity and low-latency of FSO and low-cost, high-coverage and high-availability of microwave/\acrshort{mmwave} RF. Examples of research using FSO can be found in \cite{dong2018edge, dahrouj2015cost, esmail20195g, morra2017impact}. Other technologies that lie in this frequency range are covered in Section \ref{beyond5g} because the research that concerns them is often labeled as beyond 5G and/or even \acrshort{6g}.

\begin{table*}[t]
\caption{An overview of the mentioned frequencies that can be employ for wireless backhaul and their characteristics. Wired alternatives are provided for comparison purposes~\cite{gsma5gguide}.}
\label{curstate:freq-summary}
\begin{tabular}{ |p{5em}||p{5em}|p{24em}|p{24em}| } 
 \hline
 Frequency & Range & Strengths & Weaknesses \\ 
 \hline
 \hline
 Microwave Frequencies & \textless 40GHz & exclusive use from licensing, high range, can operate without LoS & high costs of licensing, suffers from interference, low throughput \\ 
 \hline
 V-band & 57-71 GHz & no license costs, resistant to interference, higher bandwidths available compared to lower bands, quick deployment & no exclusive use from licensing, requires LoS, severely affected by oxygen absorption, limited range \\ 
 \hline
 E-band & 71-76 and 81-86 GHz & light-licensing, high throughput, resistant to interference, higher range than other \acrshort{mmwave} bands, adaptable to different conditions, quick deployment & requires LoS, severely affected by heavy rainfall, still less range than microwave \\ 
 \hline
 W-band & 92-110 GHz & slightly better than D-band in terms of attenuation and range, high bandwidth, resistant to interference & smaller contiguous regions available than D-band, short range \\
 \hline
 D-band & 110-170 GHz & light-licensing, contiguous regions available for higher channel sizes, high throughput, resistant to interference & affected by atmospheric attenuation, higher free space path loss compared to lower bands, less mature than lower bands, short range \\
 \hline
 Sub-THz Frequencies & 170GHz-1THz & unlicensed usage, quasi-immunity to interference, high bandwidth & very short range, LoS requirement \\ 
 \hline
 Fiber-optic & - & ranges up to 80km, quasi-immunity to interference, high throughput & very long deployment time, higher deployment costs than other wireless alternatives \\ 
 \hline
 Copper & - & ranges up to 15km, quasi-immunity to interference & very long deployment time, higher deployment costs than other wireless alternatives, performance not adequate for future deployments \\ 
 \hline
\end{tabular}
\end{table*}

\subsection{Topology}

For wireless backhaul, there are two topologies considered, which are point-to-point (\acrshort{ptp}) and point-to-multipoint (\acrshort{ptmp}). In PtP topology, the connection is typically LoS and highly directional antennas are used for this purpose. The advantages are that interference is often not a problem due to the connection's directivity and the whole bandwidth is available for transmission. However, as the number of connections increase, hardware costs increase considerably due to both sides requiring transceivers. Moreover, if LoS connection is not viable e.g. in parks where foliage is common, performance suffers considerably. PtP topology is also used frequently with tree topology and a hierarchical network structure. The root of the tree has fiber connectivity and data is sent from the leaves until a fiber-connected base station is found.

\begin{figure}[!b]
\centering
\includegraphics[width=3.3in]{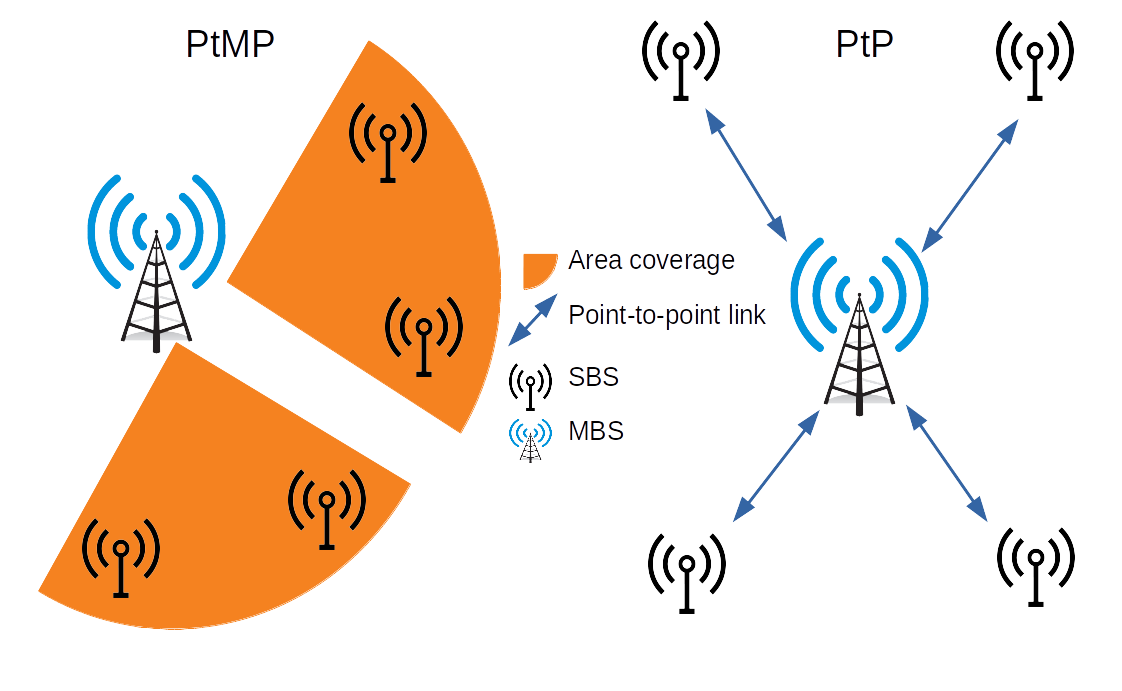}
\caption{Topology types and links on a cellular network. PtP case pairs a single transceiver to a BS, whereas PtMP gives area coverage, reducing the number of transceivers needed at the MBS.}
\label{top:ptp-ptmp}
\end{figure}

PtMP topology, on the other hand, requires only one transceiver, often on the macro base station, which is linked to the small cells. The bandwidth is shared among all small cells connected to the transceiver, which usually means that the throughput of the links will usually be lower than that of the PtP links. Setup times and number of transceivers required are also less in PtMP, meaning that the expenditure for setting up such a network is also lower \cite{gsm2018mobile}. Figure \ref{top:ptp-ptmp} exemplifies the two topologies. While the number of MBS/SBSs and their positions are the same, the links are different between two cases. In PtMP case, there is only one transceiver on the MBS supporting both LoS and NLoS connections; whereas in the PtMP case, there is one transceiver for every SBS that the MBS is associated with due to LoS connections. 

\subsection{MIMO}

The concept of using multiple antennas for communication is not a new one, being in commercial use since \acrshort{lte} \cite{lee2009mimo}. However, early MIMO systems made use of only a few (less than 10) antennas in both BS and \acrshort{ue} side \cite{swindlehurst2014mmwave}, and the technology was mainly employed to make use of multiple propagation paths. In theory, this concept can also be used to employ spatial multiplexing by serving different users instead of cooperating to serve a single user. This concept was initially known as multi-user MIMO \cite{marzetta2010noncooperative}. Later on, with the introduction of antenna arrays that house orders of magnitude more antennas, this was named massive MIMO in 5G. Massive MIMO is an important concept in wireless backhauling as it enables effective \acrshort{mmwave} usage with multiple small cells.

In 5G, with the introduction of higher frequencies, using directional antennas became a necessity for effective communication, as explained in the Section IV-A. Massive MIMO became an enabler for effective \acrshort{mmwave} frequency usage in MBSs. Massive MIMO employs a large number of small antennas used together to generate a directional beam. This approach is also known as spatial multiplexing, because the antenna array can generate multiple beams that use the same frequency. These beams do not interfere with each other if their directions are different. 

Furthermore, asymptotic arguments establish that as the number of antennas in an array approach to infinity, under certain conditions enormous gains are made. The effects of uncorrelated noise and fast fading are nullified. Cell size stops to limit throughput and number of terminals. Spectral efficiency becomes independent of bandwidth. Finally, the required transmitted energy per bit vanishes altogether \cite{swindlehurst2014mmwave}.

Another advantage of massive MIMO systems is their energy efficiency. Compared to a single-antenna BS, a massive MIMO system can theoretically achieve the same uplink rates with $\frac{1}{N_t}$, where $N_t$ is the number of antennas in the array. If imperfect channel state information (\acrshort{csi}) is assumed, the power requirement becomes $\frac{1}{\sqrt{N_t}}$, which still corresponds to significant power savings. Alternatively, the energy efficiency can be employed to extend the range \cite{swindlehurst2014mmwave}.

Massive MIMO systems improve the efficiency in a number of ways. These systems employ a high count of small amplifiers that are less complex, have low power usage, and considerably cheaper costs compared to the conventional antennas with high power usage. Moreover, thanks to the concept of beamforming and having a large number of antennas, the air latency is substantially reduced and an inherent strength against jamming(or interference) is achieved. A final advantage is that with massive MIMO usage, employing \acrshort{ofdm} results in a more efficient air channel, which reduces the physical layer signalling \cite{gupta2015survey}. These aspects result in improvements from commercial and signaling perspectives.

While the massive MIMO technology is considered as an enabler for \acrshort{mmwave} usage, it can also be used below 6 GHz frequencies. This can be advantageous since hardware components are more mature for this frequency range. However, the hardware implementation will surely be different \cite{bjornson2016mmimo}.

Massive MIMO helps wireless backhauling in the following ways. First of all, enabling \acrshort{mmwave} usage strengthens the 5G \acrshort{gnb}'s in a number of ways. With denser cell usage, a \acrshort{gnb} can serve multiple small cells or UEs with massive MIMO, employing spatial multiplexing and frequency reuse. With \acrshort{mmwave} frequencies, the performance will also be on par with what is expected in 5G networks. Furthermore, being able to extend the range with massive MIMO also helps cell-edge performance, which is an issue in the past networks. We do not highlight other characteristics of massive MIMO as it requires a detour from our original topic, but interested readers are encouraged to read \cite{bjornson2016mmimo} where massive MIMO is explained by clarifying the common misunderstandings. 

\subsection{Half- and Full-duplexing}

Due to its medium, wireless communication has to solve the medium access problem different than wired communication for efficient usage. In wired communication, this problem is arguably easier to solve since it is easier to understand whether the medium is accessed or not. If a power spike is detected on the wire, then it is concluded that someone is accessing the medium. This cannot be done on a wireless medium in this way and with this level of reliability. Furthermore, traditional antennas are able to either act as a receiver via listening the medium, or as a transmitter via generating the necessary waves. This cannot be done on the same time because an antenna can only listen to itself (since the generated wave is the closest) while generating a wave.

The aforementioned communication is half-duplex, since one party can either transmit or receive in a time instant. This is not an efficient way of communication since it cuts the potential performance by half during operation. For example, in a two-party communication, each party can only assume one role at a time, and if every party has 5 seconds worth of data to send, it takes 10 seconds in the best case to send the data. In theory, this communication could be carried out in 5 seconds if both parties were able to send their data while receiving the other party's data. 

Full-duplex communication is an important concept that can result in enormous performance gains if it can be implemented correctly. With the introduction of beamforming and directive communication, full-duplex communication can be realized with the use of spatial multiplexing. This means that irrespective of the bands used, a station can receive from one sector while transmitting from another one. Full-duplex communication can also be done in-band i.e. using the same frequency. This is harder to implement since opposite directions generate interference which makes communication difficult. However, as mentioned, potential performance gains make this research promising. 

Full-duplexing is being considered for wireless backhauling with different settings. There are examples with spatial multiplexing that are arguably easier to implement thanks to beamforming. In-band full-duplex is harder to implement because of the interference problem. This is an open research area important for future wireless communication.  

\section{Challenges of Wireless Backhaul}

Having mentioned the general concepts that are seen in most papers of wireless backhauling area, we now present the reader our survey. Further sections will present the challenges of wireless backhauling, state-of-the-art in these topics, main learning points, and open issues regarding these areas. 

For this purpose, we have highlighted six main topics: resource allocation, deployment, scheduling, performance evaluation, coverage, and security. Each of these topics have their own section with the same structure defined in the previous paragraph. These topics are especially important because their solutions are directly related to the effective realization of wireless backhauling. The specific points regarding their importance to the wireless backhauling are highlighted in the respective sections.

Wireless backhauling is an area that shares many challenges with traditional networking. Deployment, security, and coverage are examples of such challenges. Furthermore, there are challenges more specific to wireless backhauling due to different characteristics, such as scheduling of small cell backhaul traffic or deployment of small cells for high performance. Resource allocation seems to be the paramount problem for wireless backhauling as the dynamic management of resources is required to effectively use the scarce resources available. Finally, performance evaluation works are significant since these show the feasibility of wireless backhaul usage in the field without any performance losses, while also highlighting the necessary adaptations for effective field usage. 

\subsection{Resource Allocation}
\label{resalloc}

The first challenge before the implementation of any wireless backhaul system is the resource allocation problem. Resource in this context can be defined as anything that the network manages to perform its operation. Examples can be the bandwidth, energy, links and their capacities, or time-slots. Networks use these resources to optimize their performance. Frequently, the bandwidth or energy allocation has to be carefully adjusted to get the best possible performance from a wireless backhauled network. 

User association is another frequently seen aspect of the resource allocation problems. For HetNets, a UE can get service from multiple BSs at any given time since SBSs and their associated MBSs often have overlapping coverage areas. Because of this, user association becomes another decision which can be used to optimize overall network performance. In this regard, approaches may be to maximize the average sum rate or the worst performing UE. User association is also used for load balancing of various resources, meaning that suboptimal associations may take place. User association is mostly taken into account with another parameter such as transmit power allocation or spectrum allocation/scheduling. 

In this section, we gather the literature aiming to tackle the resource allocation problem; the findings are also presented in Table \ref{challenges:resalloc-summary} at the end of this subsection for readers' convenience. This section is the longest part of our survey regarding the wireless backhauling in 5G as the scope of resource allocation is quite large. Furthermore, some points are shared between the resource allocation and further sections; for example, some papers tackle the user association and scheduling problem, which makes them eligible for both sections in our paper. We mention these works in resource allocation section.

\subsubsection{State of the Art}

In \cite{bonfante2018performance}, Bonfante et al. use massive MIMO to construct a two-layer network in which small cells use wireless access and backhaul links. The two-layered system consists of \acrshort{mmimo}-BSs and small cells that use a fixed allocation of access and backhaul slots. All small cells receive their respective backhaul traffic from the base stations employing \acrshort{mmimo}. These BSs are connected to the core network with high-capacity wires and serve only the small cells, whereas the small cells only serve UEs and do not take part in backhaul of other small cells' traffic. For their system, the parameter $\alpha \in [0,1]$ defines how the available timeslots were divided between access and backhaul usage. Two different deployment scenarios are considered for small cells. In the first scenario, the small cells are randomly and uniformly distributed to the BS area; whereas in the second one, which the authors named as ad-hoc deployment, they are deliberately placed closer to UEs to increase the throughput.

Tang et al. optimize beamforming and power allocation to maximize energy efficiency in \acrshort{comp} networks doing simultaneous wireless information and power transfer (SWIPT) \cite{tang2018energy}. There are two types of users in these networks, namely the energy harvesting users (e.g. IoT devices) and information decoding users (e.g. laptops, mobile phones). Tang et al. separately consider beamforming and power allocation since the combined problem is non-convex. The authors first propose a zero forcing approach for beamforming to suppress interference for all nodes. The authors also propose a partial zero forcing approach for beamforming, which exploits interference as an energy source for energy harvesting users, while at the same time cancelling the interference of information decoding users. Simulation results show that partial zero forcing outperforms the zero forcing method. Moreover, both methods converge to their global optimals after a number of iterations. Increasing the number of small cells and information decoding users also increases the energy efficiency. Increasing the number of energy harvesting users increases the energy efficiency up to a certain number, after which efficiency sharply decreases for zero forcing, and decreases mildly for partial zero forcing. 

In \cite{zhang2017downlink}, Zhang et al. have formulated a problem to optimize the energy efficiency of wireless backhaul bandwidth and power allocation problem in a small cell. The network in question is a two-tiered network in which the MBS uses \acrshort{mmimo} to communicate with the small cells whereas small cells use OFDM to communicate with the UEs. The aforementioned problem is formulated as a nonlinear programming problem, which is shown to be nonconvex. Therefore, the authors have decomposed the problem into two convex subproblems for wireless backhaul bandwidth allocation and power allocation. Two versions of the algorithm are developed, namely the optimum iterative algorithm and low-complexity algorithm, which fixes the bandwidth allocation factor by the value calculated from the equal power allocation. It is shown that the proposed iterative algorithm gets better results in all metrics than the low-complexity one, whereas they both outperform the existing benchmark. 

Nguyen et al. study the joint design of downlink transmit beamforming and power allocation in two-tier wireless backhauled small-cell heterogeneous networks \cite{nguyen2017centralized}. Nguyen et al. first formulate the problem of maximizing the energy efficiency of the wireless backhauled small cell HetNets. The proposed power consumption model considers adaptive decoding power at each small cell access point (SAP), making the model more appropriate. The problem is non-convex and NP-hard, and Nguyen et al. propose an exhaustive search based on branch-and-bound algorithm for global optimal solution. However, this solution has high complexity, and Nguyen et al. also propose a low-complexity algorithm based on first order Taylor convex approximation (FOTCA). As the final step, Nguyen et al. also extend the problem formulation to consider the small cell selection that takes into account the impact of power to switch on/off the SAPs. Introducing this as a binary variable, the authors model this as a mixed integer second order cone programming (MISOCP) approximated problem. This problem is decomposed into subproblems by the ADMM approach and Nguyen et al. develop a distributed algorithm to solve this problem individually at every SAP. Numerical results indicate that the all proposed methods outperform the fixed power allocation scheme. Moreover, the small cell selection algorithm also outperforms others since selectively activating the SAPs results in significantly less power consumption. Second best algorithm is the adaptive decoding power. Finally, it is shown that having a high power budget usually results in the MBS consuming too much power and reducing the energy efficiency. Nguyen et al. suggest that having low budgets result in better energy efficiency since it does not allow excessive power allocation.  

In \cite{saha2018integrated}, Saha, Afshang and Dhillon explore the bandwidth partition scenarios between access and backhaul side of the network in a full wireless configuration. The authors propose a network that used sub-6 GHz frequencies for control traffic and \acrshort{mmwave} for data traffic. For the backhaul traffic of small cells, Saha et al. compare two different strategies. In the first strategy, the available bandwidth is partitioned equally to all small cells. In contrast, in the second strategy, the bandwidth is partitioned proportional to each small cell load. The authors report that the load-balanced partitioning always provides higher coverage than fixed partitioning. Furthermore, for a given partition strategy, the existence an optimal access-backhaul split  maximizing the rate coverage probability is shown. As a third point, Saha et al. show that \acrshort{iab}-enabled network outperforms the macro-only network up to a certain point, beyond which the performance gains disappear and they converge to the same point. 

Huang et al. propose a fairness-based distributed resource allocation algorithm (FDRA) \cite{huang2019fairness}. Huang et al. use stochastic geometry to model the two-tier HetNet and define the channel reuse radius based on the spectrum sensing threshold. FDRA is proposed to maximize the total throughput in the small cell tier while considering the outage probability and fairness. To model the real traffic load accurately, tidal effect and users' actual geographical distribution are jointly considered and improved FDRA is proposed to further improve the throughput in hot areas. Simulation results show that FDRA improves fairness compared to the benchmark method CSRA. In all metrics, IFDRA outperforms FDRA while FDRA in turn outperforms CSRA. 

In \cite{pham2019mobile}, Pham et al. consider the effects of resource offloading for mobile edge computing on the small cells that use wireless backhauling to a MBS. The authors highlight that the resource offloading research mostly assumes there is backhaul capacity for their purpose, but the real-life situation does not conform to that assumption. The authors decompose this joint task offloading and resource allocation into two parts. In the first part, the decision to offload the computation based on a given bandwidth factor and computation resource allocation is made; whereas in the second part, wireless backhaul bandwidth and computation resource allocation is made for a fixed offloading decision. These two sub-problems are solved individually with an iterative algorithm for the solution of the original problem. Moreover, the authors have extend the problem into multiple scenarios, such as ultra-dense networks, machine learning usage for computation offloading, and partial offloading of the computation. The proposed algorithm is compared to the given problem's two extremes where all computations are either performed locally or offloaded to the mobile edge computing server. As expected, with an increasing number of UEs, fewer offloading decisions are made, since the available resources decrease with a larger number of UEs. Furthermore, the proposed algorithm achieves lower computational overhead since its decisions are made only when offloading is advantageous. The same performance improvements are also seen in UEs' transmit power because the algorithm considers the consumed power while offloading for its decision-making.

Song, Liu and Sun consider the joint radio and computational resource allocation problem in \acrshort{noma}-based MEC in HetNets \cite{song2018joint}. Song et al. minimize the energy consumption while guaranteeing task execution latency. The optimization problem is solved using decomposition and iteratively employing alternating optimization. Simulation results show that partial offloading outperforms full offloading and local computing in terms of energy efficiency. Increasing task tolerance latency increases energy efficiency for all configurations.

In \cite{han2016backhaul}, Han et al. investigate the user association and resource allocation problem in the \acrshort{mmimo}-enabled HetNets. The network in question uses renewable energy but is also connected to the grid in case of deficiencies. Han et al. formulate a problem of network utility maximization subject to backhaul, energy, and resource constraints. The given problem is solved using first a primal decomposition, and then a Lagrange dual decomposition. In the first decomposition, the problem is divided into two parts: the lower-level problem i.e. resource allocation problem of BSs and higher level problem i.e. cell-association problem of users. The lower level problem is also divided into $N$ subproblems for individual BSs. The higher level problem is solved using the solution of resource allocation problem. To make this a distributive algorithm, the Lagrange dual decomposition is used to divide the problem into dual and master problems. In the resulting distributive algorithm, the UEs calculate and choose the optimal BS allocation at their side and report this connection to BSs; whereas BSs solve the dual problem and then broadcast the next turn's information to users. Furthermore, Han et al. design a virtual user association and resource allocation (vUARA) scheme, which reduces the communication overhead over the air interface and removes the information leaks to other users. In this scheme, after the initial resource allocations are made and user associations are reported, the measurements are reported to the radio access network controller (RANC), in which vUARA resides. RANC simulates the iterative cycles and after finding an optimal user association and resource allocation, reports this to the users and BSs. The proposed algorithm outperforms the common user-association schemes, i.e., max-\acrshort{sinr} association and range expansion association. 

Le et al. consider network and weighted sum energy efficiency of dense HetNets \cite{le2019energy}. The authors propose a resource allocation algorithm supporting both delay-sensitive users with minimum data rate requirements and delay-tolerant users with data rate fairness considerations. The problem is formulated and solved in three different ways. Mixed-integer programming, time-sharing, and sparsity-inducing formulations are employed to tackle the nonconvex functions with efficient sub-optimal solutions. All formulations make use of successive convex approximations to find the solutions iteratively. Numerical results show that the proposed formulations perform quite similar from an energy efficiency perspective in terms of static consumed power and maximum transmit power at femtocells. Mixed-integer programming formulation is shown to be superior in terms of energy efficiency to other approaches when users' QoS and number of femtocells are considered.

In \cite{omidvar2018optimal}, Omidvar et al. propose a hierarchical optimization method for HetNets with flexible backhaul. In flexible backhaul, only some network elements are connected to the core with fiber connection and the rest of the network must reach to these BSs via wireless connections with dynamic links. The authors formulate a two-timescale hierarchical \acrshort{rrm} control problem  for HetNets with flexible backhaul. For the overall problem, RRM control variables are divided in two as long-term and short-term control variables. For long-term control, flow control, routing control and discontinuous transmission control are used whereas for short-term control, link scheduling control is used. From these variables the problem is formulated. The problem is then divided into two sub-problems by using primal decomposition. The inner problem becomes the optimization of routing control and flow control under a fixed DTX control and link scheduling policy. For the outer problem, an iterative algorithm is proposed whose solution is shown to be optimal. At each subframe, the scheduling information is sent to the \acrshort{ue} and \acrshort{ue} sends back the instantaneous SINR value as feedback. At each superframe, the RRM system receives the average associated link rates of BSs and sends back the next DTX time-sharing scheme, routing and flow control updates. The proposed algorithm performs better than five chosen baseline methods in terms of network utility for different backhaul connected BS percentage and BS power. The algorithm converges faster than all baseline methods, has similar signaling overhead, is shown to be more cost-saving compared to the baseline methods and adds only little overhead in terms of CPU utilization.

Gao et al. propose a quantum coral reefs optimization algorithm for joint resource allocation and power control problem in \acrshort{d2d} HetNets \cite{gao2019joint}. Gao et al. first propose a model for cooperative D2D HetNets and then derive analytical formulas for the total throughput of this model. A combination of coral reefs optimization algorithm and quantum evolution are proposed as a novel quantum coral reefs optimization algorithm (QCROA) to optimize the resource allocation and power control problem. The strength of this approach is that the same algorithm can be used for different parameters by changing the parameter vector, which showcases the reusability of this algorithm. Simulation results show that the proposed algorithm outperforms all alternative methods in throughput under different maximum power, cellular unit and idle relay count constraints. 

In \cite{liu2016joint}, Liu et al. have considered the resource allocation problem on a two-tier network that uses \acrshort{mmimo} on the MBS and in-band full-duplex communications on small cells. User association and spectrum allocation in the aforementioned network scheme is formulated as a mixed-integer nonlinear programming (\acrshort{minlp}) problem. To solve this problem distributively and efficiently, the primary problem is divided into two sub-problems, namely the user association problem for fixed spectrum allocation, and spectrum allocation with given user association. Next, an iterative algorithm is developed to solve the original problem. The proposed algorithm is compared with the max SINR user association algorithm and shown to be outperforming it in terms of sum logarithmic rate. It is also shown that the optimal spectrum allocation outperforms any fixed spectrum allocation scheme.

Song, Ni and Sun investigate the distributed power allocation problem in non-orthogonal multiple access HetNets \cite{song2018distributed}. The power allocation problem is modeled as a two-stage Stackelberg game where MBS is the leader and SBS is the follower. MBS tries to maximize its throughput under power and minimum rate constraints whereas SBS tries to find the best response to MBS' actions. Simulations show that the proposed algorithm converges fast and is better in terms of throughput compared to Nash game, dedicated spectrum allocation, and orthogonal multiple access.

In \cite{bojic2013advanced}, Bojic et al. have proposed a dynamic resource allocation algorithm for mobile backhauled networks. The authors state that the problem with mobile backhaul resource allocation is often the assigned capacity. If this is made statically, resources are wasted in lightly loaded areas, whereas areas with heavy loads will experience congestion. For dynamic assignment, the size of the resource chunks assigned needs to be carefully considered; if the size is too large, resources will be wasted, if they are small, then too many signaling may occur. The authors propose a backhaul resource manager (BRM) that fairly allocates the resources based on the network topology, link capacities and resource allocations. BRM is linked with the network management system and/or with the backhaul nodes to receive the needed information. From the aforementioned information, the BRM will perform the capacity-aware path computation. It is shown that the dynamic algorithm outperforms SPF by 40\% and CSPF by 20-30\%.

Hu et al. propose a power allocation model to maximize energy efficiency in HetNets \cite{hu2018robust}. The problem is modeled as a probabilistic fractional programming optimization problem. Bernstein approximation and Dinkelbach method are used to transform the problem into a standard convex approximation problem. Simulation results show that the transmit power convergence is quite fast and outperforms the rate maximization-based robust power allocation algorithm in terms of energy efficiency. The proposed algorithm is slightly worse than the energy-efficiency based optimal power allocation algorithm, but the aforementioned algorithm considers perfect CSI, which is not available in the actual communication scenario.

In \cite{faruk2016energy}, Faruk et al. have thoroughly analyzed the energy consumption characteristics of macro and small cells. It is shown that wireless backhaul, especially with PtMP microwave links, results in high power consumption compared to other alternatives. BS equipment and cooling are also shown to be the most power-consuming tasks. Self-backhauling is shown to be an alternative to other wired and wireless backhaul methods due to its advantageous characteristics such as reduced deployment and maintenance costs, fast rollout and relaxed constraints compared to LoS deployments. The power consumption of small cells with fixed wireless access to MBS and self-backhauled small cells are compared. Since PtP links consume power independent from the load, they are shown to be more power-efficient than self-backhauling as the BS load increases. The break-even point is shown to be 50\% load. It is given that after 55W power consumption on fixed wireless radio links, they become less-efficient than self-backhauling solutions. This result is more evident if the number of MBSs increase. Lastly, it is shown that the implementation of PtMP links where a single PtMP unit serves several small cells results in a more power-efficient solution as the number of served small cells increase.

In \cite{wang2015backhauling}, Wang, Hossain and Bhargava have delved into the topic of radio resource management of 5G small cell backhauling. General aspects such as cell association, RAT selection, resource and interference management, cell coordination, spectrum sharing and energy-efficient backhauling are explored. Then, some existing works about these aspects are briefly analyzed. Finally, a case study is conducted with a \acrshort{mmimo}-based in-band wireless backhauling network is given. It is shown that user rates can significantly be increased with backhaul-aware cell association and bandwidth allocation schemes. 

Siddique, Tabassum and Hossain compare the performance of in-band full-duplex (\acrshort{ibfd}) and out-of-band full-duplex (\acrshort{obfd}) backhauling \cite{siddique2017downlink}. Siddique et al. formulate a problem to maximize the minimum achievable rate at the small cells in a hybrid IBFD/OBFD setting. This problem is solved in a centralized fashion by transforming the original problem into the epigraph form. The authors also solve the individual IBFD and OBFD backhauling problems, which provides useful insights into the nature of these schemes. Then, solutions are found for two distributed backhaul spectrum allocation schemes, namely maximum received signal power (max-RSP) and minimum received signal power (min-RSP). Max-RSP scheme assigns a larger amount of backhaul spectrum to closer SBSs, offering higher backhaul rates to SBSs at the cost of lowering the available access spectrum. Min-RSP scheme conversely allocates larger backhaul spectrum to the SBSs that are farther away from the MBS. Numerical results are validated using Monte Carlo simulations. These simulations show some interesting results. Hybrid allocation scheme favors IBFD for low interference and OBFD for high interference settings. For OBFD scheme, access spectrum allocation seems to be following the same pattern for lower rank (i.e. closer) SBSs as well as the higher rank (i.e. farther) SBSs, but the backhaul spectrum required is higher for higher rank SBSs due to weak backhaul signal strength. Conversely, IBFD case lower rank SBSs are limited by backhaul interference. This starts to improve with higher rank SBSs, which require higher access rates i.e. more access spectrum for higher performance. It is shown that hybrid scheme outperforms both individual schemes by prioritizing IBFD for low interference and OBFD for high interference scenarios. 

In \cite{shariat2014joint}, Shariat et al. have decomposed the overall resource allocation problem between backhaul and access links to power and sub-channel allocation. The optimization is first done between one small cell and a \acrshort{ue}, and then were generalized into multiple UEs. A novel algorithm is proposed for rate balancing that employed small cell grouping and resource slicing. The evaluation shows that joint optimization with rate balancing provides significant improvements for resource allocation between access and backhaul links. This method works not only for \acrshort{iab} systems, but it also shows significant improvements for fixed backhaul partitioning systems.

Khodmi, Benrejeb and Choukair develop a joint power allocation and relay selection algorithm for heterogeneous ultra-dense networks \cite{khodmi2018iterative}. Khodmi et al. use the increment water-filling algorithm for power allocation and a non-linear programming algorithm for resource allocation. Simulations show that multi-hop relay improves the coverage and achieves the data rate requirements of cell-edge users compared to no relay or single-hop relay architectures. 

In \cite{hao2017energy}, Hao et al. have developed an energy-efficient resource allocation algorithm for two-tiered networks that use \acrshort{mmimo} on the MBS and OFDMA-based cellular frequencies on the small cells. A hybrid analog-digital precoding scheme at MBS is proposed. From this scheme, a problem for power optimization and subchannel allocation is formulated to maximize the energy efficiency of the HetNet subject to users' \acrshort{qos} and limited wireless backhaul. The formulated problem is a MINLFP problem, and the authors made use of transformations, namely MINLFP to DCP, and then DCP into convex optimization by appropriate approximation, to solve it iteratively. In simulations, the proposed algorithm is used with two structures: fully connected structure and subarray structure, for beamforming RF chains. If subchannel allocation is not considered, then the hybrid precoding scheme with subarray structure performs the best in energy efficiency, while having the lowest throughput of three. The energy efficiency of hybrid precoding with fully connected structure is negatively correlated with the number of RF chains, whereas its throughput is positively correlated. The results are similar when subchannel allocation is considered.

Mahmood et al. propose the adaptive capacity and frequency optimization method for wireless backhaul networks \cite{mahmood2020capacity}. Conventional network planning and its shortcomings are mentioned in this work. Mahmood et al. generate synthetic time series data based on real network data. This time series data is then used to train autoregressive integrated moving average (ARIMA) and multi-layer perceptron (MLP) models. These models are then used in the automated planning algorithm. Results show that MLP models especially have very low mean absolute percentage error (MAPE) and all models have performed below the error threshold of 10. The proposed methods are shown to achieve better capacity planning and optimization as well as reduce resource wastage. 

In \cite{han2013green}, Han and Ansari have developed a system for user association and power consumption on two-tiered HetNets with green energy and on-grid power usage. The proposed algorithm works in a distributed fashion and optimizes the use of green energy and the traffic delivery latency of the network. The system tries to make use of green energy wherever possible over on-grid power. Moreover, when forced to use on-grid power, the system balances the use of on-grid power and the traffic delivery latency. The traffic arrival process is modeled to be a Poisson process; therefore, base stations realize a M/G/1-PS (processor sharing) queue. The algorithm, which the authors have named GALA, is distributed to BS and users. The BS-side algorithm measures the traffic load and updates the advertising traffic load. In contrast, the user-side algorithm selects the optimal BS based on the advertised traffic load, BS's energy-latency coefficient and BS's green traffic capacity. The developed algorithm is compared with maximum rate algorithm and $\alpha$-distributed algorithm, which consists of several optimization policies for user-BS associations to balance the flow level traffic among BSs. The GALA algorithm is shown to outperform both of the algorithms significantly in terms of power saving at the cost of negligible latency increases compared to the $\alpha$-distributed algorithm.

Ma et al. investigate the user association and resource allocation problem of MIMO-enabled HetNets \cite{ma2017backhaul}. The network in question is a two-tiered \acrshort{hetnet} consisting of a single \acrshort{mmimo} MBS and multiple small cells. Ma et al. formulate the problem of maximizing the $\alpha$-fairness network utility via user association and resource allocation. The optimization problem is formed as a mixed-integer nonlinear programming problem. Then, this problem is solved using an algorithm based on the Lagrangian dual decomposition method. Simulations show that the proposed method significantly outperforms the benchmark methods such as optimal \acrshort{ua} and equal \acrshort{ra} (where all UEs associated with a BS use the same resources) or max-SINR UA and optimal RA (UA is done solely by checking max-SINR), while falling short of the global optimum only by a very small margin.

Huskov et al. propose a smart backhaul architecture that adapts itself according to given task \cite{huskov2018experience}. The system can adapt itself to certain equipment conditions for different equipment types defined previously. This development aims to save power by reducing the amount of unused resources, while keeping network at a high performance regardless of the load. The proposed system also makes use of a concept called multiple-input distributed-output (\acrshort{mido}). MIDO aims to serve multiple devices on the same spectrum band providing interference alignment. Simulations show that the usage of MIDO with the proposed system outperforms standard MIMO systems 2.5 and 30 times for 2x2 and 8x8 configurations, respectively. However, Huskov et al. also highlight the fact that employing higher number of antennas at receiver side is unrealistic.

Zhang et al. design a two step resource allocation scheme for mmWave enabled IAB networks \cite{zhang2021backhaul}. User association and transmit power allocation problem is iteratively solved. User association problem is solved using a distributed framework based on the many-to-many matching game. For this purpose, Zhang et al. design a novel backhaul capacity and interference aware matching utility function considering the interference penalty and backhaul capability simultaneously. For power allocation, convex approximation is used. This problem is then solved iteratively to converge to a KKT point. Simulations show that the proposed algorithm outperforms the random matching algorithm by up to 71.9\% when the number of SBSs is 10.

Pu et al. formulate a resource allocation problem for mmWave self-backhauling networks \cite{pu2019resource}. The problem is modeled as a combinatorial integer programming problem. By introducing penalty factors, Pu et al. transform the problem into another equal problem which can then be solved by the Markov approximation method. The proposed algorithm maximizes the sum data rate and satisfies the minimum traffic demands of all users while satisfying the backhaul constraints of the individual SBSs. Simulation results show that the achieved data rates in each UE is slightly bigger than the required data rate, and repeating BH traffic is minimized with the proposed algorithm. In terms of spectral efficiency, the proposed algorithm outperforms the benchmark methods of OBFD and access-backhaul split significantly.

Dai et al. tackle sum-rate maximization problem via joint user association and power allocation in H-CRANs \cite{dai2017resource}. Dai et al. solve the user association problem first in a central fashion. The power allocation problem is formulated as a generalized Nash equilibrium problem and then solved distributively using variational inequality theory. Simulations show that the proposed algorithm has considerably lower computational complexity and signaling overhead than the centralized max sum-rate method.

Xu et al. give a detailed survey of resource allocation problem of 5G HetNets \cite{xu2021survey}. Xu et al. first define different types of HetNets, and the resource allocation problems related with these types. For traditional HetNets, Xu et al. highlight transmit power, user association, and bandwidth allocation optimization as general problems. For OFDMA-based HetNets, subchannel allocation is another problem due to multiple orthogonal subcarrier usage. For NOMA-based HetNets, cross-tier and co-channel interference are problems to be considered alongside user fairness. For relay-based HetNets, relay selection and transmit power allocation are recurring problems. For \acrshort{hcran}'s, inter-cell interference, signal overhead reduction and baseband unit management are the problems to be tackled. For multi-antenna HetNets, interference cancellation and beamforming design are architecture-specific problems. Integration of new techniques and technologies such as IRS and ultra-dense networks require careful consideration for resource allocation problems to attain the best performance possible. 

\begin{table*}[t]
\caption{A review of the papers on resource allocation based on their common points}
\label{challenges:resalloc-summary}
\begin{tabular}{ |P{5em}|P{6em}|P{6em}|P{6em}|P{6em}|P{6em}|P{6em}|P{6em}|P{6em}| } 
 \hline
 Papers & mmWave & User Association & CoMP & Relay Selection & Bandwidth / Spectrum / Channel / Resource Block Allocation & Power Allocation & Computation Resource Allocation & Access/ Backhaul Resource Allocation \\ 
 \hline 
 \cite{pu2019resource} & $\checkmark$ & $\times$ & $\times$ & $\times$ & $\checkmark$ & $\times$ & $\times$ & $\times$ \\
 \hline
 \cite{saha2018integrated} & $\checkmark$ & $\times$ & $\times$ & $\times$ & $\times$ & $\times$ & $\times$ & $\checkmark$ \\
 \hline
 \cite{bojic2013advanced} & $\checkmark$ & $\times$ & $\times$ & $\times$ & $\times$ & $\times$ & $\times$ & $\checkmark$ \\
 \hline
 \cite{zhang2021backhaul} & $\checkmark$ & $\checkmark$ & $\times$ & $\times$ & $\times$ & $\checkmark$ & $\times$ & $\times$ \\
 \hline
 \cite{dai2017resource} & $\times$ & $\checkmark$ & $\times$ & $\times$ & $\times$ & $\checkmark$ & $\times$ & $\times$ \\
 \hline
 \cite{han2013green} & $\times$ & $\checkmark$ & $\times$ & $\times$ & $\times$ & $\checkmark$ & $\times$ & $\times$ \\
 \hline
 \cite{ma2017backhaul} & $\times$ & $\checkmark$ & $\times$ & $\times$ & $\times$ & $\times$ & $\times$ & $\times$ \\
 \hline
 \cite{liu2016joint} & $\times$ & $\checkmark$ & $\times$ & $\times$ & $\checkmark$ & $\times$ & $\times$ & $\checkmark$ \\
 \hline
 \cite{han2016backhaul} & $\times$ & $\checkmark$ & $\times$ & $\times$ & $\times$ & $\times$ & $\times$ & $\times$ \\
 \hline
 \cite{huskov2018experience} & $\times$ & $\times$ & $\checkmark$ & $\times$ & $\times$ & $\times$ & $\times$ & $\times$ \\
 \hline
 \cite{tang2018energy} & $\times$ & $\times$ & $\checkmark$ & $\times$ & $\times$ & $\checkmark$ & $\times$ & $\times$ \\
 \hline
 \cite{khodmi2018iterative} & $\times$ & $\times$ & $\times$ & $\checkmark$ & $\times$ & $\checkmark$ & $\times$ & $\times$ \\
 \hline
 \cite{gao2019joint} & $\times$ & $\times$ & $\times$ & $\checkmark$ & $\checkmark$ & $\checkmark$ & $\times$ & $\times$ \\
 \hline
 \cite{nguyen2017centralized} & $\times$ & $\times$ & $\times$ & $\checkmark$ & $\times$ & $\checkmark$ & $\times$ & $\times$ \\
 \hline
 \cite{siddique2017downlink} & $\times$ & $\times$ & $\times$ & $\times$ & $\checkmark$ & $\times$ & $\times$ & $\checkmark$ \\
 \hline
 \cite{shariat2014joint} & $\times$ & $\times$ & $\times$ & $\times$ & $\checkmark$ & $\checkmark$ & $\times$ & $\times$ \\
 \hline
 \cite{wang2015backhauling} & $\times$ & $\times$ & $\times$ & $\times$ & $\checkmark$ & $\times$ & $\times$ & $\times$ \\
 \hline
 \cite{mahmood2020capacity} & $\times$ & $\times$ & $\times$ & $\times$ & $\checkmark$ & $\times$ & $\times$ & $\times$ \\
 \hline
 \cite{huang2019fairness} & $\times$ & $\times$ & $\times$ & $\times$ & $\checkmark$ & $\checkmark$ & $\times$ & $\times$ \\
 \hline
 \cite{le2019energy} & $\times$ & $\times$ & $\times$ & $\times$ & $\checkmark$ & $\checkmark$ & $\times$ & $\times$ \\
 \hline
 \cite{hao2017energy} & $\times$ & $\times$ & $\times$ & $\times$ & $\checkmark$ & $\checkmark$ & $\times$ & $\times$ \\
 \hline
 \cite{song2018distributed} & $\times$ & $\times$ & $\times$ & $\times$ & $\times$ & $\checkmark$ & $\times$ & $\times$ \\
 \hline
 \cite{hu2018robust} & $\times$ & $\times$ & $\times$ & $\times$ & $\times$ & $\checkmark$ & $\times$ & $\times$ \\
 \hline
 \cite{faruk2016energy} & $\times$ & $\times$ & $\times$ & $\times$ & $\times$ & $\checkmark$ & $\times$ & $\times$ \\
 \hline
 \cite{zhang2017downlink} & $\times$ & $\times$ & $\times$ & $\times$ & $\times$ & $\checkmark$ & $\times$ & $\checkmark$ \\
 \hline
 \cite{nguyen2017centralized} & $\times$ & $\times$ & $\times$ & $\times$ & $\times$ & $\checkmark$ & $\times$ & $\times$ \\
 \hline
 \cite{pham2019mobile} & $\times$ & $\times$ & $\times$ & $\times$ & $\times$ & $\times$ & $\checkmark$ & $\checkmark$ \\
 \hline 
 \cite{song2018joint} & $\times$ & $\times$ & $\times$ & $\times$ & $\times$ & $\times$ & $\checkmark$ & $\times$ \\
 \hline 
 \cite{omidvar2018optimal} & $\times$ & $\times$ & $\times$ & $\times$ & $\times$ & $\times$ & $\times$ & $\checkmark$ \\
 \hline 
 \cite{bonfante2018performance} & $\times$ & $\times$ & $\times$ & $\times$ & $\times$ & $\times$ & $\times$ & $\checkmark$ \\
 \hline 
\end{tabular}
\end{table*}

\subsubsection{Lessons Learned}

While resource allocation is by no means a problem unique to wireless backhauling, effective distribution of the resources is crucial for a performing network. With this in mind, we highlight the following points:
\begin{itemize}
    \item Access/backhaul resource split is very important in terms of effective resource usage, because employing a dynamic split makes the network adaptable to different scenarios. Employing dynamic and fast algorithms that can adapt to changing conditions is very useful for ultra-dense networks with wireless backhauling.
    \item Ultra-dense networking employs many small cells that a UE can connect to, meaning that at any time instant, multiple BSs are available for UEs. While this increases the network adaptability and redundancy, user association is important for network performance as the small cells usually do not have good range and require LoS, finding a suitable BS is important. From the network's perspective, the network can employ suboptimal user associations to keep up the average performance at a certain level or achieve a certain minimum. 
    \item Transmit power allocation is important for the wirelessly backhauled networks as high air interface usage may result in inter- and intra-tier interference. Employing spatial multiplexing with the right transmit power allocation effectively nullifies the interference from the UE side, while BSs employ beamforming in tandem with power allocation to reduce the effect of interference.
    \item With the usage of MEC in 5G networks, computation offloading is a possibility that can result in power savings and effective computation. However, while wireless backhaul can allow offloading computation-heavy tasks, there may be cases in which offloading can induce performance penalties to the air interface, resulting in power wastage from the network side. Consideration of air interface is required for computation offloading decisions in wirelessly backhauled networks.
\end{itemize}

\subsubsection{Open Issues and Challenges}

Resource management remains an open challenge for future networks. 3D networking is a new concept that significantly changes the network design. Because of this, resource allocation schemes also require a new approach for 3D networking for the best performance. Spectrum and interference management also remains an important open issue for future networks \cite{chowdhury20206g}.

With a myriad of new concepts being introduced in 5G networks and beyond, networks are inevitably getting more and more complex. While new concepts come with performance gains, they also introduce complexity to the design of a network. In terms of resource allocation, using multiple technologies increases the complexity of the problem. New models are required in this direction to allow better design; which, in turn, will allow better problem formulations that capture the requirements and result in better performance gains \cite{xu2021survey}.

Spectrum sensing is another way to increase resource efficiency. The tradeoff between dynamic resource allocation and spectrum sensing capabilities remain an open research area that can result in better spectrum and resource efficiency. 

The majority of the mentioned works do not use mmWave frequencies. Consequently, most mentioned works cannot meet the KPI demands of 5G communications in terms of throughput.  While beamforming and mMIMO are technologies that can enable better performance with mmWave usage, resource allocation problems have to consider beamforming and mMIMO usage alongside the main problem, which adds more degrees of freedom and makes the problem even harder. Due to these points, mmWave frequencies and the resource allocation schemes require a more in-depth exploration.

Machine learning-based resource allocation is also promising since it allows autonomous management of networks. BSs can capture an enormous amount of data that can be used for training a model for user association, transmit power calculation, or spectrum allocation. While these are open research areas, the training process is a problem since there aren't enough resources for training in one BS. Therefore, either distributed algorithms are required for this task, or a new paradigm must be developed in order to realize machine-learning based dynamic resource allocation.

\subsection{Deployment}

Base station deployment is a problem not specific for wireless backhauling. Nonetheless, deployment plays a critical role when wireless backhaul is employed. Here, the challenge is where to place MBSs and small cells. Wireless backhaul enables more radical deployment strategies since infrastructure (or the lack of it) has less of an impact compared to traditional backhaul. Similar to the traditional networks though, deployment directly impacts network performance. Therefore, innovative deployment schemes have to be employed to make wireless backhauling feasible. 

Mesh, star and tree topologies are usually considered while deploying networks. Tree topologies are especially used with hierarchical networks, as well as with wireless backhauled networks. In such schemes, root node of the tree has fiber backhaul capabilities, and all other nodes send their data via their parent to the root node to be backhauled to network core. Mesh topology is also suitable to be used with wireless backhauling as it promotes redundancy via multiple paths. Finally, wireless backhauling also allows dynamic deployment with mobile base stations, which improves network's capabilities while making management and deployment harder.

While the deployment of small cells is a task performed before the actual deployment, moving elements such as UAVs and nomadic BS make this problem dynamic. Because of this, finding the optimal deployment in a fast manner becomes important as this increases the network adaptability and reduces the time that the network is affected by the condition changes in a bad way.

\subsubsection{State of the Art}

In \cite{bonfante2018performance}, Bonfante et al. propose a two-layer network scheme where self-backhauling small cells serve UEs and MBSs using \acrshort{mmimo} backhaul their data. Small cells use fixed allocation of access and backhaul slots. Two different deployment scenarios are considered for small cells. In the first scenario, the small cells are randomly and uniformly distributed to the BS area; whereas in the second one, which the authors named as ad-hoc deployment, they are deliberately placed closer to UEs to increase the throughput. To reduce the inter-cell interference,  small cells in ad-hoc deployment use a more directive antenna that serve only the closest UEs, thus reducing interference. Results show that the performance increases threefold when ad-hoc deployment is used compared to uniform deployment.

McMenamy et al. investigate mmWave backhaul network flow affected by the number of hops and deployment of fiber-connected BSs \cite{mcmenamy2019hop}. The authors first determine which nodes will have fiber backhauling given the deployment while satisfying the constraint on the number of hops. Then, the authors maximize the overall flow while meeting the flow demand of every node and the constraint on the number of RF chains for every node in the network. Simulation results show that increasing the number of hops reduces the number of required fiber-connected BSs (BGWs), but also the total flow. Number of RF chains grows with the number of hops. If the number of BGWs are reduced, then the performance can be maintained by increasing the number of RF chains at the expense of power consumption.

Aftab et al. address UAV-BS placement problem with a novel machine learning based intelligent deployment mechanism \cite{aftabblock}. An intelligent long short-term memory (LSTM) model is used to predict the deployment with a feedback mechanism for self-correcting. Simulation results show that deep learning-based approach outperforms the normal deployment in terms of outage probability. ML-based deployment also achieves far better energy efficiency compared to the normal deployment. 

In \cite{fouda2018uav}, Fouda et al. have considered a scenario where \acrshort{uav} are deployed to a highly congested area where some UEs suffer low quality and some have no service at all. The network in question is a two-tiered network with \acrshort{gnb}'s that serve UEs in tandem with \acrshort{uav} small cells. The network uses in-band frequency division duplexing (\acrshort{fdd}) as the operation mode. The \acrshort{uav}s' mobile capabilities are used as the main degree of freedom to maximize the overall system sum-rate. The authors have formulated an optimization problem to find the \acrshort{uav}s' optimal 3D deployment locations, \acrshort{ue} power allocations, precoder design at \acrshort{gnb} and \acrshort{ue}-base station associations. Since these optimization parameters have different update intervals, the problem is decomposed into two parts: in the first part, average sum-rate is maximized by optimizing the \acrshort{uav} locations, access and backhaul link power allocations; whereas in the second part, precoder design and transmit power allocation is optimized to maximize the instantaneous sum-rate. Simulations show that \acrshort{uav} deployment increases the average sum-rate at the expense of performance of UEs that are close to gNBs. It is also shown that \acrshort{uav}s cannot support a high number of UEs as efficiently and distributing users between \acrshort{uav}s is key to a high-performance network.

Park, Tun and Hong propose initial deployment and trajectory optimization techniques for stable communication between high speed train (HST) and UAVs \cite{park2020optimized}. Park et al. use Soft Actor-Critic (SAC) method of reinforcement learning for optimizing the UAV trajectory and Support Vector Machine (SVM) for optimal initial UAV deployment. Numerical results show that with less HST speed, less handovers occur and UAVs can have lower operating altitudes with higher data rates. Conversely, with faster HST speeds, UAVs have to increase their altitudes to keep the connection up.

Zhang et al. formulate a joint optimization problem of UAV deployment, caching placement and user association for maximizing the QoE of users \cite{zhang2020cache}. The cache-enabling UAVs are deployed at peak hours to the hotspots to alleviate the pressure on the MBSs. The formulated optimization problem is divided into three subproblems. UAV deployment subproblem is solved via a swap matching based algorithm. Then, Zhang et al. solve the caching placement subproblem with a greedy algorithm, and user association problem with a Lagrange dual. The joint problem is solved with a novel low complexity iterative algorithm. Simulation results show that the proposed algorithm can reach the near optimal value of the exhaustive search within 0.02 mean opinion score (MOS). The proposed algorithm outperforms the random and classic algorithms in terms of UAV backhaul traffic offloading ratio, and average MOS, under varying user numbers, cache space, and UAV height.

Okumura and Hirata propose an algorithm to automatically plan the deployment of 300 GHz wireless backhaul links \cite{okumura2021automatic}. The algorithm first selects a pair of adjacent cells that can get a LoS connection. After deploying a master BS to the center rooftop, the algorithm deploys the slave BSs according the the previously mentioned pairs of adjacent cells to find a LoS path from every slave BS to the master BS. Simulation results show that this deployment scheme exceeds the standart demand of 100 Gbit/s even when the rain rate is 100 mm/hr.

In \cite{polese2020integrated}, Polese et al. have explained the integrated access and backhaul concept and explored some scenarios for its real-life usage. A thorough explanation of \acrshort{iab} is made, its architecture, network procedures, topology management, scheduling and resource multiplexing aspects are detailed. Then, an evaluation of an \acrshort{iab} \acrshort{mmwave} network is made through simulations with different configurations. For routing the backhaul traffic, wired-first and highest-quality-first approaches are tested and wired-first approach is found to be better in terms of throughput and latency. For \acrshort{iab} deployment scenarios, three alternatives are compared: namely, only donors scenario in which only some wired gNBs exist, \acrshort{iab} scenario where alongside the first scenario donors, some \acrshort{iab} nodes are available, and all wired scenario where it is assumed that the topology is the same with the second scenario but all backhaul connections are wired. All wired scenario outperforms the others as expected, but \acrshort{iab} scenario outperforms the donors-only scenario in average delays, 5th percentile throughput and average buffering in target UEs.

Hu et al. propose an intelligent UAV-BS deployment scheme based on machine learning \cite{hu2019intelligent}. Hybrid ARIMA-XGBoost model is used for predictions: ARIMA handles linear predictions and XGBoost is then applied on the residue of ARIMA. Simulation results show that the proposed hybrid prediction model has a prediction accuracy of 87\% in the worst case. The proposed deployment scheme also significantly reduces the blocking ratio, increases the capacity, throughput, and resource utilization of the network.

In \cite{dahrouj2015cost}, Dahrouj et al. have proposed a hybrid system that uses RF as well as FSO links for wireless backhauling. The hybrid usage combines the low-cost, high-availability and high-coverage nature of RF and high-capacity, low-latency nature of FSO, bringing the best of both worlds. To reduce complexity, the problem is first relaxed such that only optical fiber links are allowed. Using the found planning, the found nodes are connected with fiber or hybrid RF/FSO links. To improve reliability, the planning is made so that the network is $K$-link disjoint. This is done by clustering the BSs together and iteratively finding the links between clusters. The authors show that the solution is equivalent to a maximum weight clique in the planning graph. The results indicate that as $K$ increases, the network design is cheaper than fiber-only while maintaining the required reliability and key performance indicators (\acrshort{kpi}s).

\begin{table*}[t]
\caption{A review of the papers on deployment issues based on their common points.}
\label{challenges:deployment-summary}
\begin{tabular}{ |P{5em}|P{8em}|P{8em}|P{8em}|P{8em}|P{8em}|P{8em}| } 
 \hline
 Papers & mmWave & MIMO & Machine Learning & Multi-Hop & UAV Deployment & FSO \\ 
 \hline 
 \cite{nasr2017millimeter} & $\checkmark$ & $\times$ & $\times$ & $\times$ & $\times$ & $\times$ \\
 \hline
 \cite{polese2020integrated} & $\checkmark$ & $\times$ & $\times$ & $\times$ & $\times$ & $\times$ \\
 \hline
 \cite{mcmenamy2019hop} & $\checkmark$ & $\times$ & $\times$ & $\checkmark$ & $\times$ & $\times$ \\
 \hline
 \cite{okumura2021automatic} & $\checkmark$ & $\times$ & $\times$ & $\checkmark$ & $\times$ & $\times$ \\
 \hline
 \cite{zhang2021leveraging} & $\checkmark$ & $\times$ & $\times$ & $\times$ & $\times$ & $\times$ \\
 \hline
 \cite{bonfante2018performance} & $\checkmark$ & $\times$ & $\times$ & $\times$ & $\times$ & $\times$ \\
 \hline
 \cite{aftabblock} & $\times$ & $\times$ & $\checkmark$ & $\times$ & $\checkmark$ & $\times$ \\
 \hline
 \cite{park2020optimized} & $\times$ & $\times$ & $\checkmark$ & $\times$ & $\checkmark$ & $\times$ \\
 \hline
 \cite{hu2019intelligent} & $\times$ & $\times$ & $\checkmark$ & $\times$ & $\checkmark$ & $\times$ \\
 \hline
 \cite{liu2020fast} & $\times$ & $\times$ & $\checkmark$ & $\times$ & $\checkmark$ & $\times$ \\
 \hline
 \cite{saadat2018multipath} & $\times$ & $\times$ & $\times$ & $\checkmark$ & $\times$ & $\times$ \\
 \hline
 \cite{kulkarni2018max} & $\times$ & $\times$ & $\times$ & $\checkmark$ & $\times$ & $\times$ \\
 \hline
 \cite{dahrouj2015cost} & $\times$ & $\times$ & $\times$ & $\checkmark$ & $\times$ & $\checkmark$ \\
 \hline
 \cite{fouda2018uav} & $\times$ & $\times$ & $\times$ & $\times$ & $\checkmark$ & $\times$ \\
 \hline
 \cite{zhang2019optimizing} & $\times$ & $\times$ & $\times$ & $\times$ & $\checkmark$ & $\checkmark$ \\
 \hline
 \cite{zhang2020cache} & $\times$ & $\times$ & $\times$ & $\times$ & $\checkmark$ & $\times$ \\
 \hline
\end{tabular}
\end{table*}

Liu et al. investigate a fast deployment strategy of UAVs using deep learning \cite{liu2020fast}. Liu et al. formulate a problem of finding the optimal UAV-BS position as fast as possible, while maximizing the sum of downlink rates in the network. The authors solve this problem by designing a geographical position information learning algorithm, which employs a dueling deep Q-network. Simulation results show that the proposed algorithm finds the deployment position in seconds with little loss. While reinforcement learning requires training after any changes in the scenario, the proposed algorithm does not have such a requirement and achieves 96\% of the optimal value. For different area sizes and UE densities, the proposed algorithm performs almost indistinguishably from reinforcement learning but with less time.

Zhang et al. optimize the BS and gateway densities of an mmWave backhaul network to improve network spatial throughput \cite{zhang2021leveraging}. Low density makes the network backhaul-limited, but after a certain density, the performance gains disappear because the network becomes interference-limited due to the intra-tier interference of the newly added elements. The optimal densities are found numerically and the simulation results show that with the increasing network density, the spatial throughput first increases and then converges to a gateway density-related constant. 

Kulkarni, Ghosh and Andrews \cite{kulkarni2018max} propose a network with k-ring deployment with wireless backhauling. In k-ring deployment, a BS with fiber connection is on the centre while all other BSs are backhauled wirelessly to this BS, and then to the core. Kulkarni et al. argue that with their design, instantaneous rate performance is independent of the scheduling, which makes the system noise-limited. The authors formulate an optimization problem for a fixed k that maximizes the end-to-end rate achieved by all UEs. For this purpose, Kulkarni et al. define the max-min rate as the maximum rate that can be achieved by the worst member of a routing strategy. The analysis is done initally for \acrshort{iab}, and then extended to orthogonal access and backhaul (\acrshort{oab}), which basically means different resources for access and backhaul. For \acrshort{iab} case, Kulkarni et al. employ highway routing, which designates the $x$ and $y$ axes that are centered on the fiber backhauled BS as the highways. The data is then distributed from these highways, even if this results in more hops than a direct path. If BSs on the highways have larger antenna gains than others, this scheme is justified. Nearest neighbor routing is used as the benchmark scheme. \acrshort{iab} solution requires the global load and access rate information, which makes its solution more complex. OAB is simpler to implement since it does not require such information. Kulkarni et al. present results in four main fields. First, it is shown that with higher number of rings, the performance drops significantly since the network is backhaul-limited. Second comparison is made between full-duplex and half-duplex modes. Full-duplexing shows significant potential in terms of data rate if the self-interference is kept below -100 dB. However, Kulkarni et al. also note that these results do not seem practical. Third comparison is made between OAB and \acrshort{iab}. It is shown that OAB can slightly outperform \acrshort{iab} if the access/backhaul resource assignment is made according to the load. If this assignment is made in a fixed fashion regardless of the expected load, then \acrshort{iab} is the better choice. Finally, dual and single connectivity are compared in terms of performance. It is shown that, performance gains with dual connectivity is higher if there are load imbalances in single connectivity cases.

Zhang and Ansari investigate the backhaul-aware uplink communications in a full-duplex drone BS-aided HetNet problem with the objective to maximize the total throughput while minimizing the number of deployed BSs \cite{zhang2019optimizing}. The authors decompose the problem into three parts, namely (i) joint user association, power, and bandwidth problem, (ii) drone BS placement problem, (iii) determining the number of drone BSs to be deployed. The first problem is solved using an approximation algorithm. The second problem is solved using exhaustive search, while the third problem is solved using linear programming. Simulation results show that the blocking ratio of the proposed algorithm outperforms the benchmark methods in terms of total throughput and the blocking ratio.

Saadat, Chen and Jiang consider multi-hop backhauling for ultra-dense networks \cite{saadat2018multipath}. The authors propose a deployment scheme that allows multiple backhaul paths to MBSs, helping overcome the blockage problem to some extent and improve overall reliability. The proposed method for backhaul is called the multipath multi-hop (MPMH) backhaul. Saadat et al. assume that the small cell density is much higher than the user density, and each SBS only serves one user. User-serving SBSs relay their backhaul traffic via the closest SBS to the MBS. The backhaul routing is based on minimizing the number of hops. The proposed MPMH backhaul is compared with direct backhauling, multihop backhauling and multiple-association backhauling that employs multiple paths between sender and receiver. Simulation results show that MPMH backhaul outperforms all other alternatives regardless of the inter-site distance. It is also shown that multi-hop scheme significantly improves LoS probability. Finally, as the inter-site distance increases, multipath scheme shows significant gains over multi-hop scheme. Needless to say, MPMH backhaul incorporates the advantages of both schemes together.  

Nasr and Fahmy analyze the scalability of ring and star topologies in \acrshort{mmwave} backhaul networks \cite{nasr2017millimeter}. The network in question has one MBS with fiber connectivity and multiple small cells with star or ring topology. Mesh topology uses multiple hops to relay traffic to the MBS, whereas in star topology, every small cell has direct connection to the MBS. Simulation results show that initially, star topology performs better but as transmit SNR increases, mesh topology performance increases faster. Similar results can be observed for different small cell sizes as well. Increasing small cell numbers almost linearly increases the performance for both topologies.

\subsubsection{Lessons Learned}

We highlight the following points for the deployment problem:
\begin{itemize}
    \item UAVs are used for different purposes, such as content caching and performance improvement. UAV deployment directly affects the performance gains from such use cases. It is shown that with correct deployment, UAVs can be used as small cells with better freedom and adaptability. 
    \item With the new ultra-dense network architectures, hierarchical designs seem to be advantageous. Using tree structures allow better flow designs for backhauling, while mesh structures allow better flexibility and redundancy by employing multiple links for backhauling. Mesh structures employ every small cell as both an access node and a backhaul node for other small cells, whereas tree structures make use of small cells that perform backhauling most of the time, especially for nodes that have higher heights in the tree structure. 
    \item Machine learning approaches are widely employed with UAV deployment problems, and with solid results. Machine learning approaches are also shown to be quite fast, which is a significant step to realize a self-organizing network with UAVs. Compared to the optimization approaches using linear programming and similar formulations, machine learning approaches give flexibility to the network to adapt to different conditions rapidly. 
\end{itemize}

\subsubsection{Open Issues and Challenges}

Current research on UAVs do not make use of mmWave frequencies. In the future, UAVs can use mmWave frequencies for better performance. The deployment of mmWave-using UAVs is an open research problem that requires investigation. This problem has different facets such as vertical modeling, beam management, and antenna design. 

While UAVs and other mobile cell options provide flexibility for 5G networks, current approaches are still not fast enough to be used for self-organizing networks on a grand scale. Current formulations are often optimization problems that require considerable time for solving. Consequently, these solutions do not use the mobility capabilities to the fullest extent. Faster and more adaptable approaches are required for self-organizing networks and this problem remains an open challenge. While some machine learning approaches are performing in this area, there is still room for improvement, especially for larger-scale self-organizing networks.

UAV and mobile cell mobility also presents problems for mesh and tree topologies, as the moving cells alter the topology and adaptation to these movements are required. This problem can be solved in a central or distributed fashion. While each seem to have their own advantages, distributed approaches seem to have an edge since the cells have a high volume of base stations. The adaptability of topologies is an open challenge that needs an effective solution for a moving network with high performance. Table \ref{challenges:deployment-summary} gives a summary of the papers mentioned in this section.

\subsection{Scheduling}

Similar to the deployment problem, scheduling is a general problem as well. Here, we consider the scheduling of access and backhaul traffic, which is an essential aspect of wireless systems when both access and backhaul uses the same medium. Employing a good scheduling scheme for access and backhaul traffic is imperative to attain performances up to the network's potential.

\subsubsection{State of the Art}

In \cite{vu2016joint}, Vu et al. consider a HetNet with a MBS and its subordinate small cells that use wireless backhauling. In such a system, Vu et al. point out that the cross-tier (between MBS and small cells) and co-tier interference (between small cells) becomes a significant problem that reduces the throughput. Therefore, a joint optimization of scheduling, operation mode, interference mitigation and transmit power allocation that satisfies the available transmit power budget is considered. Small cells operate either half-duplex to reduce interference, or full-duplex to increase performance. The proposed system is compared with a traditional network in which MBS serves the UEs without any small cells. It is shown that for three configurations that are modeled at 28 GHz, 10 GHz and 2.4 GHz, the proposed network outperforms the traditional network in both achievable average throughput and average queue length.

Gupta et al. consider the multihop link scheduling problem in self-backhauled mmWave cellular networks \cite{gupta2020learning}. The authors formulate the given problem as a Markov decision process and use reinforcement learning techniques to solve it. The training is conducted for two scenarios, namely an ideal case where a central scheduler performs all scheduling decisions with full knowledge at any instant, and a more realistic case where network state feedback and scheduling decisions are limited to one per frame. For the realistic case, Gupta et al. model the scheduler as a recurrent neural network. Simulation results show that the proposed method outperforms the other benchmark methods (max-min, backpressure, DRQN, r-REINFORCE) in terms of packet delay for one-hop, two-hop, and three-hop users at high- and medium-load scenarios.

Zhang et al. propose a joint cooperative orthogonal and non-orthogonal multiple access wireless backhauling model \cite{zhang2019joint}. Using this model, the authors formulate an optimization problem to maximize energy efficiency via resource block scheduling.  Zhang et al. try to maximize the energy efficiency while meeting the given rate threshold of SBSs. Orthogonal multiple access (C-OMA) can make use of good channels in a more effective way, while non-orthogonal multiple access (C-NOMA) consumes less power with the same threshold. Using these two methods together is much more complicated than a mixed integer non-linear programming problem. Zhang et al. propose a greedy algorithm to use the two multiple access methods in tandem. Simulation results show that the combined greedy algorithm (JCC) is the best in terms of average power consumption and energy efficiency. C-OMA performs close to JCC for smaller cell radii while C-NOMA does the same for larger radii. 

Arribas et al. study the mmWave backhaul scheduling problem for mmWave MBS and microBSs \cite{arribas2019optimizing}. The problem is modeled as a mixed integer linear programming problem and the problem is NP-hard even when not considering the interference. Arribas et al. establish upper and lower bounds of the problem and formulate the problem as a simpler linear program to approximate the solution. Two heuristic methods are also provided for comparison. Experimental result show that the optimal method for small sizes perform close to the lower bound, while for larger sizes optimal cannot be solved in a reasonable time. Greedy heuristics perform closer to the upper bound, while `resched` heuristic method performs closer to the lower bound. Greedy heuristic loses its edge when multiple RF chains can be employed. Resched heuristic performs better than greedy in terms of aggregate rates. 

Li et al. perform distributed scheduling in an industrial wireless network using reinforcement learning \cite{li2020learning}. Li et al. propose a learning-based autonomous orchestration agent, namely L-AoI, for timely state updates in D2D-enabled networks. The proposed orchestrator guides distributed users to select a proper resource block and power level to deliver state updates. Since D2D users do not have global CSI or knowledge of other users' scheduling actions, the orchestrator allows each users to forecast other future decisions. Simulation results show that the proposed orchestrator outperforms Q-learning and Q-coalition methods in terms of age of information violation, average data rate, and average power consumption.

In \cite{goyal2017scheduling}, Goyal, Liu and Panwar propose a scheduling system that works on a two-tiered network with MBS and small cells. Small cells and MBSs support full-duplex operation and the scheduler decides in each time step which transmission mode will be used. The transmission mode is chosen such that the maximum gain capacity can be achieved. Moreover, queue stability is also considered while deciding to schedule as not to introduce congestion. For the overall solution, a well-known back-pressure based scheduling algorithm that is throughput-optimal is used. For the full-duplex operation, power management is also considered to minimize the interference. It is shown that the scheduler performs well to reduce the interference, so much so that the performance gains are almost double (95-85\%) compared to half-duplex only operation. However, after a certain point the bottleneck results from the weaker backhaul links. The authors also propose using directional antennas for backhaul between small cells and MBSs, strengthening the links considerably and improving overall cell throughput.

Rasekh, Guo and Madhow formulate the wireless mesh backhaul design problem as a joint routing and resource allocation optimization for mmWave picocells \cite{rasekh2019joint}. The optimal resource allocation is modeled as scheduling the time into slots in which only a subset of all available links are active. This decision is made according to the interference levels since interference is directly related to the available data rate. Simulation results show that the proposed scheduling scheme outperforms the interference-agnostic scheduling by 25\% in terms of backhaul throughput. This figure becomes higher as link SNR increases.

Niu et al. formulate the joint optimization problem of concurrent transmission scheduling and power control into a MINLP problem \cite{niu2017energyeff}. A contention graph is used to model the concurrent transmissions. Then, a maximum independent set (MIS)-based algorithm is used to allocate flows into different pairings. This algorithm controls the interference between links and schedules as many links into each pairing as possible, exploiting concurrent links to its fullest potential. This algorithm is shown to achieve better performances than \acrshort{tdma} scheme. However, concurrent transmission scheme with maximum transmission power (CTFP) outperforms this scheme at the cost of higher energy consumption. 

Hu and Blough explore the relay selection and scheduling problem for \acrshort{mmwave} relay-assisted backhaul \cite{hu2017relay}. Hu and Blough propose a new \acrshort{mmwave} backhaul network architecture where base stations serve as mesh nodes in the backhaul and are connected via multi-hop paths. Moreover, the authors also propose a linear-time scheduling algorithm achieving max throughput when there is no secondary interference and a relay-selection algorithm to select a path with no secondary interference. An algorithm is proposed that finds the minimum scheduling length by employing the maximum demand sum of two consecutive links. This algorithm is proven using induction. For relay selection, the authors employ depth-first search using the number of hops and considering only LoS paths. Numerical results and simulations show that even for short distances from the MBS (< 200m), relays are required as LoS paths are not available. It is shown that throughput is 10 Gbps at the worst case (800-1000m). While minimum number of hops deliver adequate performance, going over this number improves even further, but after a certain number of extra hops (3) the gains plateau. It is also shown that the proposed algorithm results in a throughput gain of 23\% to 49\%.

Forouzan et al. consider the joint user scheduling, operation mode selection, and power allocation in a two-tier cellular network with full duplex capable small cells \cite{forouzan2021distributed}. The authors design a distributed resource allocation scheme determining user scheduling, operation mode of access/backhaul links, and the transmission powers to maximize the overall network throughput. The user scheduling and operation mode problems are solved locally for each cell. The power allocation problem is solved using successive convex approximation method. Numerical results show that the proposed algorithm outperforms IBFD- or OBFD-only approaches in terms of average network sum rate. Increasing the number of SBSs also increases the average network sum rate.

\begin{table*}[t]
\caption{A review of the papers on scheduling based on their common points.}
\label{challenges:sched-summary}
\begin{tabular}{ |P{5em}|P{12em}|P{12em}|P{12em}|P{12em}| } 
 \hline
 Papers & mmWave Scheduling & Machine Learning & Full-duplexing via Scheduling & Interference Management \\ 
 \hline 
 \cite{sahoo2017millimeter} & $\checkmark$ & $\times$ & $\times$ & $\times$ \\
 \hline
 \cite{yuan2020optimal} & $\checkmark$ & $\times$ & $\checkmark$ & $\times$ \\
 \hline
 \cite{hu2017relay} & $\checkmark$ & $\times$ & $\times$ & $\times$ \\
 \hline
 \cite{rasekh2019joint} & $\checkmark$ & $\times$ & $\times$ & $\times$ \\
 \hline
 \cite{vu2016joint} & $\checkmark$ & $\times$ & $\times$ & $\times$ \\
 \hline
 \cite{gupta2020learning} & $\checkmark$ & $\checkmark$ & $\times$ & $\times$ \\
 \hline
 \cite{arribas2019optimizing} & $\checkmark$ & $\times$ & $\times$ & $\checkmark$ \\
 \hline
 \cite{li2020learning} & $\times$ & $\checkmark$ & $\times$ & $\times$ \\
 \hline
 \cite{forouzan2021distributed} & $\times$ & $\times$ & $\checkmark$ & $\times$ \\
 \hline
 \cite{goyal2017scheduling} & $\times$ & $\times$ & $\checkmark$ & $\checkmark$ \\
 \hline
 \cite{yuan2020optimal} & $\times$ & $\times$ & $\checkmark$ & $\times$ \\
 \hline
 \cite{niu2017energyeff} & $\times$ & $\times$ & $\times$ & $\checkmark$ \\
 \hline
 \cite{zhang2019joint} & $\times$ & $\times$ & $\times$ & $\checkmark$ \\
 \hline
\end{tabular}
\end{table*}

Yuan et al. consider the scheduling problem for networks using \acrshort{mmwave} small cells \cite{yuan2020optimal}. Due to \acrshort{mmwave} characteristics, it makes perfect sense to make use of spatial reuse via scheduling multiple links. All small cells (named mmBSs) have only one RF chain available, effectively making the communication half-duplex. Yuan et al. formulate the problem of maximum throughput fair scheduling for backhaul networks. This problem is solved by the ellipsoid algorithm. This optimization is scheduling oriented, and one downside is that it can require a high runtime if the network has a large number of mmBSs. Due to this, the authors also propose an edge-coloring based approximation algorithm. This method works by computing the link time first, and then scheduling based on edge-coloring. The proposed method is shown to be applicable to access and backhaul networks, as well as multiple RF chains at each node. Numerical evaluations show that the proposed EC-based method achieves 70\% to 90\% of the optimal throughput with 100x better scalability. With multiple RF chains, EC-based method achieves 85\% to 90\% of the optimal throughput while using 2\% to 20\% of the time. 

Yuan et al. investigate scheduling algorithms under different interference models \cite{yuan2020optimal}. The authors design a scheduling algorithm to optimize mmWave backhaul efficiency for full-duplex and half-duplex radios assuming pairwise link interference or not, and realistic or maximum single-user spatial multiplexing . For the full-duplex no-interference model, Yuan et al. provide a polynomial-time optimal algorithm for the maximum throughput fair scheduling problem (MTFS). Adding pairwise link interference or half-duplex mode makes the problem NP-hard, but the authors provide a special case of half-duplex MTFS problem which is solvable in polynomial time. For NP-hard problems, the authors provide approximation algorithms which transform the MTFS problem into a fractional weighted vertex coloring problem. Numerical evaluation results show that mean throughput increases as the number of RF chains increase. Full-duplex approximation algorithms are shown to achieve 90\% performance of the optimal algorithms, while this figure decreases to 70-60\% for half-duplex approximation algorithms. 

Sahoo, Yao and Wei present a multi-hop relay scheduling algorithm to maximize the overall system performance for \acrshort{mmwave} backhauling \cite{sahoo2017millimeter}. The network in question has small cells that operate with PtP \acrshort{mmwave} backhaul and fiber-connected MBSs. A dynamically adaptable time division duplexing \acrshort{tdd} scheme is used for scheduling. A non-linear integer programming problem is formed to optimally schedule the backhaul traffic according to bandwidth, power, and half-duplex traffic. The authors propose a heuristic algorithm to solve this NP-hard problem. In the first scheduling policy, local traffic load at each BS is considered. In the second scheduling policy, links with higher channel quality are prioritized. Individually, the first policy (Load-iBS) tries to transmit maximum aggregate traffic, whereas the second policy (Link-iBS) tries to maximize the overall network throughput via maximizing the traffic over high-quality links. A third policy (TO-iBS) is proposed to use the aforementioned concepts in tandem to achieve better performance. Load-iBS is used to determine transmitting and receiving nodes, and then Link-iBS is used for power reallocation to improve the active P2P link quality. Simulation results show that TO-iBS outperform both Load-iBS and Link-iBS and performs very similar to the exhaustive search scheduling. However, it is also much easier to implement than the exhaustive search, making the advantages evident. 

\subsubsection{Lessons Learned}

Scheduling research on wireless backhauling is heavily invested on the mmWave link scheduling and spatial reuse. We highlight the following learning points for this area:
\begin{itemize}
    \item With the combined use of mmWave, beamforming, and mMIMO, spatial multiplexing can be easily achieved with solid results and performance gains. If the traffic of an element, be it MBS, SBS, or UE, is scheduled in a specific way to send and receive traffic without intra-tier interference, the data rate is effectively doubled. Good scheduling algorithms are very important in mmWave systems to increase performance.
    \item Employing spatial multiplexing effectively makes the communication full-duplex. Because of this, scheduling research with mmWaves often focus on how to make the communication full-duplex by reducing or outright removing intra-tier interference. Consequently, effective scheduling is one way to implement full-duplex communication. Research on full-duplex communications focus on multiple areas to make it feasible, but scheduling is a proven way with performance gains for implementing full-duplex communication. 
\end{itemize}

\subsubsection{Open Issues and Challenges}

Machine learning is very promising for scheduling problems, but it is not applied as widely as it can be. 5G networks are employing a high number of small cells, which makes centralized management algorithms suboptimal since they incur too many control messages. Distributed machine learning approaches are very promising to solve the scheduling problem for ultra-dense networks. This is an open research problem for scheduling.

Currently, scheduling is widely employed to implement full-duplex mmWave communications. While spatial multiplexing can be implemented with good scheduling, interference management without spatial multiplexing is not thoroughly explored. Different interference management techniques without spatial multiplexing is an open research area that can help with adaptability of 5G networks when spatial multiplexing cannot be employed. 

Scheduling is an area that is explored for multiple contexts in 5G networking such as wireless backhauling, industrial networks, and IoT. Currently, full-duplex and mmWave seem to be the common points of scheduling research in these areas. For future research, we highlight the machine learning as the main point to increase the effectiveness of scheduling in 5G networks. To end this section, we present a summary of mentioned papers and their common points at Table \ref{challenges:sched-summary}.

\subsection{Performance Evaluation}

In this section, we will include works that evaluate the performance of existing works alongside novel concepts. These works have been essential to show the feasibility and gains of wireless backhauling, and therefore are included here.

Note that almost every work mentioned in our paper evaluates the performance of its proposed method. However, the works that we mention here focus solely on evaluating the performance of some concept. Other works formulate a problem and then show the performance of their solution; these papers are not mentioned here. 

\subsubsection{State of the Art}

In \cite{polese2018end}, Polese et al. implement a network with \acrshort{iab} architecture in ns-3 that operates in \acrshort{mmwave} frequencies and then validate its results through simulations. The authors extend the \acrshort{mmwave} module in ns-3 to support advanced \acrshort{iab} functionalities such as \acrshort{iab} nodes, single- and multi-hop control procedures for these nodes, a look-ahead backhaul-aware scheduler and the simulator making use of these implemented features. The authors verify that \acrshort{iab} architecture results in significant improvements for cell-edge users. However, fine-tuning for the network is required as this performance improvement at cell-edge also results in the performance degradation for the users close to the gNBs that had no performance issues before.

Moon et al. propose an OFDM-based E-band wireless backhaul system \cite{moon2020ofdm}. The network that Moon et al. design uses 1024 QAM modulation and E-band frequencies to achieve 25 Gbps throughput. For experimentation, an FPGA implementation is made and experiments show that the setup achieves 75\%, 84\%, and 96\% resource utilization for LUT, DSP, and BRAM, respectively.

In \cite{bonfante2018performance}, Bonfante et al. analyze \acrshort{ue} performance in \acrshort{mmimo} self-backhauling small cell network. As a benchmark, Bonfante et al. compare their network to a standard network configuration using only MBSs with \acrshort{mmimo}. The results indicate that increasing small cells in a given area results in only marginal improvements as this also increases the inter-cell interference. For the ad-hoc deployment, i.e., deploying small cells closer to UEs, reducing the deployment distance to UEs significantly increases the performance. Backhaul links are found to be the main bottleneck when ad-hoc deployment is used. Furthermore, while allocating backhaul and access slots, cell-edge and average users are shown to have different optimal allocation percentages, showing a conflict of interest between cell-edge and center. Finally, Bonfante et al. compare the performance of different antennas. Directed antennas perform better than isotropic antennas since directiveness reduces the overall inter-cell interference and provides a larger antenna gain. 

Wang and Si design a wireless mobile ad hoc network (MANET) based on dedicated short-range communication (DSRC) technology and GPRS for mobile terminals \cite{wang2019system}. When the mobile terminals cannot reach a BS, then DSRC is used for mobile backhauling via the neighboring nodes. Simulation results show that without Wi-Fi interference, maximum packet loss is less than 2\%, whereas with Wi-Fi interference this value is around 6-10\%. As the number of hops increase, packet loss also gradually increases. For one-hop, packet loss is kept within 2\%. For three-hops, packet loss remains below 3\% for a packet rate of more than 5s, but the loss sharply increases if the packet rate is below 5s.

Zhang et al. investigate the effect of gateway density on a mmWave backhaul network in terms of spatial throughput \cite{zhang2021leveraging}. The backhaul traffic is conveyed from gateways to BSs through mmWave links. Zhang et al. show that considering backhaul-RAN coupling, network spatial throughput is greatly dependent on the gateway density. While deploying more gateways provides benefits for backhaul capacity, BS-generated interference overwhelms the benefits of spatial reuse gain after a certain point, resulting in exponential degradation of spatial throughput. Monte Carlo simulations show that given a gateway density, the activated BS density increases to a gateway density-dependent saturation. As the activated BS density grows, both signal power and interference increases. After a certain point, interference power overwhelms the signal power in dense BS regime. In terms of beamforming angle, absence of side lobes reduces the transmission probability when antenna miscommunication occurs, it also makes fewer gateways interfere with the intended transmission. This phenomenon offsets the transmission probability reduction and boosts network capacity. 

Hu et al. propose a novel UAV-BS intelligent deployment scheme based on machine learning and evaluate its performance on a real-world dataset \cite{hu2019intelligent}. Hu et al. propose a hybrid model consisting of two models, namely ARIMA and XGBoost. It is shown that the proposed hybrid scheme makes more accurate forecasts compared to the its individual submodels when compared to the actual data. Moreover, with the proposed deployment scheme, the blocking ratio becomes zero most of the time, whereas with random deployment the network cannot reduce this effectively and consequently, the load balancing performance suffers. 

In \cite{esmail201928ghz}, Esmail et al. study the effect of dust on FSO communication. A network using \acrshort{mmwave} communications is chosen as a benchmark performing on the same conditions. Esmail et al. prepare an experimental setup emulating a dust storm, through which the communication signals are sent. RF signal is not affected by dusty conditions as expected, whereas FSO signals are hampered by the same conditions. The reason of this observation is that dust particles have comparable sizes to FSO signal wavelengths, whereas \acrshort{mmwave} wavelengths are higher by a couple orders of magnitude. As a final note, Esmail et al. combine the FSO and RF to form a hybrid system that combines high bandwidths of FSO and dust-resistance of RF. The hybrid system uses RF when visibility is less than 50m due to bit-error rate (\acrshort{ber}) of FSO is more than RF below this threshold. RF link performance in dusty conditions is also analyzed, whose results indicate 200m visibility to be the threshold for maximum performance, beyond which there are no gains/losses based on dusty conditions.

Shi et al. evaulate the performance of joint optimization algorithm proposed in \cite{wang2016joint} for 3.5 and 30 GHz frequencies \cite{shi2018channel}. Simulations show that for 30 GHz, more than 80\% of the resources are allocated for backhaul, whereas for 3.5 GHz, this allocation is only 45\%. The sum of logarithmic user rates are also higher for 3.5 GHz due to high path loss at 30 GHz. 

Yin et al. study the MIMO-IAB systems using mmWave with TDM operation for single- and multi-user scenarios \cite{yin2020evaluation}. Simulation results show that the throughput of single-user system with IAB has 20\% better throughput in a certain range of time allocation coefficient. Below this value, there are simply not enough users to effectively use IAB, whereas above this value the bottleneck reduces the performance. For multi-user single RF chain, the throughput gains hare higher and the time allocation coefficient range is wider compared to the previous case. For multi-user dual RF chain, performace gains are only marginal and the time allocation coefficient range is much narrower. However, for ideal antenna array model, the results are much better in terms of throughput (40\% gain) and time allocation coefficient range (more than four times wider).

Solihah, Nashiruddin and Nugraha evaluate the performance of a wireless backhaul system using 10 Gbit-capable symmetric passive optical network (XGS-PON) \cite{solihah2021performance}. The authors implement mobile backhaul in a cellular network in Indonesia using the aforementioned technology. The PON system can deliver 10 Gbps upstream and downstream simultaneously, which makes it ideal for 5G systems. Results show that the system satisfies all ITU-T G.9807 recommendations but the upstream nominal line rate, which it misses by only 4\%. The proposed system can successfully handle jumbo frames of 2000 to 9000 bytes and up to 4096 virtual workgroups. The performance of the proposed system is shown to be adequate for 5G systems, with the exception of sensitivity value which did not meet the ITU standards.  

Bishnu, Holm and Ratnarajah propose a full-duplex enabled IAB network operating in 28 GHz band with large-scale array systems and evaluate its performance in terms of bit-error-rate and spectral efficiency \cite{bishnu2021performance}. Bishnu et al. evaluate the performance of fully loaded multi-cell, multi-user IAB network in the presence of additive white gaussian noise (\acrshort{awgn}) and intra-cell interference. Simulation results show that residual self-interference (SI) limits the BER performance but this is mitigated if the SI to desired signal power ratio (SIDR) is high. If there are other users in the vicinity of the desired user, these users cause interference and degrade the BER, which highlights that effective scheduling is required for reliable communication. The authors also implement analog- and digital-domain interference cancellation and show that if the ratio of SI signal after analog-domain cancellation to the desired signal is high, then this leads to better channel estimation, which in turn leads to better interference cancellation in digital-domain, finally improving BER and overall performance. 

Mesodiakaki et al. evaluate the performance of different frequency bands in a multi-hop small cell network employing wireless backhaul \cite{mesodiakaki2016energy}. Mesodiakaki et al. consider V-,E- and D-band solutions and give a comprehensive review about their capabilities and current maturity levels. Evaluations are done in two scenarios. In scenario A, Mesodiakaki et al. focus on a single backhaul link to highlight the performance of each frequency band. Results show V-band presents highest attenuation due to oxygen, but this solution is still good in ranges less than 200m. D-band presents higher rain attenuation compared to other bands, but below 200m this attenuation is only 5 dB, making short-range usage possible. In the ideal case, D-band with 58 dBi antenna gain presents the best performance with longer ranges, but this is not supported by the state-of-the-art case since this technology is not yet mature enough. In the state-of-the-art case, E-band presents the best performance. In scenario B, Mesodiakaki et al. consider a multi-hop small cell network using different frequency bands. D-band achieves best performance in terms of energy efficiency and network throughput. This is also evident in state-of-the-art case, but Mesodiakaki et al. note that D-band performance is highly correlated with available bandwidths.

\begin{table*}[!h]
\caption{A review of the papers on performance evaluation based on their common points.}
\label{challenges:perfeval-summary}
\begin{tabular}{ |P{5em}|P{8em}|P{8em}|P{8em}|P{8em}|P{8em}|P{8em}| } 
 \hline
 Papers & mmWave & Ad Hoc Networks & ML & FSO/PON & MIMO & IAB \\ 
 \hline 
 \cite{esmail201928ghz} & $\checkmark$ & $\times$ & $\times$ & $\checkmark$ & $\times$ & $\times$ \\
 \hline
 \cite{mesodiakaki2016energy} & $\checkmark$ & $\times$ & $\times$ & $\times$ & $\times$ & $\times$ \\
 \hline
 \cite{zhang2021leveraging} & $\checkmark$ & $\times$ & $\times$ & $\times$ & $\times$ & $\times$ \\
 \hline
 \cite{moon2020ofdm} & $\checkmark$ & $\times$ & $\times$ & $\times$ & $\times$ & $\checkmark$ \\
 \hline
 \cite{polese2018end} & $\checkmark$ & $\times$ & $\times$ & $\times$ & $\times$ & $\checkmark$ \\
 \hline
 \cite{bishnu2021performance} & $\checkmark$ & $\times$ & $\times$ & $\times$ & $\times$ & $\checkmark$ \\
 \hline
 \cite{estupinan2020performance} & $\checkmark$ & $\times$ & $\times$ & $\times$ & $\times$ & $\times$ \\
 \hline
 \cite{shi2018channel} & $\checkmark$ & $\times$ & $\times$ & $\times$ & $\checkmark$ & $\times$ \\
 \hline
 \cite{wang2019system} & $\times$ & $\checkmark$ & $\times$ & $\times$ & $\times$ & $\times$ \\
 \hline
 \cite{hu2019intelligent} & $\times$ & $\times$ & $\checkmark$ & $\times$ & $\times$ & $\times$ \\
 \hline
 \cite{solihah2021performance} & $\times$ & $\times$ & $\times$ & $\checkmark$ & $\times$ & $\times$ \\
 \hline
 \cite{yin2020evaluation} & $\times$ & $\times$ & $\times$ & $\times$ & $\checkmark$ & $\checkmark$ \\
 \hline
 \cite{bonfante2018performance} & $\times$ & $\times$ & $\times$ & $\times$ & $\checkmark$ & $\times$ \\
 \hline
 \cite{madapatha2020integrated} & $\times$ & $\times$ & $\times$ & $\times$ & $\times$ & $\checkmark$ \\
 \hline
\end{tabular}
\end{table*}

Estupinan, Jaramillo-Ramirez and Puerta evaluate the performance of OFDM and single-carrier \acrshort{qam} in mmWave channels for fixed wireless access (FWA) \cite{estupinan2020performance}. Results show that single-carrier QAM performs particularly better for higher-order modulations in terms of BER. In terms of phase noise under 64QAM modulation, SC-QAM outperforms OFDM by a significant margin. SC-QAM is also less affected by faulty antenna elements.  

Madapatha et al. analyze the performance of \acrshort{iab} networks compared to the partially fiber-connected networks \cite{madapatha2020integrated}. A comprehensive analysis of the \acrshort{iab} concept according to 3GPP is made. Later, the authors compare the performance of an \acrshort{iab} network to a (partially) fiber-connected one. The \acrshort{iab} network is a two-tier HetNet with multiple MBS and SBSs. A \acrshort{ue} can connect only to one MBS or SBS. Simulations results show that \acrshort{iab} networks can achieve the same performance as the fiber networks with a small increment in \acrshort{iab} nodes. However, the authors claim that these results will be better in the real world since the network planning will be thorough and the access/backhaul resource split can be made dynamically (this split is fixed in the simulation). The authors also highlight the significant cost reduction and increase in network flexibility that comes along with using SBSs instead of fiber. Lastly, if the network density is high, blockage density/length does not affect the performance much. Tree foliage affects coverage density more severely than rain and blockage in low/moderate SBS densities. 

\subsubsection{Lessons Learned}

We highlight the following points from the performance evaulation papers:
\begin{itemize}
    \item The IAB architecture is very promising for future usage. Its performance is shown to be not only adequate for daily use, but with mmWave usage and integration of new technologies, it can easily be an integral part of future 5G networks.
    \item While mmWave is strong in terms of throughput, its limitations are showcased in multiple works. The range is smaller compared to the sub-6 GHz frequencies, and the communication is more sensitive to weather conditions, free space path loss, and blockage. Because of this, some consideration on mmWave usage is required to make it as effective as possible with minimal drawbacks. Effective deployment strategies, beamforming, intelligent reflecting surfaces, and mMIMO seem to be some example technologies that can significantly boost mmWave performance. 
    \item The wireless ecosystem is gaining multiple alternatives with different characteristics. For example, various communication mediums such as mmWave, microwave, or FSO can be used. These mediums have different advantages and drawbacks. Consequently, the combination of multiple technologies may result in a system that is more robust than its individual parts. This is shown for hybrid FSO/mmWave systems in multiple works and hybrid approaches will be used more in the future 5G networks.
\end{itemize}

\subsubsection{Open Issues and Challenges}

While there are numerous papers showing the performance of wireless backhauling with mmWave frequencies, current works mostly focus on the lower parts of the spectrum. While there is an idea about the performance of higher bands such as V-, E- or D-bands, this area is still not yet explored. Performance of IAB networks with higher frequency bands is an open research area requiring more attention. 

While the (relatively) older concepts and technologies such as FR1-FR2 mmWave ranges, IAB usage and hybrid systems are shown to be performing, newer concepts such as cell-free networking, integrated terrestrial and satellite networks, or THz communications do not have such performance evaluation works. Newer technologies need to be shown as performing as well with such works, and this is still an open research area since there are a multitude of concepts being developed for beyond 5G systems. Performance evaulation works on these new concepts can hasten the research and draw interest on these new areas. Table \ref{challenges:perfeval-summary} summarizes the papers mentioned in this section. 

\subsection{Coverage}

In 5G, HetNets, and small cell deployments such as femto- or picocells are envisioned to be employed for expanding cell coverage. As we mentioned in previous sections, these technologies employ wireless backhauling and their performance is directly linked with the wireless backhaul quality. Because of this, coverage enhancements via wireless backhaul are frequently analyzed in the literature. 

\subsubsection{State of the Art}

Lukowa et al. evaluate the performance of IAB with central scheduling \cite{lukowa2020coverage}. The authors consider IAB with mmWave for coverage extension. Lukowa et al. also propose two different TDD schemes, namely Split TDD, and Flexible TDD. In Split TDD, data flows are split in the time domain for all cells, which avoids downlink-to-uplink interference. In Flexible TDD, any BS can be scheduled for downlink or uplink flow in any time slot, allowing effective multiplexing of downlink and uplink slots. Beamforming is used to cancel out the interference and increase the coverage. Power allocation for backhaul links are also controlled to protect the uplink access from strong backhaul interference. Results show that while Split TDD achieves high downlink throughput, Flexible TDD outperforms it even though interference is arguably higher. The deployment of relay nodes also provides significant throughput gains in the $5^{th}$ percentile users for both downlink and uplink. 

Qu, Li and Zhao study the coverage problem in device-to-device relay networks \cite{qu2019coverage}. Qu et al. formulate the coverage problem to maximize the overall system downlink rate. The proposed problem is transformed into a 0-1 integer programming problem and the authors propose a novel algorithm that uses a greedy approach to find the optimal solution. Simulation results show that the relay node covers a larger area as it moves outward from the cell center. This is because the required bandwidth for device-to-device communication gets less and the allocated resources can serve more UEs. As the relay location grows, spectral efficiency goes up as the relay covers more UEs, but then goes down as the backhaul link deterioration overpowers the coverage increase. The proposed scheme is also shown to outperform the existing SINR-based scheme. 

Singh and Singh use full-duplex amplify and forward relays to enhance the coverage probability and the transmission capacity of D2D enabled cellular networks \cite{singh2018coverage}. The authors derive closed form expressions for coverage probability and transmission capacity as a function of D2D user density and SIR threshold. Singh et al. also analyze the effect of D2D user density, relay node density and D2D link distance on the coverage probability and transmission capacity. Numerical results show that for cellular users, coverage probability is improved with relay nodes as they enhance the SIR at receiver. For lower SIR threshold, coverage probability is higher and vice versa. For D2D users, same phenomenon can be observed albeit with higher coverage probabilities than the cellular users in all cases. The transmission capacity also increases with the introduction of relay nodes for both cellular and D2D users, whereas it decreases with higher SIR thresholds. The transmission capacity initially increases for higher relay node densities, but after a certain density value, any increase results in a decrease in the transmission capacity since the gains are overpowered by the interference introduced with the high relay density. 

In \cite{sharma2017joint}, Sharma, Ganti and Milleth have analyzed the coverage of a two-tier network, in which small cells are connected to the MBSs wirelessly and use in-band full-duplex communication. MBSs also serve the UEs, meaning that they have to manage the resource split for access and backhaul. Coverage formulation is made for both IBFD and FDD for both macro and small cells. These formulations are then compared in simulations, which are shown to be on-par with the numerical results found. FDD is shown to have a better coverage probability than IBFD for different configurations of small cell signal-to-interference thresholds and small cells per macro cell ratios. The authors have also introduced bias to the system to connect to small cells even if MBS gives better SIR values. For high bias values, coverage as well as average rate significantly decreases. IBFD suffers from inter-tier interference and low coverage but provides higher rates.

Khan et al. use UAVs to outsource network coverage according to a desired QoS requirement  \cite{khan2020trusted}. In the proposed system, UAVs belong to the operators and the operators may use other operators' UAVs to provide coverage. The authors propose a reputation-based auction mechanism to model the interaction between the outsourcing operators and the serving UAVs. To enforce the service level agreement, Khan et al. propose a blockchain-based system using support vector machines (SVM) for real-time autonomous and distributed monitoring of the UAV service. Simulation results show that increasing the number of monitoring nodes also increases the probability of true classification and reduces the false alarm probability. Ergodic capacity is shown to be positively correlated with the UAV cost for deviation. Furthermore, if the UAV moves towards the area that requires coverage, service quality increases but this also increases the cost since the old service area gets worse service. 

Kamel, Hamouda and Youssef investigate the uplink coverage and ergodic capacity of mMTC considering an ultra-dense network environment \cite{kamel2019uplink}. Kamel et al. model the small-scale fading using a general $\alpha-\mu$ channel model and the Rayleigh fading, and consider the direct MTC access mode where MTC nodes directly connect to small cells. Analysis on the proposed system model reveals that the impact of high density of small cells in UDNs relaxes the requirements on the maximum transmit power, which in turn reduces the complexity, and increases the battery life and cost savings. Simulation results show that the uplink coverage is significantly high at relatively low coverage thresholds. 80\% of the time, MTC nodes experience an SINR of at most 0 dB. As the number of MTC nodes increase, the coverage performance declines since the limited resources are allocated to a higher number of nodes. Higher power truncation threshold results in a significantly higher coverage performance. As the power truncation threshold gets smaller, the effect of MTC node density on the coverage performance also decreases. This is because the network becomes noise-limited at low power truncation threshold. Increasing the small cell density results in a significant coverage performance gain, but after a certain density, the coverage probability reaches a maximum. Increasing the MTC node density at low small cell density results in a sharp decrease in coverage probability, but this decrease becomes milder at higher small cell densities.

Zaidi et al. propose a network where UAVs are used as decode-and-forward relays to extend the coverage \cite{zaidi2019exploiting}. NOMA is applied on two different set of users, namely the ground users and the UAVs. UAVs are used to forward the data to cell-edge users. Simulation show that the outage probability is reduced with increasing SNR and reducing the value of threshold data rate. However, while increasing the transmit power of the UAVs reduces the outage probability, this is not possible as UAVs are power-limited. Increasing the altitude of UAVs also increases the outage probability. Finally, it is shown that for the same parameters, NOMA outperforms OMA systems by a significant margin. 

Jo et al. employ a joint method of "iteration and convex optimization based on power control (PC)" and "advanced range expansion (RE) technique" to overcome the trade-off between the throughput maximization and coverage optimization in 5G ultra-dense HetNets \cite{jo2019self}. The authors derive the transmission power of optimal coverage in terms of minimal handover failure rates. Jo et al. try to minimize the sum of handovers occuring too early or too late. Too early handover is defined as when a handover occurs due to higher signal power from a target BS but actually the received signal is enough to keep up the connection alive. Too late handover is defined as when handover never occurs until the receiver power from serving BS decreases below a certain threshold that is not high enough for connection. Jo et al. propose an iterative power allocation algorithm for the derived power problem. Numerical results show that the proposed algorithm shows 2-5\% better fairness performance in idle mode coverage compared to the benchmark methods. Moreover, RLF rate is shown to converge to zero for the proposed algorithm whereas benchmark methods do not converge. The proposed method also achieves almost the same result as the power control algorithm that only focuses on maximizing the throughput. 

\begin{table*}[!h]
\caption{A review of the papers on coverage issues based on their common points.}
\label{challenges:coverage-summary}
\begin{tabular}{ |P{5em}|P{8em}|P{8em}|P{8em}|P{8em}|P{8em}|P{8em}| } 
 \hline
 Papers & mmWave & Coverage Modeling & HetNets & UAV & ML & IBFD \\ 
 \hline 
 \cite{lukowa2020coverage} & $\checkmark$ & $\times$ & $\times$ & $\times$ & $\times$ & $\times$ \\
 \hline
 \cite{azimi2018coverage} & $\checkmark$ & $\checkmark$ & $\times$ & $\times$ & $\times$ & $\times$ \\
 \hline
 \cite{qu2019coverage} & $\times$ & $\checkmark$ & $\times$ & $\times$ & $\times$ & $\times$ \\
 \hline
 \cite{singh2018coverage} & $\times$ & $\checkmark$ & $\times$ & $\times$ & $\times$ & $\checkmark$ \\
 \hline
 \cite{kamel2019uplink} & $\times$ & $\checkmark$ & $\times$ & $\times$ & $\times$ & $\times$ \\
 \hline
 \cite{jo2019self} & $\times$ & $\times$ & $\checkmark$ & $\times$ & $\times$ & $\times$ \\
 \hline
 \cite{khoshkholgh2019coverage} & $\times$ & $\times$ & $\checkmark$ & $\checkmark$ & $\times$ & $\times$ \\
 \hline
 \cite{sharma2017joint} & $\times$ & $\times$ & $\checkmark$ & $\times$ & $\times$ & $\checkmark$ \\
 \hline
 \cite{shokry2020leveraging} & $\times$ & $\times$ & $\times$ & $\checkmark$ & $\checkmark$ & $\times$ \\
 \hline
 \cite{zaidi2019exploiting} & $\times$ & $\times$ & $\times$ & $\checkmark$ & $\times$ & $\times$ \\
 \hline
 \cite{khan2020trusted} & $\times$ & $\times$ & $\times$ & $\checkmark$ & $\checkmark$ & $\times$ \\
 \hline
 \cite{siddique2017downlink} & $\times$ & $\times$ & $\times$ & $\times$ & $\times$ & $\checkmark$ \\
 \hline
\end{tabular}
\end{table*}

Shokry et al. use UAVs as a cell-free network to provide coverage to vehicles on a highway with no infrastructure \cite{shokry2020leveraging}. Shokry et al. formulate the UAV trajectory decisions as a Markov decision process and use deep reinforcement learning to learn the optimal trajectories of the deployed UAVs to maximize the vehicular coverage. Deep deterministic policy gradient is adapted to solve the continuous control task. Simulations show that the proposed method consistently outperforms random dispatching, fixed dispatching and hovering methods. The proposed method provides the desired coverage with fewer UAVs compared to other methods. The authors also show that the proposed method inherently reduces the energy consumption as well since adding an energy penalty results in a reduction of only 16\% in energy consumption.

Khoshkholgh et al. \cite{khoshkholgh2019coverage} propose a large-scale aerial-terrestrial HetNet and specifically focus on its coverage performance. The authors evaluate the coverage probability as the complementary cumulative distribution function of the SIR ratio. From there, Khoshkholgh et al. conduct Monte Carlo simulations. Results indicate that increasing the percentage of aerial BSs do not always result in a coverage increase. Moreover, high-rise environments can support a higher aerial-to-ground BS ratio since high-density blockages in such environments result in the dominance of NLOS interference. Furthermore, the authors also report that due to the aggressive interference in \acrshort{uav} communication, densifying the network is especially destructive, regardless of the environment. However, this effect can be mitigated by replacing some ground BSs while introducing aerial BSs. Finally, adjusting the height at which aerial BSs reside can increase the coverage probability, especially at sub-urban environments in which the overall performance is lowest. 

Azimi-Abarghouyi et al. \cite{azimi2018coverage} perform coverage analysis for finite wireless networks. The authors model the network using stochastic geometry and Poisson point processes. The given setup is highlighted as being useful for \acrshort{mmwave} communications, indoor, and ad hoc networks. The coverage probability is mathematically derived with a lower bound as well. Numerical results indicate that there is an optimal distance of the user in terms of coverage probability. The tightness of the bounds are validated with the numerical results. Finally, it is shown that coverage probability is improved when the path loss exponent is larger. 

Siddique, Tabassum and Hossain analyze the performance of spectrum allocation schemes for IBFD and OBFD backhauling \cite{siddique2017downlink} and the coverage rate for these schemes. The authors propose two schemes, namely maximum received signal power (max-RSP) and minimum received signal power (min-RSP). It is shown that IBFD and OBFD schemes work well in different environments in terms of coverage. For min-RSP, SBSs that are farther from the MBS have higher coverage rates while closer ones have lower coverage rates. Conversely, for max-RSP, the closer SBSs have higher coverage rates. In terms of duplexing, IBFD scheme has better overall coverage rate than OBFD scheme with both algorithms. In max-RSP case, OBFD backhauling outperforms IBFD for the farthest SBSs only. Whereas in min-RSP case, OBFD backhauling outperforms IBFD only for the closest SBSs.

\subsubsection{Lessons Learned}

We highlight the following points for network coverage with wireless backhauling:
\begin{itemize}
    \item In terms of coverage, small cell usage is the best solution to cell-edge performance issues to date. Employing SBSs to extend the coverage is a solid approach that will be employed frequently in future networks.
    \item While mmWave communications have less range than the traditional sub-6 GHz frequencies, ultra-dense networking is made possible with mmWave usage. In contemporary networks, deploying new BSs to improve coverage often bring their own problems such as inter-cell interference. Using mmWave frequencies with beamforming and mMIMO mitigates most of these problems, and allows new deployments to extend network coverage.
    \item In terms of coverage, machine learning usage is very promising for future networks for a multitude of tasks such as mitigation of coverage holes or interference. While the performance of these ML algorithms is solid in terms of network KPIs, more research on such ML algorithms are required to make these algorithms perform faster, allowing better response times to changing network conditions. 
\end{itemize}

\subsubsection{Open Issues and Challenges}

Coverage analysis is a tool that researchers have frequently used to show the performance gains of small cell usage. Moreover, the chosen papers act as proof that various concepts such as IBFD or aerial-terrestrial networks perform well. Though coverage analysis is a general problem of networking, it is nevertheless important to show that the performance of newer concepts are adequate. Table \ref{challenges:coverage-summary} summarizes the papers that we have covered in this section. 

Coverage enhancements are often made at the expense of capacity or throughput. While coverage enhancements are very important for ubiquitous connectivity targets for future cellular networks, these enhancements need to be able to satisfy the demands of 5G networks. This is also due to the inherent characteristics of the frequencies used for wide coverage; lower frequencies have better range but worse capacity. This tradeoff is still an open research problem and its solution will directly allow the operators to increase the coverage as well as the performance with the existing hardware and infrastructure.

Network deployment is an area closely linked with coverage. Massive coverage gains can be made at the table with the correct network deployment designs. Small cells are at the heart of this phenomenon, and the points that we mentioned for network deployment in Section VII also hold for coverage to some extent.

Although we will mention the rural connectivity in a later section, bringing ubiquitous coverage while meeting the demands of 5G use-cases is a very challenging task. Doing this in a rural context is even harder. While wireless backhauling makes ubiquitous connectivity possible, there are challenges specific to rural connectivity in terms of both coverage and performance. Therefore, ubiquitous coverage is still an open challenge. 

Satellites, UAVs, and HAPSs are also other network elements whose use are made feasible with wireless backhauling. Currently, their main use-case is ubiquitous connectivity and expanding the coverage of existing networks. Consequently, the challenges related to satellites, UAVs, and HAPSs are also in some sense open challenges related to coverage. 

Another point with wireless backhauling is that the traditional coverage models are often 2D models. With mmWave usage and integration of UAVs, HAPSs, and satellites, these models will no longer be accurate. Because of this, new 3D models for coverage are required for efficient network designing \cite{borralho2021survey}.

Infrastructure sharing is another concept that can result in enhancement of coverage quality and availability \cite{borralho2021survey}. This concept not only reduces deployment and management costs, but it also aids in coverage/capacity balance. Network slicing and other technologies that can aid in infrastructure sharing in an effective manner should be investigated in the future. 

\subsection{Security Aspects of Wireless Backhaul}
\label{wbsecurity}

Irrespective of backhauling, security is a concern throughout the networking area. In wired networks, it is theoretically possible to secure the communication by securing the wires and the overall environment. However, this is not possible for wireless communications and in some sense also counter-intuitive since it would require an enormous amount of effort. Moreover, security issues present a conflict of interest for the network operators since implementing such measures not only increases the CapEx and OpEx, but it also means that network performance may be reduced due to implemented security measures. Therefore, a delicate balance is required for implementing security applications in networking.

For the backhaul networks, there are various security requirements. Confidentiality and integrity of data are the basic security requirements of backhaul networks. Since the wireless channel is available to everyone due to its nature, transmitted data can be received by third parties by eavesdropping attacks, or even modified and retransmitted by man-in-the-middle attacks. Mutual authentication can be used to verify the network elements to increase the transmission security, securing the individual links. Authentication also regulates the access to the network and prevents unwanted users from connecting. While the aforementioned problems are also general security problems, the exposure is higher for wireless backhauling as the medium itself is not secure and cannot be secured. Therefore, alternative methods are required to strengthen the network security while using wireless backhauling. 

Network availability is a requirement for seamless service. Availability can be hampered by denial of service attacks, which prevents legitimate users from connecting to the network and/or accessing services. The effect of such attacks on network availability should be minimized. All transmissions between any two elements of the network must be protected by session keys. While the keys must make sure that the transmission is secure from any tampering, the cryptographic operations should be as fast as possible since the required time for these operations directly incurs latency on the transmission and hampers network performance especially for use-cases like eMBB (by reducing the transmission speed) or URLLC (by increasing latency). 

The keys used to secure the communication must also be fresh to prevent any replay attacks. The freshness is especially important as the medium is exposed and if the keys are not fresh, the attacker can capture a sizeable amount of data from the wireless interface and use this data to compromise all communication made with that key. Key freshness helps secure the network and prevents spoofing attacks. Finally, since wireless services often require handovers or migrations to keep the signal strength high, it is especially important that migrations are securely done in order not to compromise the users or other network elements.

\subsubsection{State of the Art}

A comprehensive work on this topic is given by Choudhary, Kim and Sharma \cite{choudhary2019security}. In this work, authors mention various key security requirements. Mutual authentication is required between the terminal and the backhaul hub in order to foil any attempts of camouflage attacks \cite{basin2018formal, moreira2018cross}. Perfect forward secrecy is required in such a way that a compromised private key must not compromise any session keys as well. Cryptographic operations must be lightweight to enhance the backhaul device life as well as satisfy high-speed transmission and low latency \cite{kornaros2018hardware}. The usage of Physical Unclonable Functions (PUF) can significantly increase the security of the system by strengthening the randomization in the cryptographic functions \cite{bordel2018physical, tajik2018physical}. Choudhary et al. also highlight the tradeoffs between high performance and security in the contemporary 5G networks. Moreover, operators also have to consider the disruption caused by any available security implementation to their businesses. Future solutions should keep the operator perspective in mind while providing adequate security functions for the networks.

Cao et al. also give a survey of 5G network security, albeit not necessarily related to wireless backhauling \cite{cao2019survey}. Cao et al. thoroughly explain 5G architecture in general as well as related to security. Regarding wireless backhauling, handover and mobility security aspects in 5G are detailed and the handover procedure is shown to be vulnerable to jamming and replay attacks. Moreover, handover procedure cannot ensure backward security in its current form. Some potential solutions are highlighted by the authors; we encourage the interested readers to check out the paper. Cao et al. stress the need to improve the handover security in order to meet the low latency requirements for 5G. Cao et al. also analyze IoT, D2D, V2X, and network slice security in detail.

Khan et al. analyze the cutting-edge technologies of 5G from a security perspective \cite{khan2019survey}. Similar to \cite{cao2019survey}, Khan et al. do not directly consider wireless backhaul security, but some of the mentioned concepts are related. Regarding wireless backhaul, Khan et al. make a comprehensive analysis physical layer security. Jamming, eavesdropping, and wiretapping are highlighted as attacks that can target the physical layer. MmWave and mMIMO security is highlighted as important facets since these concepts are integral to 5G. Various works related to security in mmWave, mMIMO and other mentioned concepts are surveyed.

Li et al. consider a cache-enabled network and its security in 6G networks \cite{li2020physical}. Li et al. propose a model consisting of UEs, BSs and eavesdroppers. By using a two-hop transmission scheme, Li et al. make efficient use of caching resources while also increasing the secure transmission probability. It is shown via simulation results that increasing BS density increases the security by increasing the hit probability. 

Communication methods that make use of directivity such as mmWave or FSO are inherently secure to eavesdropping, jamming, or interception. Kafafy et al. try to maximize total transmitted rate of a parallel mmWave and FSO link in the presence of an eavesdropper \cite{kafafy2020secure}. The attack surface here is the scattering of mmWave transmission. With a novel power and rate allocation scheme, Kafafy et al. maximize the total information rate between two links while satisfying constraints on the maximum secrecy outage probability. Kafafy et al. also show that increasing the rate of mmWave link decreases the network security, while this problem is not present for FSO. The inherent nature of FSO alongside mmWave/RF is considered by various works in the literature from a security perspective \cite{lei2020secure, lei2017secrecy, wang2019privacy, kafafy2020topology}.

Hao and Qiu develop a MIMO cross-layer secure communication algorithm based on interference strategies \cite{hao2020mimo}. A physical channel-based bi-directional authentication scheme is proposed, which reduces computational complexity of the initial authentication. Convolutional neural network (CNN), recurrent neural network (RNN), support vector machine, and k-nearest neighbor algorithms are used for training the authentication system. In terms of accuracy, 4-layer CNN and 3-layer RNN are the best performers. Certification accuracies of BT and TD schemes are also analyzed and BT is found to be more accurate overall. 

\begin{table*}[t]
\caption{A review of the papers on security based on their common points}
\label{challenges:security-summary}
\begin{tabular}{ |P{5em}|P{8em}|P{8em}|P{8em}|P{8em}|P{8em}|P{8em}| } 
 \hline
 Papers & MEC Security & Node Selection-related Security & Physical Layer Security & FSO Security Analysis & Caching Security & Network Security Surveys \\ 
 \hline 
 \cite{ranaweera2021survey} & $\checkmark$ & $\times$ & $\times$ & $\times$ & $\times$ & $\times$ \\
 \hline
 \cite{kundu2021ergodic} & $\times$ & $\checkmark$ & $\times$ & $\times$ & $\times$ & $\times$ \\
 \hline
 \cite{kim2021moth} & $\times$ & $\checkmark$ & $\times$ & $\times$ & $\times$ & $\times$ \\
 \hline
 \cite{xie2020survey} & $\times$ & $\times$ & $\checkmark$ & $\times$ & $\times$ & $\times$ \\
 \hline
 \cite{hao2020mimo} & $\times$ & $\times$ & $\checkmark$ & $\times$ & $\times$ & $\times$ \\
 \hline
 \cite{lei2020secure} & $\times$ & $\times$ & $\times$ & $\checkmark$ & $\times$ & $\times$ \\
 \hline
 \cite{lei2017secrecy} & $\times$ & $\times$ & $\times$ & $\checkmark$ & $\times$ & $\times$ \\
 \hline
 \cite{wang2019privacy} & $\times$ & $\times$ & $\times$ & $\checkmark$ & $\times$ & $\times$ \\
 \hline
 \cite{kafafy2020topology} & $\times$ & $\times$ & $\times$ & $\checkmark$ & $\times$ & $\times$ \\
 \hline
 \cite{kafafy2020secure} & $\times$ & $\times$ & $\times$ & $\checkmark$ & $\times$ & $\times$ \\
 \hline
 \cite{li2020physical} & $\times$ & $\times$ & $\times$ & $\times$ & $\checkmark$ & $\times$ \\
 \hline
 \cite{zhang2021data} & $\times$ & $\times$ & $\times$ & $\times$ & $\checkmark$ & $\times$ \\
 \hline
 \cite{cao2019survey} & $\times$ & $\times$ & $\times$ & $\times$ & $\times$ & $\checkmark$ \\
 \hline
 \cite{khan2019survey} & $\times$ & $\times$ & $\times$ & $\times$ & $\times$ & $\checkmark$ \\
 \hline
 \cite{choudhary2019security} & $\times$ & $\times$ & $\times$ & $\times$ & $\times$ & $\checkmark$ \\
 \hline
\end{tabular}
\end{table*}

Kundu and Flanagan investigate the effect of multiple source nodes with wireless backhauling on enhancing the secrecy through source node selection \cite{kundu2021ergodic}. The authors provide the solution of ergodic secrecy rate in closed-form. For high SNR values, the slope of ergodic secrecy rate asymptote increases as backhaul reliability improves. The solution also provides a generalized analysis of ergodic secrecy rate under different conditions such as optimal transmit node selection, optimal cooperative relay selection, and optimal transmitter antenna selection. 

Zhang et al. propose an algorithm to add noise to user preference data for increased security while caching \cite{zhang2021data}. Zhang et al. use a local differential private mechanism to protect the users' true preference, while also enabling the CP to estimate the frequency distribution of different contents from the noisy content preference information. A caching revenue maximization problem is formulated and results show that without privacy, the proposed method outperforms the baselines, while with privacy the revenue decreases.  

Kim et al. propose a novel protocol to secure handovers for mobile terminals in 5G and beyond \cite{kim2021moth}. The proposed protocol is shown to satisfy the confidentiality, integrity, authentication, key management, and perfect forward secrecy. Kim et al. also optimize the handover procedure with the proposed protocol. The proposed protocol is shown to be better than the existing 5G security protocols in terms of computational overhead and security performance. 

Xie, Li and Tan give a comprehensive analysis on Physical-Layer Authentication (PLA) \cite{xie2020survey}. PLA is an important concept that can provide security against active attacks (e.g. impersonation, substitution) as well as passive attacks (e.g. eavesdropping), which is an improvement over the existing physical layer security protocols. PLA can be used to authenticate the device itself, being used in tandem with upper layer security. The mentioned work gives a comprehensive literature survey for PLA in terms of works on passive and active PLA. 

Ranaweera, Jurcut and Liyanage survey the security landscape of multi-access edge computing \cite{ranaweera2021survey}. After a brief summary of MEC and its predecessors, Ranaweera et al. highlight the MEC-specific security aspects. Low-resource IoT devices are bringing the risk of compromising the communication channels used in MEC operations, which can then compromise the edge apps and the infrastructure itself. Jamming and denial of service attacks can reduce the availability of MEC systems. The aforementioned threat vectors are related to access network, which is in turn related to wireless backhaul. Ranaweera et al. also mention other threat vectors related to other concepts such as mobile edge network, core network, or the architecture; however, these are not necessarily related to wireless backhauling and hence not mentioned in detail in our work.

The findings of this section are summarized in the table \ref{challenges:security-summary}. 

\subsubsection{Open Issues and Challenges}

To end the section, we wish to mention the open research areas in wireless backhaul security. The traditional architecture is changing to accommodate more SBSs and serve a higher volume of UEs in a cell. Because of this density increase, availability is more important as the attack surface has increased. We have to mention that while the network performance is the prime objective in all research, reducing latency and increasing the capacity and quality of connections should not be done at the expense of network security. Lightweight cryptography is an open research area that can keep up with the performance requirements while increasing the network security. 

As an alternative approach, ultra-dense networks can also make use of decentralized or distributed security. This can reduce the latency in security applications as the decisions can be directly taken by network elements without requiring approval by another network entity. SDN/NFVs can increase the network security via individual policies fitting the applications as well as monitoring the network for any signs of attacks or weaknesses. Nevertheless, distributed security applications and SDN/NFVs are open research areas that can increase network security. 

While the distributed network management significantly increases the performance, it also creates a larger attack surface. Caching is an advertised approach to increase network performance, but secure caching is essential to keep the network secure. While the caching approaches in order to increase network performance is an open research area, security must be taken into consideration in these works. 

With the advancements in wireless sensor networks and IoT, networks will incorporate an enormous volume of these devices compared to UEs. However, their security is a concern as these small devices can be used for or be targeted by attacks. This is an open research area with similar characteristics to wireless backhauling since IoT devices operate with wireless backhauling. However, objectives and problems are not fully shared between wireless backhaul security and IoT security. Literature in this area is growing rapidly; examples can be seen in \cite{sikder2021survey,ren2020potential,butun2019security, cao2019survey, khan2019survey}.  Machine learning approaches are also widely used for IoT security \cite{hussain2020machine,al2020survey}. 

\section{Opportunities and Applications}

\begin{figure*}[]
\centering
\includegraphics[width=4.2in]{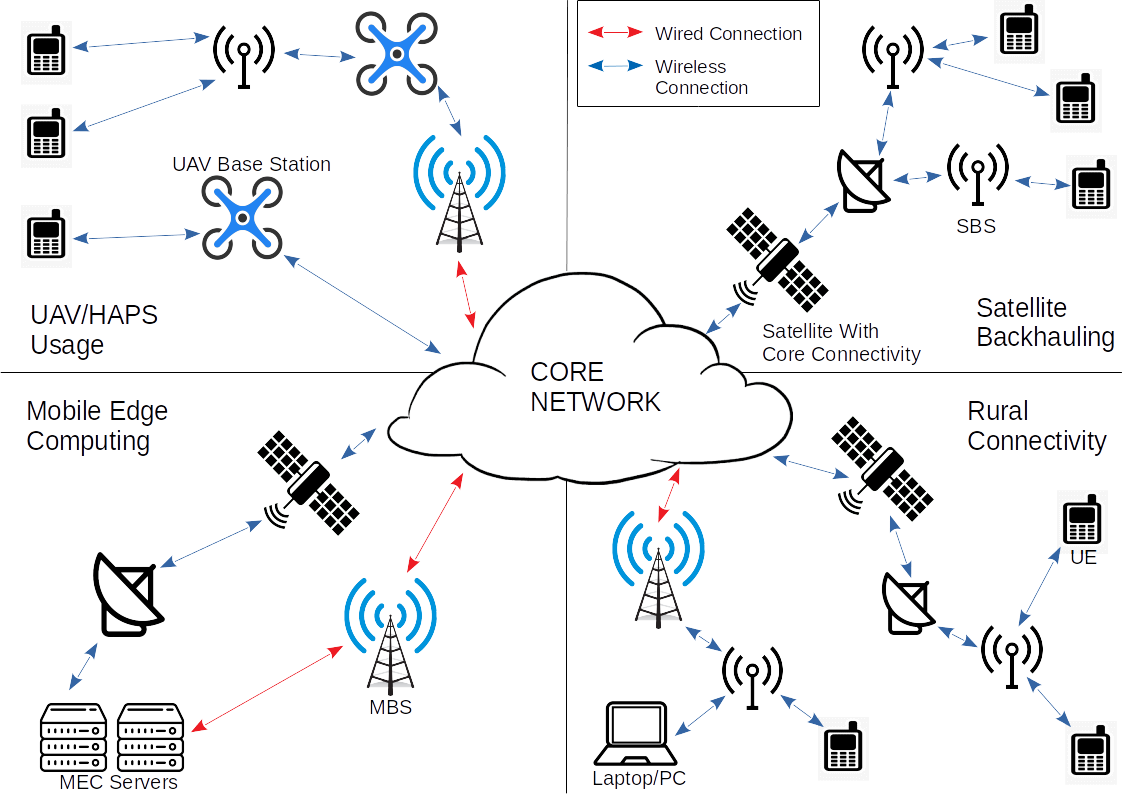}
\caption{Overview of wireless backhaul applications mentioned in this section is given in this figure under four areas. Note the use of different base station types such as \acrshort{uav}, satellite, SBS and MBS.}
\label{oppo:opportunities}
\end{figure*}

This section will overview some of the concepts that are made possible by advances in wireless backhauling. We will first introduce the \acrshort{haps} and \acrshort{uav} usage in the researched networks. This introduction is necessary as these technologies are very important in the areas we will mention due to their characteristics. We will then overview works on mobile edge computing (\acrshort{mec}), satellite backhauling, and rural connectivity. The emergence of wireless backhaul strengthens mobile edge computing use-cases since it provides an opportunity to bring edge nodes closer to base stations. Similarly, the advancements in wireless backhaul provide new opportunities for rural connectivity. Finally, although satellite backhauling has been in use for quite some time, it was not competitive in terms of cost and latency. With the advancements in wireless backhauling in 5G, however, satellite backhauling also becomes a competitive alternative. Figure \ref{oppo:opportunities} gives an overview picture of the mentioned concepts and how they perform together.

\subsection{HAPS and UAV Usage in the Network}

In this section, we will mention the \acrshort{haps} and \acrshort{uav} usage in wireless backhauling and different ways that these technologies are used in the research. We will not mention all works that use \acrshort{haps}/\acrshort{uav} as some of them are linked closer to their main topic (e.g. rural connectivity, satellite backhauling, deployment, or coverage) than the technology itself. Therefore, this section will not be a comprehensive collection of all existing works, but the technology itself.

\acrshort{haps} and \acrshort{uav} are aerial platforms that can arbitrarily change their locations and act as mobile network components. Their high altitude enable them to maintain line of sight for reliable communications. Moreover, they can move closer to the UEs to allow transmission with lower power. Mobility of these solutions introduce completely new opportunities to improve the effectiveness of the networks or solve old problems efficiently enough to be considered by the service providers. However, just like the wireless backhaul, \acrshort{haps}/\acrshort{uav} usage have challenges of their own such as node placement, air-to-ground channel modeling and resource management \cite{bekmezci2013flying, al2014modeling, orfanus2016self}. 

\subsubsection{State of the Art}

Azizi et al. propose to use profit maximization as a utility function to solve the joint radio resource allocation, air base station (ABS) altitude determination, and user association in a heterogeneous network of ground and air base stations \cite{azizi2019profit}. Since the given problem is a MINLP, it is decomposed into three subproblems, which are iteratively solved until convergence. One of these subproblems is an INLP, which is slightly challenging compared to the other two subproblems which are continuous. This INLP is solved without any relaxation methods. The proposed method has linear complexity in terms of users, which is a significant improvement compared to the similar methods having cubic or higher complexities. Results show that macro aerial base stations (MABS) can operate at higher altitudes than small aerial base stations (SABS) and cover a larger area. However, lower operating altitude means lower path-loss for SABSs. If all of the optimization variables are used jointly (altitude determination, power allocation, sub-carrier allocation, and user association), a network profit increase of 47\% can be achieved. Moreover, compared to having a single ground base station, employing two MABS and two SABS almost doubles the overall network profit. These performance improvements are more evident in suburban \& urban areas, and decrease for dense \& high-rise urban areas.

Jaffry et al. cover the moving networks in 5G in their work \cite{jaffry2020comprehensive}. In this work, non-terrestrial networks supported backhaul is highlighted as a future direction in this research area. \acrshort{uav}/\acrshort{haps} are envisioned to give wide area coverage with wider area and near-LoS coverage. \acrshort{uav}s' low cost also allow them to act as relays with ease of deployment and extended operation times. These \acrshort{uav}s can provide backhaul links to ground-based networks. While this provides a new degree of freedom, there are also challenges related to three-dimensional channels that these \acrshort{uav}s will have to use. Even though \acrshort{uav}/\acrshort{haps} are mentioned only as a future direction, this work is very comprehensive in the area of land-based moving networks with challenges, use-cases and applications and any we encourage all interested readers to consult \cite{jaffry2020comprehensive} for these topics. 

\begin{table*}[t]
\caption{A review of the papers on HAPS/UAV usage based on their common points}
\label{applications:uavhaps}
\begin{tabular}{ |P{5em}|P{8em}|P{8em}|P{8em}|P{8em}|P{8em}|P{8em}| } 
 \hline
 Papers & Machine Learning & Trajectory Optimization & Caching & Coverage & Surveys on UAV/HAPS & Deployment \\ 
 \hline 
 \cite{aftabblock} & $\checkmark$ & $\times$ & $\times$ & $\times$ & $\times$ & $\checkmark$ \\
 \hline
 \cite{park2020optimized} & $\checkmark$ & $\times$ & $\times$ & $\times$ & $\times$ & $\checkmark$ \\
 \hline
 \cite{hu2019intelligent} & $\checkmark$ & $\times$ & $\times$ & $\times$ & $\times$ & $\checkmark$ \\
 \hline
 \cite{liu2020fast} & $\checkmark$ & $\times$ & $\times$ & $\times$ & $\times$ & $\checkmark$ \\
 \hline
 \cite{shokry2020leveraging} & $\checkmark$ & $\times$ & $\times$ & $\checkmark$ & $\times$ & $\times$ \\
 \hline
 \cite{cao2020deep} & $\checkmark$ & $\times$ & $\times$ & $\times$ & $\times$ & $\times$ \\
 \hline
 \cite{khamidehi2021trajectory} & $\times$ & $\checkmark$ & $\times$ & $\times$ & $\times$ & $\times$ \\
 \hline
 \cite{wu2020joint} & $\times$ & $\checkmark$ & $\times$ & $\times$ & $\times$ & $\times$ \\
 \hline
 \cite{zhang2020cache} & $\times$ & $\times$ & $\checkmark$ & $\times$ & $\times$ & $\checkmark$ \\
 \hline
 \cite{wang2021content} & $\times$ & $\times$ & $\checkmark$ & $\times$ & $\times$ & $\times$ \\
 \hline
 \cite{khan2020trusted} & $\times$ & $\times$ & $\times$ & $\checkmark$ & $\times$ & $\times$ \\
 \hline
 \cite{jaffry2020comprehensive} & $\times$ & $\times$ & $\times$ & $\times$ & $\checkmark$ & $\times$ \\
 \hline
 \cite{mozaffari2019tutorial} & $\times$ & $\times$ & $\times$ & $\times$ & $\checkmark$ & $\times$ \\
 \hline
 \cite{kurt2021vision} & $\times$ & $\times$ & $\times$ & $\times$ & $\checkmark$ & $\times$ \\
 \hline
 \cite{dao2021survey} & $\times$ & $\times$ & $\times$ & $\times$ & $\checkmark$ & $\times$ \\
 \hline
\end{tabular}
\end{table*}

Wang et al. investigate the successful content delivery performance in integrated UAV-terrestrial networks \cite{wang2021content}. In the proposed model, caching-enabled UAVs are used to offload the bursty traffic from the terrestrial cellular networks. UAVs use self-backhauling and share the spectrum resources with the terrestrial network. Wang et al. derive a closed-form expression for the achievable rate of the mmWave backhaul link for UAVs under general and noise-limited cases. Then, the authors analyze the minimum cache hit probability to achieve a certain backhaul rate requirement. Finally, using stochastic geometry, Wang et al. also analyze the successful content delivery probabilities with an approximated LoS model. Numerical results show that adding UAVs with LoS increases the successful content delivery probability. Self-backhauling with mmWave is shown to achieve adequate results, but with increased UAV density the achievable rate is reduced. Minimum cache hit probability is positively correlated with the backhaul rate and UAV density. Caching-enabled UAV network also outperforms the terrestrial-only network in terms of successful content delivery probability. Increasing UAV density increases the successful content delivery probability up to a certain point, after which the probability decreases because of the LoS interference. The optimal UAV height is also shown to decrease with higher UAV densities. 

Wu et al. study the joint optimization of user scheduling and trajectory for UAV networks with NOMA usage \cite{wu2020joint}. K-means clustering is used to partition the users. The problem of joint scheduling and trajectory is a non-linear non-convex optimization problem, and block coordinate descent method is used to divide the original problem into two blocks, namely the multi-user communication scheduling block and UAV trajectory block based on NOMA. Wu et al. propose an iterative algorithm to alternately solve the subblocks. Simulation results show that the proposed NOMA method outperforms OMA, NOMA simple circular trajectory and NOMA static position methods in terms of system max-min rate. 

Dao et al. study the aerial radio access networks (ARAN) in their survey \cite{dao2021survey}. Dao et al. first describe the ARAN architecture and its fundamental features related to 6G networks. Then, the authors analyze ARANs from several perspectives such as energy consumption, latency, transmission propagation, and mobility. Dao et al. mention energy refills, network softwarization, mobile cloudization, data mining, and multiple access methods as technologies that enable the success of ARANs. Event-based communications, aerial surveillance, smart agriculture, urban monitoring, health care, and networking in underserved areas are given as the application areas of ARANs. As a final note, Dao et al. highlight the open research areas and trends towards 6G ARANs.

Cao, Lien and Liang propose a deep reinforcement learning scheme for intelligent multi-user access to non-terrestrial BSs \cite{cao2020deep}. Cao et al. use a centralized agent deployed in non-terrestrial BSs to train a deep Q-network (DQN), and UEs independently make their access decision based on DQN's input. The authors use a UE-driven scheme, which eliminates the need to retrain the system when the number of UEs is changed. Cao et al. also design a new long short-term memory (LSTM) network to capture the time-dependent feature of non-terrestrial BSs. Simulation results show that system throughput increases with a higher number of deployed UEs/non-terrestrial BSs. The proposed deep reinforcement learning scheme outperforms RSS-based, Q-learning-based, UCB learning-based and random methods in terms of system throughput. The proposed scheme also performs significantly less handovers than other methods.

Mozaffari et al. present a tutorial on UAVs in wireless communications \cite{mozaffari2019tutorial}. In their tutorial, Mozaffari et al. present the potential applications of UAVs, key research directions on these potential applications, and open problems of these research directions. Coverage and capacity extension, disaster-related deployments, connectivity enhancement, 3D MIMO-, IoT-, and cache-enabling are highlighted as potential applications for UAVs as base stations. Flying backhaul is also highlighted as another potential application to enable cost-effective, reliable, and high-speed wireless backhaul connectivity for terrestrial networks. Air-to-ground channel modeling, optimal deployment, trajectory optimization, cellular network planning and provisioning, resource management and energy efficiency, and drone UE usage are highlighted as challenges and open problems. Centralized optimization, optimal transport theory, machine learning, stochastic geometry, and game theory are proposed as the analytical frameworks to enable UAV-based communications.

Khamidehi and Sousa investigate the trajectory optimization problem for multi-aerial base station (ABS) networks \cite{khamidehi2021trajectory}. The objective of the optimization problem is to maximize the minimum data rate of cell-edge users under ABSs power, backhaul link capacity, and collision avoidance constraints. ABSs are first partitioned into clusters with a modified K-means approach so that UEs are served by their associated ABS. Then, the optimization problem is divided into three sub-problems, namely power allocation, joint ABS-user association and subchannel asignment, and trajectory optimization. An iterative method with successive convex approximation is used to efficiently solve these subproblems. Simulation results show that increasing the maximum flight time also increases the max-min data rate since ABSs have more time to reduce the distance to the cell-edge users. With increased maximum flight time, ABSs also adjust their trajectories so that they can visit all associated users. Backhaul capacities are given as the bottleneck of the given problem, but after a certain increase in backhaul capacity, the bottleneck becomes the maximum transmit power constraint. Constraining the propulsion power of ABSs also increases the service time; without this constraint, ABSs cannot finish their mission with the available power.

Kurt et al. present a vision for future HAPS networks and state-of-the-art literature review \cite{kurt2021vision}. Kurt et al. first present HAPS use-cases for next generation networks. HAPS-Mounted super macro base station (HAPS-SMBS) is proposed as a complementary solution to terrestrial systems. These HAPS-SMBSs are envisioned to support data acquisition, computing, caching, and processing. Use-cases for HAPS-SMBSs are given as supporting IoT devices, backhauling SBSs, covering unplanned user events, operating as aerial data centers, filling coverage gaps, supporting and managing aerial networks, supporting intelligent transport systems, and handling LEO satellite handoffs and providing seamless connectivity. Kurt et al. present a general view of HAPS system and its subsystems. Then, the authors give an overview of the channel models, radio resource management, interference management, and waveform design for HAPS systems. Kurt et al. also highlight the contributions of machine learning in design, topology management, handoff, and resource allocation problems of HAPSs. Finally, the authors present the challenges and open issues in two groups, namely next-generation (10 years) and next-next-generation (20 years) and provide possible roadmaps. 

It is worth mentioning that machine learning approaches are widely used for UAV problems. UAVs form a distributed network and with machine learning, UAV networks can effectively become self-organizing. In the previous sections, we have mentioned numerous works using machine learning for UAV-related problems \cite{aftabblock,park2020optimized,hu2019intelligent,liu2020fast,shokry2020leveraging}. In the future, employing fast machine learning methods and distributive approaches will result in highly flexible UAV networks that require little to no management at all.

\acrshort{haps}/\acrshort{uav} especially shine in rural settings and satellite backhauling. As such, next sections have numerous mentioned works that make use of \acrshort{haps}/\acrshort{uav}. We do not mention them here since \acrshort{haps}/\acrshort{uav} is used as a component in those works to enable satellite backhauling or rural connectivity. A summary of the papers related to UAV/HAPS usage can be found in Table \ref{applications:uavhaps}.

\subsubsection{Open Issues and Challenges}

Although there are many use-cases related with UAVs and HAPSs, there are also open issues prohibiting wide usage. For example, regulatory aspects for UAV/HAPS operation are not uniform and dependent on the location. Furthermore, frequency usage is also crucial for UAV/HAPS since it directly affects performance. As we mentioned before, higher frequencies have different licensing schemes, which may impose limitations on UAV/HAPS operations. Handling the regulatory aspects regarding UAV/HAPS is a challenge that needs to be solved. 

The mentioned use-cases in \cite{kurt2021vision} have different requirements and hence different problems. For instance, HAPS operating as a data center have different requirements than a HAPS extending coverage. Since there is a high diversity in terms of use-cases, system models also have different challenges, and integrating multiple of these use-cases in one design remains an important open research area.

Radio resource management with UAV/HAPS also remains an open challenge. Low computational overhead is desired with UAV/HAPS systems and machine learning approaches are promising in this regard. Moreover, serving different 5G use-cases such as URLLC and mMTC together is another challenge in terms of radio resource management. In such systems, objectives are completely different and changing the service type or supporting multiple use-cases at the same time requires a specific design. UAV/HAPS operations have to support multiple use-cases and switch between them seamlessly as needed, and radio resource management seems to be the key concept to realize this switch.

Network stability is another problem with UAV/HAPS usage. Since aerial systems have potentially high mobility and use aerial ad hoc technologies for interconnection, guaranteeing stability in such a dynamic network is difficult. Hierarchical systems formed between satellites, terrestrial networks and UAV/HAPSs are helpful in this regard, but highly dynamic topologies bring challenges to keep up the existing state of the network and high performance. To alienate the network stability problem and handle the dynamic nature of UAV/HAPS networks, routing protocols and network organization designs seem to be directions to handle different mobility patterns, traffic characteristics, and implement redundancy. To summarize, network stability remains an open research area for UAV/HAPS integration into contemporary networks.

Channel models for 3D designs are also not developed enough and requires investigation. Air-to-air and air-to-satellite links require further elaboration as the channel model directly affects system design and performance. Having accurate channel models can pave the way for the researchers to make performance evaluations to show that UAV/HAPS systems can perform well in their mentioned use-cases. Thus, channel modeling and performance evaluation remain open research challenges.

\subsection{Satellite Backhauling}

Communications using satellites is by no means a new concept, being around for a long time for video and audio broadcasting over large areas \cite{seymour2011history}. For cellular communications, satellite backhauling was seen as the enabler to deliver service to large rural or remote areas where the installation of base stations was not logical for reasons such as high latency and low available bandwidths \cite{chia2009next}. Because of these reasons, satellite backhauling did not receive widespread attention and use, apart from niche use-cases such as post-disaster services. However, satellite backhauling allows rapid deployment and extensive coverage, making this technology still a worthy candidate for research and development. 

There are also challenges from the management side of the network. For example, owning and maintaining the satellite links is not only an expensive endeavor, but it also may not be in the interests of a network operator. The regulations for ownership is also different for different countries, meaning that tailored solutions are required for different locales. Data transportation is also expensive if satellites are used, e.g., per Mbit costs of up to \$350 (as of 2019) have to be considered, whereas a base station easily supplies 150 Mb/s speeds with only 20 MHz bandwidth \cite{turk2019satellite}. As can be seen, the difficulties before satellite backhauling are not entirely technological. Commercial aspects such as who the contractor will be for capital expenditure (\acrshort{capex}) and operating expenses (\acrshort{opex}) of the network, or how network management will be considering multiple contractors are still open questions. Different perspectives have to be considered, and the solutions have to take into account various stakeholders if this approach is to be used effectively.

The characteristics of our atmosphere gives us different alternatives on where to deploy the satellites. Low Earth Orbit (\acrshort{leo}) is the closest region where the satellite constellations reside. This orbit is informally defined as the range below 2000km \cite{mcdowell2018edge}. After LEO, Medium Earth Orbit (\acrshort{meo}) begins. The upper limit is the geostationary orbit, which has an universal definition as the orbit 35786 km above the equator \cite{mcdowell2018edge}. Research is often focused on LEO and geostationary orbit (\acrshort{geo}) areas. In LEO, there is the possibility of forming constellations of satellites that act together to provide coverage from the orbit. 

An example of the projects that make use of the LEO is SpaceX's Starlink. Starlink's satellites are orbited at 550km, which make them visible to the naked eye \cite{mcdowell2020leosat}. These satellites consist of a flat panel that houses the communication antennas, propulsion system and the solar panels. At maximum, up to 30000 satellites are envisaged. This infrastructure will be then used to deliver ubiquitous internet access on a global scale, redundancy in terms of alternative paths, and ubiquitous edge caching/computing \cite{tataria20206g}. Apart from their commercial uses, these constellations can also serve the scientific community. Satellites create enormous data from their sensors, and while the institutions cannot bring this data to the ground for processing, satellite constellations can be employed for this purpose \cite{levchenko2020hopes}.

Satellite networks are almost always proposed to be used in tandem with terrestrial networks or other types. For example, space-air-ground integrated network (\acrshort{sagin}) is a concept that aims to create a solid structure by combining the strength of different network types while covering their inherent weaknesses. The satellite segment provides wide coverage, the air segment enhances the capacity wherever necessary, and the ground segment supports high-rate data access with dense deployment. Nevertheless, this integration is difficult since the challenges of these segments are also combined. In the end, the network design and integration are of great significance in this concept to maximize the performance. Liu et al. bring their focus on this concept in \cite{liu2018sagin} and survey the existing literature. Liu et al. analyze works on the physical layer, mobility management, and system integration in SAGINs. High latency in satellite networks, traffic offloading in integrated networks, management of HetNets, channel modeling, and gateway selection are given as some of the future challenges. 

Nevertheless, with new technologies being investigated for 5G, satellite backhauling also came back into prominence. The technological advances that made wireless backhauling more effective helped satellite backhauling as well. For instance, non-terrestrial networks and radio technologies for them are highlighted by European Technology Platform NetWorld2020 \cite{nw20202021smartnetworks} as an important concept considered for the groundwork of beyond 5G networks. For this purpose, research on this area is expected to be active in the upcoming years.

\subsubsection{State of the Art}

In their work \cite{watts20145g}, Watts and Aliu give an overall picture of the projected usage of satellite backhaul in 5G networks. The authors stress \acrshort{sdn} as an enabler to manage a large network consisting of terrestrial and satellite nodes. Traffic offloading capabilities can optimize the usage of available resources. Satellite nodes can also supplement the terrestrial network during peak-times and replace them in case of failures. To make this a reality, functional split and dynamic deployment to realize this split are the prerequisites. The broadcast and overlaying capabilities of satellites can be enablers for some 5G use-cases. The satellite operation stakeholders are also highlighted, especially the relationship between the satellite service provider and mobile service provider is named the critical link for such an operation. Finally, benefits for maintaining such an operation are availability, coverage, reliable service, and service level latency. 

In \cite{turk2019satellite}, a satellite backhauled network is compared with a traditional terrestrial network. Two satellite networks are proposed: in the first network, satellites directly served UEs, whereas in the second one, the serving gNBs used satellites for their backhaul links. Turk and Zeydan use the second network scheme. The experimental results showed that the satellite network achieved higher channel quality metrics compared to the terrestrial network. For resource usage metrics, while the satellite network achieved higher performance, its standard deviation was also higher. The authors have given the intrinsic delay and jitter of the satellite link as the reason for higher standard deviations. For user plane communication, the authors have highlighted acceleration techniques to improve the performance by mitigating the delay of satellite links. For control plane communications, retransmission and acknowledgment timers have to be calibrated according to the delay and jitter characteristics of satellite links for better performance. Finally, for synchronization, instead of SYN messages, externally installed GPS data was used to correctly synchronize the network since satellite link delay was disruptive on this. Finally, the authors make some design recommendations such as usage of DTLS for end-to-end security, integration of 5G CN inside the satellite gateway to provide adaptive configuration, \acrshort{gnb} authentication, and user and control plane acceleration, to name a few.

A simulation environment is proposed for space-air-ground integrated networks (SAGIN) \cite{cheng2020comprehensive}. The proposed environment integrates all three segments of the network simultaneously, implements the network controllers inside the environment for controlling both edge and core network, and implements various interfaces to allow usage of different mobility simulation software and allow customization of network control functions. The differences between the proposed system and other simulation software are also highlighted. The simulation platform is composed of three layers: the SAGIN infrastructure layer handles the physical environment and network nodes, both their generation and mobility. The network protocol and module layer implements the communication and network protocols used for the simulations. Finally, the network control and application layer deploys and evaluates the network according to the aforementioned two layers. ns-3 is employed as the platform's core simulator, which also supplies the models and libraries for satellite, \acrshort{uav}/\acrshort{haps}, and \acrshort{lte}/WiFi/WAVE communications. VISSIM is used to generate vehicle mobility, and STK is used for satellite orbiting movements. Network controllers are deployed to both edge and cloud to collect information and control the network in real-time. Finally, a case study is given to exemplify the operation of the proposed platform. A RAT selection and control scenario is investigated. The platform analyzes the throughput and control delay of heterogeneous access technologies (\acrshort{lte}, LEO, and \acrshort{uav}). In addition, different strategies of RAT selection are also compared within the platform.

Hu, Chen, and Saad propose a joint resource management framework for radio access, satellite backhaul, and terrestrial backhaul links, while maximizing data rates \cite{hu2019competitive}. In this framework, data rates of satellite backhaul are first optimized based on fixed user association and sub-channel allocation schemes. Then, the problem of sub-channel allocation and user association at terrestrial backhaul and radio access links is formulated as a maximum weighted bipartite matching problem and solved using the Hungarian method. Finally, Walrasian equilibrium can be reached in the problem formulation. This approach is compared with a random connection and Hungarian method without equilibrium methods. Simulations show that the proposed method outperforms the other approaches by 6.2 and 2.5 times, respectively. 

Vazquez, Blanco, and Perez-Neira propose two optimization frameworks for dealing with phase-only beamforming optimization in space-ground communication \cite{vazquez2018spectrum}. The proposed system considers two cases: in the first one, 17.3-20.2 GHz frequencies house both satellite forward links and fixed wireless links between terrestrial entities. In the second case, 27.5-30 GHz frequencies house both satellite backhaul links and terrestrial wireless links. Terrestrial links (named FS links by authors) are assumed to have single antennas, whereas the satellite links are assumed to have multiple antennas. In the first case, FS terminals have to restrict the created interference to the satellite link. In contrast, in the second case, FS terminals have to mitigate the interference that they receive from the satellite link. Both fully-digital and fully-analog beamforming solutions are presented; while the fully-analog solution requires only one RF chain, the fully-digital solution requires $N$ RF chains on transmitting and receiving sides. The transmit side optimization problem is shown to be a non-convex quadratically constraint quadratic programs (QCQP) and is relaxed to a convex problem with the semidefinite program relaxation (SDR) method and is solved via interior point methods. For phase-only beamforming optimization problems, the authors propose two novel solutions that also have relatively low computational complexity. For the transmit beamforming problem, the SDP method, which is the first proposed solution, obtains impressive results in terms of spectral efficiency as well as performance loss at the expense of CPU time required. QCQP method, which is the second proposed solution, performs comparatively worse than the aforementioned method but with very little CPU time. For the receive beamforming problem, the results are similar in terms of performance loss but CPU usage of QCQP methods gets significantly higher than the SDP method.

Artiga et al. have used multi-antenna techniques to improve the spectral efficiency of a hybrid satellite-terrestrial network \cite{artiga2017spectrum}. Two multi-antenna techniques were considered. In PtP, the approach is to optimize the gain in the desired direction while keeping the interference in non-intended recipients below a certain threshold. The considered problem is non-convex due to phase-only control of analog beamforming, and the authors solve this via novel convex relaxation techniques. The approach considers single-stream transmission, which is known to be efficient in low-rank MIMO channels found in LoS conditions. For dense cells, the model will also support multi-stream transmission if necessary. This approach's performance compared to a benchmark is shown to be 3-3.7 and 9 times for single- and multi-stream, respectively. The PtMP technique solves the limitations of sending different streams to multiple users due to its large separation. Two methods are proposed for the PtMP technique, namely one based on hybrid analog-digital beamforming and another one that is low-cost and based on multi-active multi-parasitic antenna arrays. The proposed schemes are shown to be 3.55 and 1.1 times better performing than the benchmark, respectively. Hybrid carrier allocation is also explored, complementary, or alternatively to multi-antenna techniques. In hybrid carrier allocation, the goal is to improve the worst link's maximization in terms of interference. The problem of joint carrier allocation of terrestrial and satellite segment is intractable since the coupling between these two is non-linear due to interference. Therefore, carrier assignment for satellite backhaul network is solved first, and then an iterative algorithm is used to solve the carrier assignment problem of the terrestrial backhaul network. The spectral efficiency of the proposed method is shown to be 1.9 times higher than the benchmark method.

Artiga et al. detail the hybrid satellite-terrestrial network described in the previous paper in \cite{artiga2016terrestrial}. The objective of such a network is given as improving the aggregated throughput and resilience against link failures or congestion while reducing energy consumption and guaranteeing efficient use of already scarce spectrum. Satellite nodes are integrated into the backhaul network just like normal terrestrial nodes that have a direct connection to the core network but with different physical layer features. The network is capable of reconfiguring its topology to adapt to the requirements at any instant. Satellite nodes are integral to this capability as they add various redundant paths to the network, which can be used to offload the congestion of the terrestrial links. This reconfiguration capability also reduces the network's expenditures as the installation and maintenance of the network becomes much easier. In addition, aggressive frequency reuse is employed between terrestrial links and terrestrial-satellite links to make use of the spectrum as efficiently as possible. Terrestrial backhaul nodes use smart antennas that enable reconfiguration and interference cancellation. An entity called the Hybrid Network Manager (HNM) is proposed to coordinate the reconfiguration decisions. Simulations show that the network's reconfigurability allows the reduction of end-to-end latency by employing multiple satellite links and multiple gateways as needed. For throughput, it is shown that even when satellite backhauling is employed, throughput is still maintained, albeit not as efficiently as it was with only terrestrial network. It is also shown that the throughput of the network is positively correlated with satellite links.

In \cite{palattella2019enabling}, satellite backhaul is seen as an enabler to the Internet of Everything, Everywhere (IoEE). The working principles of Long Range (LoRa) networks are explained along with the regional regulations governing such IoT devices and their communications. Since satellite systems offer ubiquitous connectivity, reliability, and cost-effectiveness, the concept is seen as an enabler for IoEE. Two different configurations are proposed for satellite backhauled LoRa networks: direct, indirect access. Direct access implies that the LoRa gateways are on the satellites, but satellites are currently not capable of supporting hundreds of millions of connections all at once. In the indirect access case, devices are connected to a LoRa gateway, which is backhauled using satellite. The authors focus on this case since it has reduced installation complexity and fewer costs associated with it. Synchronization between end-devices and gateways, gateway selection, and cross-layer optimization are given as the open challenges in this area.

Lagunas et al. consider a multi-hop wireless backhaul network composed of several terrestrial stations, some of which have satellite connections to receive backhaul from \cite{lagunas2017carrier}. Both terrestrial and satellite links operate at 17.7-19.7 GHz band, and the carriers can be assigned to multiple links but not full-duplex on a single link. Carriers cannot be assigned to multiple satellite links. The interference between satellite and terrestrial links have to be managed. The objective is to jointly design the terrestrial and satellite carrier assignment with minimal interference. Since the terrestrial and satellite segments are intractable, the problem is divided into subproblems. First, the carrier assignment problem for satellite backhaul network is solved, and then an iterative algorithm handles carrier assignment for the terrestrial backhaul network. Simulation results show a 10.5\% increase with respect to benchmark carrier allocation of the terrestrial-only backhaul network.

Völk et al. propose a network architecture with base stations backhauled by satellites \cite{volk2019satellite}. The proposed network setup has the advantage that the network operator can manage satellite modems and the satellite hub in its network all by itself. Employing these concepts allows seamless integration of this architecture into 5G cellular networks. A satellite emulator is also developed to simulate the pure satellite link in cases where a real satellite link is not available. This emulator also evaluates the building blocks of the architecture without requiring real satellite capacity. Two over-the-air demonstrations were made with the testbed that is formed by the aforementioned modules. In the first demonstration, satellite integration is made into the standard 3GPP network architecture. The satellite communications (SATCOM) network is integrated as if it is a RAN node, which also demonstrated the capabilities of SDN, network function virtualization (\acrshort{nfv}), and edge node integration to SATCOM. Five different devices connected to this integrated network and exchanged information. In the second demonstration, the on-the-move capabilities of the 5G network are shown with a SATCOM-equipped van. This van can set up a standalone cell whenever it had an LoS connection to a satellite, which can be used in disaster relief and temporary deployments. In cases of shadowing where the truck passed high-build vehicles, reconnecting to the network took a few seconds to a minute. Other than this, network functionalities worked as expected. 

\cite{hu2019joint} combines \acrshort{uav} networks with satellite backhauling. Hu, Chen, and Saad model the joint backhaul and access resource management problem across radio access, satellite, and terrestrial backhaul links as a competitive market in which data is seen as the good that must be exchanged to maximize the profit. The proposed network consists of drone base stations (DBS) and SBSs that serve the users, whereas MBSs and LEO satellites backhaul their data to the core. In the modeling, users request downlink data service with high data rates and DBS/SBSs seek to satisfy these requests. On the backhaul side, DBS/SBSs request a high rate backhaul service while MBSs/LEO satellites seek to provide this with minimal power consumption. A Walrasian equilibrium is shown to exist, meaning that the performance if optimal at that point. A distributed and iterative algorithm based on dual decomposition that reaches this equilibrium, i.e., the optimal solution, is developed. This dual decomposition is solved by the heavy ball method. The simulations show that the access link and backhaul link data rates are improved by 2.1 and 3.4 times compared to random allocation, respectively. It is also shown that in the absence of satellites, the network performance only marginally improves as new DBS/MBSs are added, compared to the presence of satellites where the introduction of more DBS/MBSs has a significantly better impact.

\begin{table*}[t]
\caption{A review of the papers on satellite backhauling based on their common points}
\label{applications:satellite}
\begin{tabular}{ |P{5em}|P{8em}|P{8em}|P{8em}|P{8em}|P{8em}|P{8em}| } 
 \hline
 Papers & Multi-Antenna Techniques & Satellite Backhauling as Enabler for Use-cases & MEC with Satellite Backhauling & Hybrid Satellite-Terrestrial Networks & SDN with Satellite Backhauling & Resource Allocation \\ 
 \hline
 \cite{vazquez2018spectrum} & $\checkmark$ & $\times$ & $\times$ & $\times$ & $\times$ & $\times$ \\
 \hline
 \cite{artiga2017spectrum} & $\checkmark$ & $\times$ & $\times$ & $\times$ & $\times$ & $\times$ \\
 \hline
 \cite{palattella2019enabling} & $\times$ & $\checkmark$ & $\times$ & $\times$ & $\times$ & $\times$ \\
 \hline
 \cite{watts20145g} & $\times$ & $\checkmark$ & $\times$ & $\times$ & $\times$ & $\times$ \\
 \hline
 \cite{sturdivant2018systems} & $\times$ & $\checkmark$ & $\times$ & $\times$ & $\times$ & $\times$ \\
 \hline
 \cite{cheng2019space} & $\times$ & $\checkmark$ & $\checkmark$ & $\checkmark$ & $\times$ & $\times$ \\
 \hline
 \cite{ge2019qoe} & $\times$ & $\times$ & $\checkmark$ & $\times$ & $\times$ & $\times$ \\
 \hline
 \cite{cheng2020comprehensive} & $\times$ & $\times$ & $\times$ & $\checkmark$ & $\times$ & $\times$ \\
 \hline
 \cite{turk2019satellite} & $\times$ & $\times$ & $\times$ & $\checkmark$ & $\times$ & $\times$ \\
 \hline
 \cite{artiga2016terrestrial} & $\times$ & $\times$ & $\times$ & $\checkmark$ & $\times$ & $\times$ \\
 \hline
 \cite{di2019ultradense} & $\times$ & $\times$ & $\times$ & $\checkmark$ & $\times$ & $\times$ \\
 \hline
 \cite{volk2019satellite} & $\times$ & $\times$ & $\times$ & $\times$ & $\checkmark$ & $\times$ \\
 \hline
 \cite{watts20145g} & $\times$ & $\times$ & $\times$ & $\times$ & $\checkmark$ & $\times$ \\
 \hline
 \cite{hu2019competitive} & $\times$ & $\times$ & $\times$ & $\times$ & $\times$ & $\checkmark$ \\
 \hline
 \cite{lagunas2017carrier} & $\times$ & $\times$ & $\times$ & $\times$ & $\times$ & $\checkmark$ \\
 \hline
 \cite{hu2019joint} & $\times$ & $\times$ & $\times$ & $\times$ & $\times$ & $\checkmark$ \\
 \hline
\end{tabular}
\end{table*}

Cheng et al. propose a SAGIN edge/cloud computing architecture for offloading computing-intensive applications for IoT \cite{cheng2019space}. The authors propose a joint resource allocation and task scheduling approach to efficiently allocate resources to virtual machines and schedule the offloaded tasks. Cheng et al. also formulate the offloading decision making in SAGIN as a Markov decision process and use deep reinforcement learning (DRL) to learn the optimal offloading policy on-the-fly. Simulation results show that increasing the available computation capacity decreases the average delay whereas increasing the number of tasks increases increases the delay. The proposed method for VM computing and task scheduling performs quite similar to the brute-force method, which shows its efficiency. Computational complexity of the proposed is shown to be very low with high performance results. DRL-based method for IoT computation offloading is shown to have lower cost compared to the greedy and random algorithms and converge quite fast. The proposed DRL-based method also performs more than four times better than the greedy method in terms of energy consumption while having less than half weighted delay. Finally, the proposed DRL-based method employs satellite-cloud offloading more than UAV-edge since its delay is lower, whereas greedy method does not show any affinity towards one offloading source.

Ge et al. present a case where mobile edge computing and satellite backhauling is combined to bring the best of both concepts together \cite{ge2019qoe}. The framework consists of three stakeholders, namely the content provider (\acrshort{cp}) that wants to deliver its content to end-users, 5G mobile network operator (\acrshort{mno}) that makes this delivery from the CPs servers to end-users, and satellite network operator (SNO) that leases its satellite backhauling capabilities. This framework allows effective use of mobile edge by changing the content delivery methods and mediums in a context-aware manner. Furthermore, to meet the QoE requirements, the delivery can use multicast-based delivery as necessary. Ge et al. highlight that this is the first system that uses 5G core network and satellite backhaul to support 4K HTTP-based live streaming applications with QoE assurance. Bottleneck-Bandwidth and Round-trip Time (BBR) is chosen as the TCP congestion control mechanism for delivery through the satellite link. BBR is a rate-based congestion control mechanism and is thus not affected by the packet loss and the satellite link's high latency. Besides, the satellite link is shown to have relatively stable latency (560.03ms) and bandwidth with low jitter (13.02ms standard deviation). Multicast capabilities of satellites are also employed in the case where multiple MEC servers request the same payload. Results show that shorter segment sizes result in more jitter because more requests are made when segment sizes are short, which is prone to channel fluctuation in terms of latency and packet errors. It is also shown that the bottleneck is file size; the rate is positively correlated with the segment size. Finally, employing multiple parallel TCP streams result in congestion, which reduces throughput, meaning that caution is advised when using multiple streams with a satellite link.

Di et al. explore the data offloading opportunities to ultra-dense LEO satellite networks integrated into the terrestrial network \cite{di2019ultradense}. Di et al. consider a multi-tier network consisting of SBSs, some of which have satellite backhauling capability, and MBSs. An optimization problem is formulated for maximizing the sum rate and number of accessed users subject to the backhaul capacity constraints. This optimization problem is decomposed into two parts using Lagrangian dual decomposition, namely maximization of the sum rate and number of accessed users, and total backhaul capacity maximization in satellite networks. Simulation results show that the proposed system outperforms traditional backhauling systems in terms of backhauling delay after loads of 10 Mbps. The proposed system's sum rate is also shown to be better than that of traditional terrestrial network or non-integrated terrestrial-satellite networks. 

Sturdivant and Lee demonstrate a use-case of satellite backhauling in aerospace industry. In their work \cite{sturdivant2018systems}, Sturdivant and Lee propose to deploy IoT systems on commercial airliners. These systems will significantly improve the health monitoring process and maintenance of the air fleets. Real-time monitoring is enabled using satellite backhauling. The data is first gathered at the aircraft manufacturers, which then relay this data to third parties such as airline operators and service providers. This work shows a solid use-case of satellite backhauling concerning multiple stakeholders.

To summarize, satellite backhauling can be a feasible alternative especially with the introduction of new LEO constellations and new technologies. Aside from its niche uses such as providing ubiquitous coverage, satellite backhauling is envisaged to be an integral part of future networks. The importance of satellite backhauling will be even more important with concepts such as integrated terrestrial and space networks. However, this is not yet the reality and the aforementioned technologies must mature to see use in the commercial environment. Table \ref{applications:satellite} presents the papers mentioned in this subsection with a grouping of their topics.

\subsubsection{Open Issues and Challenges}

While satellite backhauling is a strong concept in 5G compared to the previous cellular network generations, there are still some challenges ahead for wide usage. The integration of satellites with UAV/HAPS systems bring some challenges. For example, current routing protocols cannot be used for vertical systems and new routing techniques are required for seamless operation. These new routing protocols also need to take into account different PHY-layer infrastructures used in SAGINs. Furthermore, satellite and UAV/HAPS mobility are inherently different, whose difference can result in disconnections. Because of this, coordinating the movements of UAV/HAPS systems with satellites is also an open research problem. 

SAGINs frequently use the hierarchical topologies as they make sense in their operation. However, design of this hierarchy is crucial for a high-performance network. The deployment of network elements, which PHY system to use for communication, and mobile trajectories of deployed aerial and space elements are all aspects that have to be considered for a high-performance SAGIN. Since this problem has multiple facets, it is quite difficult to model and solve. Therefore, SAGIN deployment using hierarchical topologies remains an open challenge that must be tackled to realize SAGINs with high performance.

Delay-tolerant networking is a concept proposed for space communications since IP-based communications are not fitting for long-range space communications. However, for satellite backhauling, delay-tolerant networking can be employed to better-integrate the LEO and GEO satellites, which have inherent high delays. Managing this high delay is still an open research problem, and solving this problem will effectively allow the satellite nodes to be Internet nodes without any considerable difference. 

It is also possible to use satellites to directly serve the users. However, this is still an open research area as realizing satellite-BSs with high performance is currently beyond reach. On-board processing and implementation of L1, L2 and L3 is required for low-latency operation. Power management is another problem related with the satellite-BSs. Solving these problems will allow the realization of employing satellites as individual base stations that have comparable performance to that of the terrestrial BSs.

Similar to the UAV/HAPS related issues, employing GEO and non-GEO satellites at the same time also spawns coordination problems. Due to scarce spectrum resources, spectrum management between different orbits is required. Furthermore, these satellites have to use the spectrum in a way that does not generate interference for other orbits. For beam-based communications, side-lobe and in-line interference are problems that require management. Thus, spectrum and interference management for GEO and non-GEO satellites is an open research area. 

As we mentioned in previous paragraphs, satellite operations are quite dynamic in nature. Consequently, dynamic and fast resource management is crucial for satellite systems. Resource management systems on satellites have to meet the QoS requirements, provide coverage on its service area and allocate resources accordingly, and manage its mobile links to other network elements at the same time. While not mandatory, automation of management is also desired since it is a valuable step towards self-organizing SAGINs. Developing such resource management systems is an open challenge for satellite networks. 

\subsection{Rural Connectivity}

Even though the quality of cellular connection improves with every generation, the same cannot be said for the coverage area. In 2017, there were still 3.8 billion offline people, and of this 3.8 billion, 1.2 billion lacked any broadband connection\cite{cruz2018enabling}. The vast majority of this unconnected mass live in rural areas in developing countries. Another work gives figures of the Internet penetration across continents or regions from May 2019, which were 89.4\% for North America and 86.8\% for Europe, whereas 51.8\% for Asia and 37.3\% for Africa \cite{yaacoub2020key}. This strengthens the view that rural connectivity is a problem, especially for developing countries where technology is not readily available or affordable.

The rural connectivity problem is not a new one either, which was often neglected by the network operators due to mediocre return on investments (\acrshort{roi}). The sparsity of the population does not pay for the investments made to bring service to them. The deployment of infrastructure can be twice as expensive, whereas revenue opportunities can be as much as ten times lower \cite{cruz2018enabling, handforth2019closing}.  Furthermore, since the areas in question are enormous compared to the urban dense networks, the challenge is quite different. For example, some of the regions do not even have a power grid. The vastness of rural regions or the difficulty of reaching them due to unfavorable terrain or distance also make wired fiber connection too expensive, if not outright unavailable. 

For rural developments, the problem is often the coverage and range from the core instead of capacity. Thus, \acrshort{mmwave} and \acrshort{fso}-based newer solutions do not particularly remedy the rural connectivity problem directly. These technologies are useful only to bring backhaul to rural areas. However, their usage frees lower frequency ranges that can be used in rural context to enable connectivity with their far-reaching capabilities. Moreover, coverage problem is a commercial one that aims to maximize the ROI of rural deployments by bringing service to as many people as possible. 900 MHz frequencies allow 2 to 2.7 times more area coverage than 1800 MHz frequencies, which significantly reduces CapEx per deployed area \cite{cruz2018enabling}.

Three key aspects of rural connectivity problem is highlighted by \cite{handforth2019closing}, namely base stations, backhaul and energy. Base stations are given to be the most commercially developed, having different alternatives that employ different technologies like air balloons, traditional macrocell sites, and small-cells. Backhaul innovation is considered to be a long-term one compared to other alternatives. Existing backhaul solutions are focused on expenditure saving, reliability and adaptability to new technologies. These solutions are based on various delivery methods, such as traditional base stations, satellites, and \acrshort{haps}. Nevertheless, only low-frequency microwave, cellular relay, and satellite-based solutions are currently viable in today's conditions. PtMP solutions, lightly licensed or even license-free spectrum usage, and using multiple frequencies in tandem are highlighted as potential points to reduce rural network expenditure. 

Finally, the \acrshort{gsma} report on closing the global coverage gap \cite{handforth2019closing} focuses on the energy problem as the largest cost increase for rural deployments compared to their urban counterparts. Lack of dependable power supply forces the network operators to use diesel generators, which are prone to theft and transportation challenges. Renewable energy is seen as an alternative to replace these generators; current deployments use these sources in tandem, renewable as the main source and diesel as its backup.

Because of these reasons, wireless backhauling was the primary solution to bring service. Base stations could be backhauled to the core in single or multiple hops via long-range microwave links. If this was not possible, satellite backhauling was considered using \acrshort{vsat} technology. However, these approaches had their shortcomings: satellite backhauling was often accompanied by high latency and high initial costs, whereas microwave backhauling required installation of sometimes multiple repeaters or BSs and spectrum feed, which increased costs significantly. 

Rural connectivity problem also has social and commercial aspects requiring extensive work. New business models and innovations are required as low population densities and less developed/non-existent infrastructures challenge the investors in commercial sense. Cooperation between different network operators and governments is important to make this a sustainable business \cite{handforth2019closing, saarnisaari20206g}. 

The work \cite{yaacoub2020key} deeply delves into this topic and gives different perspectives.  Technologies for backhaul, as well as fronthaul connectivity, are thoroughly analyzed with their pros and cons. The mentioned backhaul technologies are fiber, microwave, FSO, air-based solutions such as \acrshort{haps}, drones, and satellites. Fiber is shown to be a feasible solution for some cases, e.g., plains or desert areas. For microwave usage, frequency licenses and platform leasing or installation costs are shown to be limiting commercial factors. For FSO, two alternatives were considered: Terrestrial FSO is similar to microwave backhaul solution, but instead of microwave, FSO is used. The second alternative, namely Vertical FSO, assumes usage of \acrshort{haps} such as \acrshort{uav} or drones for places where erection or usage of towers is not practical. These \acrshort{haps} use FSO to backhaul the traffic, and they are also less prone to weather conditions. FSO usage between satellites for backhaul is also considered and its advantages are highlighted, such as requiring less power and providing high bandwidth. However, FSO for satellite-ground communication is also considered, and some challenges such as attenuation, absorption, and beam divergence loss are also highlighted. \acrshort{haps} are given as an alternative to costly wired infrastructures, and various alternatives using FSO or \acrshort{mmwave} for backhauling are given. Finally, satellites are considered, which is already summarized in the previous subsection. The cost analysis for these aforementioned alternatives shows that while fiber is initially more expensive than all other alternatives, in the long run it gets cheaper compared to all except 10km microwave links, vertical solar-powered FSO and Google's Loon, which uses hot air balloons. The work also analyzes other topics such as services/applications, case studies, foundations/initiatives, and future directions. However, since they are not in line with our topic, we do not cover those aspects.

In \cite{fiorani2014green}, Fiorani et al. have considered the energy consumption of wireless access networks in rural settings under different configurations between the years 2010-2021. Macro-BS and small cell-BS deployments are considered for the access side, whereas microwave, satellite, and long-range PON are considered for the backhaul side. The methodology followed to assess the energy consumption consists of four steps. In the first step, traffic forecast is made according to the number of villages, number of people in them and their device characteristics i.e. laptop, tablet or mobile. In the second step, the area demand retrieved from the first step is used along with the chosen architectures to design the wireless network to serve the area. The second step outputs the wireless network power consumption and peak traffic statistics of BSs and the number of BSs that need to be deployed. In the third step, backhaul dimensioning is made according to the available output from the second step and the chosen backhaul method. For microwave backhauling, a two-tier network design is employed, in which the BSs are connected to the hubs that provide microwave backhaul to the network core. For satellite backhauling, each BS site is equipped with a very small aperture terminal (\acrshort{vsat}) antenna and accesses the data in TDMA fashion. A satellite gateway connects this network to the network core. For LR-PON backhauling, an alternative to traditional PON solutions is used that provides comparatively larger total split, longer reach and higher bit rates by utilizing optical amplification. Each BS is connected to an optical network unit (ONU), that connects to the network core via local exchange sites. The results highlight some points: if small cells are selected, the network is always coverage limited, whereas macro deployment results in limited capacity. In the long run, small cells consume significantly less power than macro deployments since the capacity demand goes up exponentially. Using small cells with microwave or satellite backhaul results in a significant increase in power consumption, whereas this trend cannot be observed for macro deployments. LR-PON has negligible impact on power consumption, and therefore is the best performing candidate.

For rural connectivity, Saarnisaari et al. \cite{saarnisaari20206g} introduce the term ``digital oasis". Saarnisaari et al. envisage a rural connectivity hub that can work outside the electric grid or without communication fibres. In this concept, the digital oasis has a robust and easy-to-use infrastructure available. The monetization model is quite different compared to traditional networks since rural communities may not have the economic strength of their urban counterparts. Satellites are proposed to backhaul the digital oasis' data. Saarnisaari et al. also highlight a need to work on nationwide rules in use today since they are prohibitive on rural scenarios. In commercial sense, frequency leasing costs are too expensive compared to the ROI; from the performance perspective, allowed transmit powers can be increased to extend the communication range. This work also highlights that the challenges of rural connectivity problem is different than that of \acrshort{embb} and higher data rates. In addition, the technological challenges aim to improve the efficiency of existing systems such as satellites, as well as develop new concepts to integrate rural areas into the connectivity grid. 

Chiaraviglio et al. state that while 5G will most probably not cover rural areas, cheaper technologies that are enabled by 5G such as base stations on \acrshort{uav} can be used to bring coverage to rural areas \cite{chiaraviglio2018optimal}. In their work, Chiaraviglio et al. formulate the problem of minimizing the installation costs of an \acrshort{uav}-based 5G architecture. Costs considered in the problem are the following: site installation costs, radio equipment acquisition costs, \acrshort{uav} costs, costs for placing optical fiber link between sites, and installation costs for solar panels and batteries. The envisioned architecture is a softwarized one in which low-level functionalities are deployed on the \acrshort{uav} whereas high-level virtual functionalities are installed on ground sites. \acrshort{uav} can provide coverage to an area or recharge themselves at a ground station in each discretized time slot. This work assumes that backhaul connection is available. The formed optimization problem is run on a real-time scenario and Chiaraviglio et al. show that the proposed \acrshort{uav} network is 2.5 times cheaper than the traditional 5G network solution.

Showing a different aspect of the previous work, Amorosi et al. formulate the problem of minimizing energy consumption of \acrshort{uav} while ensuring coverage, energy consumption and battery level constraints \cite{amorosi2018energy}. Amorosi et al. call this problem RuralPlan. The network architecture is similar to the previous paper, where \acrshort{uav} carry remote radio head (RRH) and some parts of the base band unit (BBU) for low-level operations. Remaining operations are carried out by dedicated ground sites. Time is discretized into slots for problem formulation and \acrshort{uav} can take one of the four available actions in each time slot. These actions are recharging the \acrshort{uav}, staying idle at a ground site, moving the \acrshort{uav} from one location to another, and providing coverage to a given area. These actions are modeled by a multiperiod directed graph. The problem is modeled through mixed integer linear programming (MILP). RuralPlan is compared with the benchmark problem MaxCov which aims only to maximize coverage. The energy consumption under the problems are found to be 32 kWh and 86 kWh, respectively. These figures are from a solution with limited time, meaning that the algorithm performs well even when execution time is limited. RuralPlan decreases the energy consumption by increasing the number of actions that do not consume energy. Finally, Amorosi et al. highlight development of faster algorithms, varying the set of available \acrshort{uav}, and evaluating QoS requirements of users as future work.

Rural connectivity is a problem that interests both the scientific community as well as the network operators. From the scientific perspective, we are much closer than before to solve the problems of rural setting, especially with the introduction of new technologies such as \acrshort{haps}/\acrshort{uav} and better performing satellite backhauling. However, from a commercial standpoint, these advances alone do not justify the rural deployments. New CapEx/OpEx models or other concepts are required to make rural deployments feasible for the operators by increasing the revenues and decreasing the expenditures. 

\subsection{Mobile Edge Computing}

MEC is defined as a new technology providing an IT service environment and cloud computing capabilities at the edge of the mobile network \cite{hu2015mobile}. The aim of this technology is reducing latency, ensuring highly efficient network operation and service delivery, and offering an improved user experience. This concept is seen as a key emerging technology for 5G networks. MEC is very similar to NFV. Instead of focusing on network functions, MEC focuses on virtualizing applications, enabling them to run at the network edge.

As will be seen in this subsection, MEC research often assumes that computation is a resource that can be assigned similar to bandwidth or power. However, MEC approach emphasizes that UEs often do not have as powerful computational capabilities as computer. Moreover, employing this computational capability in a \acrshort{ue} results in significantly higher power consumption, which is a precious resource already for UEs. Therefore, carrying the computation-heavy tasks to the network edge seems a sensible approach. 

The challenges before the real-world application of MEC is twofold. First is the location of deployment. Second is how to handle the volume of data that these applications will generate on edge. Finally, smart usage of offloading to MEC servers is required, because this decision means spending air resources to preserve computation and power of \acrshort{ue}. This subsection contains work on all the mentioned problems. 

The problem of computation offloading and interference management is also investigated for traditional networks as well as ones that employ wireless backhauling, (see \cite{wang2017computation}), but we only include the ones with wireless backhauling.

In \cite{ge2019qoe}, Ge et al. develop a system that uses MEC and satellite backhauling together. For 4K HTTP-based video streaming, CPs use MEC to manage their operations with VNFs deployed to the MNO's MEC servers. This allows CPs to manage their systems while requiring less interaction with the MNOs. Content delivery is usually unicast between the CP's servers and MEC servers, but if multiple MEC servers request the content, then multicast-based protocols are also supported. The MEC servers can cache the content before users request them. If this is done, then users are served from the cache immediately. Otherwise, \acrshort{vnf} forwards the requests to the origin server and retrieves it. The streaming client requests segments from the MEC server in a periodic manner. MEC server tries to balance this in such a way that it holds any requested segments in its cache before they are requested. However, holding these requests also creates latency in the user's end. Therefore, a balance is needed to minimize the number of held segments while assuring that all requested segments are locally available. Note that this approach realized application-layer multicast since the MEC server downloads the segments only once and then serves all clients from this copy regardless of the time of their requests. Two bitrates (10 and 20 Mbps) with three segment lengths (2s, 5s, and 10s) were used for the experiments. Results show that holding more segments result in better aggregated throughput in the client side, despite reducing the throughput of single links. It is also shown that this framework significantly reduces initial startup delay (from 10-20s to 2.6-2.7s). Finally, as segment lengths increase, latency also increases, but fewer segments are held back. The results are similar to higher bitrates, albeit more segments need to be held, and more latency is incurred.

Tan et al. investigate the virtualized FD-enabled small cell networks framework with edge computing and caching for two kinds of heterogeneous services \cite{tan2018virtual}. These services are high-data-rate service and computation-sensitive service. In the proposed scheme, the wireless network is virtualized by introducing two roles, namely infrastructure provider (InP) and mobile virtual network operator (MVNO). InP owns the physical cellular network and radio resources, which are small cells equipped with edge computing and caching. MVNO then leases these resources from InP, creates, operates, and assigns virtual resources. Tan et al. formulate a joint virtual resource allocation problem, where user association, power control, caching and computing offloading policies, and joint resource allocation are taken into account satisfying UEs' QoE requirements. The initial problem is non-convex and mixed discrete, which makes it challenging to find the global optimum. Therefore, Tan et al. reformulate the problem into a convex one, and then propose a decoupling to transform the problem into a global variable consensus optimization problem. Finally, Tan et al. propose the alternating direction method of multipliers (ADMM) algorithm to solve the problem distributively. Simulation results show that the proposed algorithm outperforms the caching-only and FD-only approaches, while being outperformed slightly by the centralized version of the algorithm in terms of system utility and cost values defined by the authors. 

In \cite{dong2018edge}, Dong et al. propose \acrshort{uav} technology to enable rapid event response and flexible deployment of an edge computing-empowered radio access network architecture. The authors have proposed a system that separates the control and user planes. The control plane is managed through an \acrshort{mbs} that has fiber-based connection to the core network. The user plane consists of the small cells along with primary and secondary \acrshort{uav} that are wirelessly connected to the core. These \acrshort{uav} use free-space optical laser to communicate with each other and the small cells as this method is affected less by the weather attenuation than the traditional RF. This architecture is proposed to be used during unusually high demands or after natural disasters or system failures where the system's repair is projected to take a long time. Due to the mobility of \acrshort{uav}, the resource allocation is a significant problem.  \acrshort{uav} trajectories are considered a resource by the authors as these directly affect the performance of the overall system. When a \acrshort{uav} experiences congestion, the authors propose that other \acrshort{uav}s' trajectories can be altered to relieve congestion and increase the number of the available links on that area. The trajectory alteration and designs also have to consider the \acrshort{uav}s' power usage as the propulsion of \acrshort{uav}s is the dominant power consumer.

Pham et al. propose an optimization framework of computation offloading and resource allocation for MEC with multiple servers \cite{pham2018decentralized}. This framework solves the user association and subchannel assignment (computation offloading part), and transmit power allocation for offloading users and computation resource allocation at MEC servers (resource allocation part). This problem is NP-hard, and authors use matching theory via employing two matching games for subproblems of computation offloading. Transmit power allocation problem is solved using a bisection method, and computation resource allocation problem is solved via the KKT optimality conditions. Using these methods enable distributed but suboptimal solution of the general problem. The proposed algorithm has a linear complexity in terms of numbers of users, servers and subchannels, which is a significant improvement considering that exhaustive search has an exponential complexity. Simulation results are compared with two extremes i.e. local computation only or offloading only and the HODA algorithm \cite{lyu2016multiuser} in which each cell independently decides offloading. The proposed algorithm outperforms HODA in terms of the percentage of offloading users and computation overhead. 

MEC is an important concept for 5G networks that is enabled with high throughputs. Wireless backhaul acts as an enabler for this concept since it allows high throughput in both directions if needed. Because of this, MEC also share some of the challenges of wireless backhauling, such as resource allocation or deployment. Nevertheless, with the direction that networks are evolving to, MEC will be an interesting concept that is enabled with wireless backhaul.

\section{Standardization Efforts on Wireless Backhaul}
\label{standardization-sect}

Some documents are trying to govern the backhaul usage in future networks, such as \cite{isg20185g, 3gpp2019iab, 3gpp.38.874}, and we will go over them in this section to shed light on how the governing bodies see the future of wireless backhauling in 5G.

\cite{3gpp2019iab} is specifically about the \acrshort{iab} usage in NR. This work highlights the opportunities presented by massive MIMO, larger bandwidth, and beamforming to use integrated access and backhaul for 5G networks to reduce dependence on wired networks. It is stressed that \acrshort{iab} is very beneficial for NR rollout and initial growth phases. 

\cite{isg20185g} presents some wireless backhaul/X-Haul scenarios for 5G networks. Wireless backhaul is seen as an enabler for new use cases as well as an enhancement for the existing use cases with the new frequencies introduced from V-band to D-band. The authors envision small cell usage in 5G NR networks' initial deployments only in dense urban scenarios. Various function split architectures are considered along with centralized or distributed RAN. It is noted that high layer split (i.e., split between PDCP and high-RLC layers) presents equal capacity and latency requirements as the traditional backhaul. 

The network topology is predicted to be tighter with the introduction of new fiber links to the existing stations. Furthermore, the introduction of new nodes closer to the existing nodes is possible with wireless backhauling. Mobility and LoS are challenges for small cell backhaul. PtMP is an attractive solution compared to the existing PtP connectivity in the macro sites due to the cost reduction it brings. For the frequencies, it is projected that future networks will rely on V-band and above since there is abundant bandwidth to be used in that range. However, authors note that a combination of a higher and a lower frequency band in a single logical link is the best approach to enjoy the advantages of both bands.

\cite{3gpp.38.874} is the study item for integrated access and backhaul usage in 5G. This study item lists the requirements for the development of \acrshort{iab} in 5G. Moreover, it also lists the potential architectures of \acrshort{iab} networks and how they can operate. In the first group of architectures, centralized/distributed unit (\acrshort{cu}/\acrshort{du}) split is employed. \acrshort{ue}'s are connected to the DU's that handle the radio functions, whereas all DU's are connected to a CU at the last element of the backhaul chain that handles the core network-related functions. The adaptation layer handles the connections between DU and CU's. The second group of architectures employ no such split, and \acrshort{iab}-nodes connect to \acrshort{iab}-donors (i.e., nodes with wired backhaul capabilities) via nested tunneling. 

Physical layer aspects such as backhaul link discovery, scheduling, resource allocation, synchronization, cross-link interference management, and spectral efficiency enhancements are also mentioned. Radio protocol aspects and backhaul considerations are also detailed in this work. The aforementioned adaptation layer is highlighted in this section for architectures alongside other user- and control-plane considerations of both architecture groups.

Backhaul considerations are detailed in the ninth section of \cite{3gpp.38.874}, and this section highlights the strengths of the \acrshort{iab} concept. Two topologies are considered in the study, namely spanning tree and directed acyclic graph (DAG). DAG can be used to introduce link- and route redundancies to increase the overall reliability. Different architectures require different route redundancy mechanisms; for example, for the first architecture group, a DU may connect to multiple CU's via different routes for redundancy. For the second architecture group, the redundancy can be via different \acrshort{iab} nodes to the same or different IP domains. It is highlighted by the document that \acrshort{iab}-nodes and \acrshort{iab}-donors might have different QoS considerations. For example, maximum bitrates can be enforced at the \acrshort{iab}-donors only. Topology discovery and adaptation procedures are also detailed for all architectures.

In conclusion, the deployed \acrshort{iab} architecture is recommended to support in-band and out-of-band scenarios, multi-hop backhauling, topology adaptation, and standalone (SA, 5G only network) and non-standalone (NSA, \acrshort{lte} and 5G networks performing in tandem) mode for \acrshort{ue} as well as the \acrshort{iab}-node. Note that these requirements are features frequently highlighted in contemporary research. 

A recent work by Madapatha et al. also gives a comprehensive review about the \acrshort{iab} concept \cite{madapatha2020integrated}. In this work, the authors give a review of the \acrshort{iab} concept from 3GPP perspective. The architecture, spectrum used for \acrshort{iab}, and the radio link are detailed according to the 3GPP documents. Finally, Release 17 NR \acrshort{iab} advancements are briefly mentioned.

Since \acrshort{iab} is a new development for the 5G networks, recent documentation is about its feasibility and operation to attain the best possible performance. The aforementioned documentations also highlight the strengths of wireless backhauling in general and the \acrshort{iab} concept specifically. These developments are expected to be in the backbone of the 5G and beyond-5G networks.

\section{Wireless Backhaul Beyond 5G}
\label{beyond5g}

In this section, we will briefly mention various technologies and concepts that are not yet mature enough to see widespread use. However, not only they have the potential to improve the performance of the cellular networks as a whole, but they can also serve as enablers for wireless backhauling in beyond 5G networks. Moreover, we will also mention how we envisage the adaptation of existing aspects of wireless networks, such as frequency usage and evolving directions of MIMO, \acrshort{iab} and extraterrestrial networks.

\subsection{Frequency Use Beyond 5G}

Initial deployments of 5G networks will only make use of frequencies up to V-band (57 GHz) \cite{3gpp.38.101-2}. In Release 17, it is expected to extend this support to up to E-band, including the V-band to the supported frequency range \cite{peisa20205g}. This means that even the E-band and D-band on which numerous researches are being conducted, will not see practical use. However, from a theoretical perspective, Terahertz frequencies are already being seen as an enabler for beyond 5G or even 6G networks.

Terahertz (THz) frequencies provide ultra-high bandwidth with ultra-low latency, but the range is significantly reduced due to free space loss by Friis' law \cite{zhao2020survey}. Nevertheless, the potential is very high with indoor usage of THz frequencies, such as personal networks, data centers, and virtual reality applications. For indoor and fixed configurations, THz communication can provide the KPIs of many use cases that require high throughput with low latency, such as holograms and VR/AR. For outdoor or cellular use, mMIMO and beamforming will be indispensable for THz communications \cite{chowdhury20206g}. However, current technology is still short of enabling THz frequencies' effective use since the transceivers and photonics are not yet mature enough.

In the document ``Smart Networks in the context of NGI"  \cite{nw20202021smartnetworks}, THz communication is highlighted as a groundwork research for beyond 5G networks, whereas optical wireless communication is highlighted as an evolutionary research topic. These show that frequency range is expected to go even higher in a drive for higher data rates.

Optical wireless technology is another technology that will be used alongside RF-based communications. These include visible light communications and FSO. These communications can provide very high data rate, low latency and secure communications \cite{chowdhury20206g}. As we mentioned in previous sections, hybrid FSO-RF networks combine the strengths of both communication technologies, increasing the overall network performance. Furthermore, FSO can be used on its own to increase remote connectivity since it can support communication even at a distance of more than 10.000km \cite{chowdhury20206g}. 

THz technology usage in space communications is analyzed in \cite{mehdi2018thz}. The norm in space communication is to use FSO, but Mehdi et al. propose Schottky diodes as an alternative frequency source to reach THz frequencies. The advantages of THz communications over FSO are given as better beam accuracy tolerance due to wider beams, less weight and power requirements, immunity to ground-based interception, no spectral noise or atmospheric attenuation, and lack of band regulation.

Open research areas for THz communications include a new design for the transceiver architecture due to high propagation loss and atmospheric absorption characteristics. The aforementioned propagation characteristics also affect beamforming and mMIMO performance of THz communications and efficient beam management remains a challenge. Furthermore, the channel modeling of THz frequencies are quite complex and since this band does not have any perfect channel model available, development of such models remain an open research area. Finally, health and safety concerns regarding THz band need to be addressed \cite{chowdhury20206g}. 

\subsection{Extremely Dense Networks}

In beyond 5G and 6G, the systems are expected to be based on extremely dense network infrastructure with elements like access points \cite{rajatheva2020scoring}. Wired fiber backhauling challenges are still a reality, meaning that wireless backhaul will be an integral part of such systems. Combined with the development of higher frequencies beyond the \acrshort{mmwave}, these network elements are expected to use high capacity PtP links that use THz or even visible light bands. Furthermore, with the integration of terrestrial and satellite networks, coverage efficiency will be increased, and the network will be able to remedy any high-load situations adaptively. These expectations can be made real using effective wireless backhaul.

Moving elements such as \acrshort{uav} or autonomous vehicles will also aid the densification of the beyond 5G networks. However, this also introduces new challenges such as new blockage dynamics, improving reliability, and availability. Moreover, distributed management is also a requirement to unleash the potential of moving access points. Extreme network densification can also be used to manage the resource demand much better since it gives the capability to handle user requests as one-on-one relationships. In the extreme case, users can also be involved in the scheme as network elements providing services. However, this is not very realistic since operators and users inherently have conflict of interests. Andreev et al. also give simulations for such a use-case involving moving access points as network elements in an extremely dense network  \cite{andreev2019future}. Network outage probability is significantly reduced with moving access point usage. Moreover, using more moving access points results in less need to use stationary \acrshort{ap}, which reduces operator expenditure. All different configurations of APs result in improvements over the baseline case where no moving access points are used.

Extremely dense networks come with their own challenges, which are also dependent on the individual use cases requiring the densification. For example, in automation and manufacturing, high reliability, low-latency and high data rates are required \cite{chowdhury20206g}. The network has to provide these services for hundreds, or even thousands of the connected elements. Service requirements remain an open challenge in this area, though we have to mention that this is not a specific challenge but a general one related to 6G communications. Moreover, management of extremely dense networks remain an open challenge as the signalling overhead will be an issue by itself due to high volume of connected elements. 

\subsection{Integrated Access and Backhaul}

In \cite{rajatheva2020scoring}, Rajatheva et al. highlight \acrshort{iab} as an infrastructure-level enabler for Terabit/s broadband connectivity goal in 6G. The large and contiguous available bandwidths in \acrshort{mmwave} and beyond create an economically viable opportunity for backhauling, compared to the already congested sub-6 GHz frequencies. Since \acrshort{mmwave} and higher frequencies have shorter ranges compared to the traditional \acrshort{lte} access frequencies due to their inherent traits, this will push for denser networks. Getting better service from a denser network is directly linked with a good quality backhaul, which will mostly have to be wireless. Finally, similar in 5G, the advances in MIMO technologies that the authors have dubbed as ultra-massive MIMO, the same frequencies can be used both for access and backhaul thanks to spatial multiplexing. 

It has to be noted that \acrshort{iab} still gives worse performance than the fiber backhaul in terms of performance and this will not change in the foreseeable future even with \acrshort{mmwave} usage \cite{rajatheva2020scoring}. Simulations and some experiments also show this trend. However, it has to be noted that, first of all, it gives a significant coverage advantage compared to the fiber. Moreover, compared to the fiber, its installation is much easier; this characteristic alone gives \acrshort{iab} an edge in flexibility and roll-out times \cite{rajatheva2020scoring}. A small cell is cheaper compared to fiber cabling, as highlighted in the previous sections of this work. Finally, it needs to be taken into account that in practice, the network operators can make \acrshort{iab} network deployments much better than the simulations, meaning that the performance gap will not be as wide.

FSO is a concept that can help for better IAB performance. FSO systems provide high data rates with low latency, which is helpful for keeping up with 6G KPIs. Using FSO can allow IAB systems to be deployed more remotely, as well as increasing the overall performance \cite{chowdhury20206g}. Hybrid FSO/RF systems can also be used for IAB, which allows reliability and increased tolerance to atmospheric effects.

\subsection{Ultra-Massive MIMO}

The ultra-massive MIMO (UM-MIMO) remedies THz communications' range problem with bigger antenna arrays, such as 1024x1024. These bigger arrays can significantly boost signal strength by steering and focusing the beams in space and frequency \cite{akyildiz20206g}. Interestingly, these arrays can also be used with lower frequencies since the frequency is inversely proportional to its spatial multiplexing capability \cite{rajatheva2020scoring}. This means that UM-MIMO has the potential to remedy both ends of the spectrum; in sub-6GHz frequencies, it can be used to increase the effectiveness of small cells and access networks, whereas in \acrshort{mmwave} frequencies and beyond, it can be used as an enabler for higher bandwidth usage.

Perhaps one of the biggest challenges related to the antenna arrays is to build an array that is cost- and power-efficient while having high performance. For this purpose, Jamali et al. propose using hybrid arrays that have active elements alongside intelligent surfaces as passive elements \cite{jamali2020intelligent}. Employing passive elements reduces the power consumption considerably, while somewhat reducing the performance. However, if the losses related to the intelligent surfaces such as phase shift loss, taper loss or aperture loss can be handled correctly, then the performance gains can be achieved with such hybrid systems.

For UM-MIMO usage in beyond 5G, channel models for higher frequency bands such as THz frequencies are needed as the mmWave channel models have different characteristics. Although there are some developments in this direction such as \cite{wang2021general}, new channel models for UM-MIMO is an open research area. Moreover, while UM-MIMO is seen as an enabler for THz frequency usage, most works are theoretical and empirical studies and performance evaluations are required to show the performance gains with the UM-MIMO systems. Performance evaluation and physical implementation are also open research problems for UM-MIMO.

\subsection{Integrated Space and Terrestrial Networks}

6G networks are also expected to initiate the integration of terrestrial and space networks, with the mission of bringing ubiquitous global connectivity \cite{chowdhury20206g}. This is by no means a new concept, but it was practically impossible until the recent technological advances have shown a theoretical possibility. Currently, existing space networks are inhibited by limited bandwidth and considerable transmission delays. In an integrated terrestrial and space network, there will be three layers. The existing ground network will be linked to the space network that consists of geostationary, medium-Earth, and low-Earth orbit satellites via an airborne network. This airborne network will comprise \acrshort{uav}, aircraft, balloons, and high-altitude platforms. This airborne network will be the middleman that enables high-speed and flexible operation in any condition. However, current aerial networks use microwave links limited in bandwidth, and hence cannot meet these demands. 

The developments in wireless backhaul will be an enabler for the future link between terrestrial and space networks. Especially, \acrshort{mmwave} seems a solid choice for this use since it can provide the necessary bandwidth and also provide multi-channel transmission and spatial reuse. Nevertheless, current \acrshort{mmwave} technology is not advanced enough to do this. The reasons are the difficulties in IBFD communications, shorter communication distances than required, and inadequate levels of achieved spatial multiplexing, to name a few \cite{rajatheva2020scoring}. With the development of \acrshort{mmwave} technologies, coupled with three-dimensional network design and optimization, the integrated terrestrial and space network will be a cornerstone of future developments.

THz frequencies are also considered for inter-satellite communications since atmospheric attenuation does not affect the communication in that medium \cite{akyildiz20206g}. In that use case, beam alignment is not necessary for THz communications compared to optical links, meaning that the system's robustness increases in case a satellite drifts out of its orbit. Moreover, since THz frequencies have much higher bandwidths available and have improved link performance, this technology will improve the wireless backhaul capabilities of extraterrestrial systems.

A challenge before the integration of terrestrial and space networks is that the conventional networks are modeled in 2D fashion whereas this integration requires the third dimension to be taken into account as well. 6G HetNets are expected to provide 3D coverage, the integration of terrestrial networks, UAV networks, and satellite systems will realize the global coverage and seamless access \cite{chowdhury20206g}. 3D networking is still an open research area for 6G in general as well as integrated space and terrestrial networks in particular. 

In 6G networks, satellite integration will bring new challenges and there are open research areas. Kodheli et al. mention cooperative satellite swarms, internet of space things, flying base stations as satellites, dynamic spectrum management, satellite network automation and resource management, and machine learning applications as some research areas for beyond 5G satellite networks \cite{kodheli2020satellite}.

\subsection{Cell-Free Networking}

Another proposed concept in 6G is cell-free networking \cite{rajatheva2020scoring, akyildiz20206g}. In this scheme, instead of one massive antenna array with hundreds of antennas at the MBS, numerous low-cost access points with few antennas are used. These APs cooperate to serve the UEs together, meaning that a \acrshort{ue} will be close to at least some of them with very high probability. To enable this cooperation, APs connect to one or multiple CPUs using fronthaul links. This approach is shown to improve the 95\%-likely per user throughput by five to ten times with respect to small-cells \cite{akyildiz20206g}. With the advances in wireless backhaul, APs in this configuration can do their work more effectively, making this theoretical scheme a viable option for ultra-dense networks. 

Cell-free networks can operate using mmWave or higher frequencies, but the problem of beam selection and control becomes a problem since a high number of APs will inevitable have beam conflicts that has to be solved. Yetis et al. \cite{yetis2021joint} propose using machine learning for beam selection, which is very effective in terms of performance but also fast and efficient in terms of complexity. 

For cell-free networking to be a practical alternative to massive MIMO and small cell schemes, some of its inherent problems have to be solved. For example, removing the cell structure changes how the initial access procedure will be executed. In cell-free networks, various APs will be in the range of the \acrshort{ue}. Problems here are how APs will identify themselves, broadcast the necessary information to the UEs, and synchronize the operation between them when multiple APs serve the same \acrshort{ue}. Furthermore, coordination among APs is required to reduce the amount of \acrshort{csi} that they have to share for seamless operation. Having APs that delegate as much of their operation as possible to the CPU massively boosts cell-free networks' performance at the cost of complexity and fronthaul signaling \cite{rajatheva2020scoring}. 

There is also a limit to the number of users that can be served with cell-free massive MIMO systems. However, this is not taken into consideration in previous works and is an open research problem \cite{akyildiz20206g}. Finally, location optimization of APs is also another open topic as AP placements are often made according to traditional cell schemes that may not perform up to this new paradigm's potential. The effect of random scatterers, users, and AP locations on a cell-free massive MIMO system's performance is not analyzed and remains an open research problem.

Nevertheless, challenges must be tackled before this scheme can be a viable option, such as initial access without cells and massive volume of cell data exchanges between APs and CPUs. While deep learning and non-orthogonal multiple access is proposed in \cite{yu2020non} for wireless backhaul links, uplink transmission is still an open research area in cell-free networking. Multi-connectivity and heterogeneous radios are seen as enablers in cell-free networking, but these are still open research areas requiring further works \cite{chowdhury20206g}.

\subsection{Intelligent Reflecting Surfaces}

In the previous sections, we have explained that due to the inherent nature of \acrshort{mmwave}, the communications often require LoS. In 6G, the work on intelligent surfaces is promising to improve the performance of LoS systems, especially in urban areas. Intelligent reflecting surfaces (\acrshort{irs}) are software-controlled passive arrays that have an immense potential towards energy-efficient green communication \cite{chowdhury20206g}. IRSs allow customizing the propagation of radio waves in such a way that data rate is increased without any increase in power consumption \cite{di2019smart}. Moreover, IRSs can also modify the signal phase and signal power. IRSs are also low-cost, due to reasons such as easy deployment, using low-cost passive components, and not requiring backhaul or any energy source \cite{zhao2020survey}. 

A practical application of this concept can be seen at \cite{papasotiriou2020performance}, where Papasotiriou et al. try to establish LoS communication in the presence of an obstacle in transmit path. For this purpose, reconfigurable-intelligent-surface (RIS) is used to reflect the beam to maintain communication. Papasotiriou et al. implement a system with RIS and results show that the performance is highly sensitive to changes in antenna gains and elevation angles, so much so that a $3.5^{\circ}$ change results in 70\% reduction in TX/RX power radiation patterns. Path gain also acts similarly, especially for high antenna gains at TX/RX side. 

IRSs can also be used together with other technologies such as \acrshort{haps}/\acrshort{uav}. Alfattani et al. \cite{alfattani2020aerial} suggest a number of use-cases for IRSs on \acrshort{haps}/\acrshort{uav}. Reconfigurable smart surfaces (RSS) can be mounted on a swarm of \acrshort{uav} to reflect the waves originated at the ground base stations (gBS) to improve the overall service. This can be used to introduce spatial diversity as well as improve data rate by employing different transmission paths. Alfattani et al. also propose to use a tethered balloon equipped with RSS (TBAL-RSS) that can be configured as a low-cost access point to connect or enhance the user capacity. Large surface area of TBAL means that it can accommodate a high number of reflectors. 

Kurt et al. also propose mounting RSS on HAPS (HAPS-RSS) in \cite{kurt2021vision}. These HAPS-RSSs can accomodate a large number of reflectors and do not suffer from noise or self-interference, meaning that they can support full-duplex communication. The power consumption requirements of HAPS-RSS is also quite low, meaning that it can have a prolonged flight duration. These aspects make HAPS-RSS an effective alternative to relays. 

Al-Jarrah et al. use IRSs in a mesh network with wireless backhauling \cite{al2021performance}. The MBS is connected to the core network with fiber cables, whereas the SBSs are connected to this MBS wirelessly. MBS and SBSs are configured in a mesh topology, and SBSs make use of IRS panels to have LoS to their next hops. SNR distribution, signal error rate, and outage probability analysis are made for the mentioned setup. Monte Carlo simulations show that increasing the number of reflectors in an IRS panel considerably enhances the signal error rate. Furthermore, increasing the number of reflectors also reduces the outage probability; increasing the hop count cannot be feasibly done with a small number of reflectors as the outage probability becomes too high.

While IRS is seen as an emerging technology enabling significant performance gains, Gong et al. mention the following open challenges and future research directions in this area \cite{gong2020toward}. Since IRSs are composed of passive elements, channel sensing and estimation is a problem without dedicated hardware. Furthermore, a practical protocol is required to govern the information exchange between an active transmitter and the IRS. Use of IRSs in HetNets require dynamic and fast reconfiguration. For future research, machine learning approaches can be used for passive beamforming. IRS-assisted D2D, mmWave and THz communications are also important research topics as IRSs are inherently helpful for extending the coverage via LoS connectivity. Finally, IRSs can also be used to convey information about the network in smart wireless sensing applications.

\section{Conclusion}

Wireless backhauling is an established concept seeing frequent use in cellular networks. However, with the introduction of the 5G and new technologies such as \acrshort{mmwave} frequencies and beamforming, wireless backhauling evolves to a new degree that acts as an enabler for new use-cases like rural connectivity, efficient satellite backhauling for ubiquitous connectivity, and mobile edge computing. However, wireless backhauling has its own challenges to be an efficient technology. We highlighted the challenges of resource allocation, scheduling, coverage, power management, and deployment. We have briefly mentioned the recent standardization efforts on wireless backhauling, especially on the \acrshort{iab} concept. We have touched upon security in wireless backhauling, highlighting the challenges specific to the multi-hop nature of small cells. Finally, we have mentioned the concepts that are researched for beyond-5G networks. These new technologies will make the wireless backhauling even stronger and place it in the beyond-5G networks' backbone.

\section*{Acknowledgment}
This study is partially supported by Turkcell Technology within the framework of 5G and Beyond Joint Graduate Support Programme coordinated by Information and Communication Technologies Authority. The authors would like to thank the Turkcell engineers and scientists for fruitful discussions on wireless backhaul.

\ifCLASSOPTIONcaptionsoff
  \newpage
\fi

\end{document}